\documentstyle[aas2pp4,epsf,astrobib]{article}
\let\oldfootsep=\footnotesep
\def\pms#1#2{$^{+#1}_{-#2}$}
\def\dgamma#1{ {d\Gamma \over d\that}({#1}) }

\def\simlt{\hbox{ \rlap{\raise 0.425ex\hbox{$<$}}\lower 0.65ex\hbox{$\sim$} }}
\def\simgt{\hbox{ \rlap{\raise 0.425ex\hbox{$>$}}\lower 0.65ex\hbox{$\sim$} }}

\def\that{{\hat t}}

\def\VEV#1{\left\langle #1\right\rangle}
\def\avethat{\VEV{\that}}

\def\pac{Paczy{\'n}ski}
\def\etal{{\it et al.}}

\def\msun{M_\odot}

\def\umin{u_{\rm min}}

\def\avethat{\VEV{\that}}

\def\Amax{A_{\rm max}}

\def\that{{\widehat t}\,} %% leave a bit of space cos the hat is big.

\def\Nexp{N_{\rm exp}}
\def\Nobs{N_{\rm obs}}

\def\leaderfill{\leaders\hbox to 1em{\hss-\hss}\hfill}
\def\etal{{\it et al.\ }}

\def\kms{{\rm\, km/s}}
\def\kpc{{\rm\, kpc}} 
\def\pc{{\rm\, pc}}

\def\ten#1{\times 10^{#1}} 
\def\msun {{ \rm \, M_\odot}} 
\def\msunpcth{ \msun  {\rm pc}^{-3}}
\def\msunpctwo{ \msun  {\rm pc}^{-2}}

\def\Amax{A_{\rm max}} 
\def\umin{u_{\rm min}} 
\def\tmax{t_{\rm max}} 
\def\eff {{\cal E}}

\def\lsim{\mathrel{\mathpalette\@versim<}}
\def\gsim{\mathrel{\mathpalette\@versim>}}
\def\spose#1{\hbox to 0pt{#1\hss}}
\def\simlt{\mathrel{\spose{\lower 3pt\hbox{$\mathchar"218$}}
     \raise 2.0pt\hbox{$\mathchar"13C$}}}
\def\simgt{\mathrel{\spose{\lower 3pt\hbox{$\mathchar"218$}}
     \raise 2.0pt\hbox{$\mathchar"13E$}}}

\begin{document}

\title{The MACHO Project: Microlensing Results from 5.7 Years of LMC
Observations}

\author{
      C.~Alcock\altaffilmark{1,2},
    R.A.~Allsman\altaffilmark{3},
      D.R.~Alves\altaffilmark{12},
    T.S.~Axelrod\altaffilmark{4},
      A.C.~Becker\altaffilmark{6},
    D.P.~Bennett\altaffilmark{10,1},
    K.H.~Cook\altaffilmark{1,2},
      N.~Dalal\altaffilmark{2,5},
    A.J.~Drake\altaffilmark{1,4},
    K.C.~Freeman\altaffilmark{4},
      M.~Geha\altaffilmark{1},
      K.~Griest\altaffilmark{2,5},
    M.J.~Lehner\altaffilmark{11},
    S.L.~Marshall\altaffilmark{1,2},
    D.~Minniti\altaffilmark{1,13},
    C.A.~Nelson\altaffilmark{1,15},
    B.A.~Peterson\altaffilmark{4},
      P.~Popowski\altaffilmark{1},
    M.R.~Pratt\altaffilmark{6},
    P.J.~Quinn\altaffilmark{14},
%    A.W.~Rodgers\altaffilmark{4},
    C.W.~Stubbs\altaffilmark{2,4,6,9},
      W.~Sutherland\altaffilmark{7},
    A.B.~Tomaney\altaffilmark{6},
      T.~Vandehei\altaffilmark{2,5},
      D.~Welch\altaffilmark{8}
	}
\begin{center}
{\bf (The MACHO Collaboration) }\\
%\lastrev   % Author and date. 
\end{center}

%%%%%% AAAAAAAAAAAA 
\begin{abstract} 
% \vspace{-10mm}
\rightskip = 0.0in plus 1em
%% Abstract in here. 

We report on our search for microlensing towards the Large Magellanic Cloud (LMC).
Analysis of 5.7 years of photometry on 11.9 million stars in the LMC 
reveals 13 -- 17 microlensing events.
A detailed treatment of our detection efficiency shows that
this is significantly more than the $\sim$ 2 to 4 events expected from 
lensing by known stellar populations.
The timescales ($\that$) of the events range from 34 to 230 days.
We estimate the microlensing optical depth towards the LMC
from events with $2 < \that < 400$ days to be
$\tau_2^{400} = 1.2 ^{+0.4}_{-0.3} \ten{-7}$, with an additional 20\% to 30\%
of systematic error.
The spatial distribution of events is mildly inconsistent with LMC/LMC disk
self-lensing, but is consistent with an extended lens 
distribution
such as a Milky Way or LMC halo.
Interpreted in the context of a Galactic dark matter halo, consisting partially
of compact objects, a maximum likelihood analysis gives a MACHO halo fraction
of 20\% for a typical halo model with a 95\% confidence interval of 8\% to 50\%.
A 100\% MACHO halo is ruled out at the 95\% C.L. for all except our most
extreme halo model.
Interpreted as a Galactic halo population,
the most likely MACHO mass is between $ 0.15 \msun$ and $ 0.9 \msun$, depending
on the halo model, and the total mass in MACHOs out to
50 kpc is found to be $9^{+4}_{-3} \ten{10}\msun$, 
independent of the halo model.
These results are marginally consistent with our previous results,
but are lower by about a factor of two. This is mostly due to Poisson
noise because with
3.4 times more exposure and increased sensitivity
to long timescale events, we did not find the expected factor of 
$\sim 4$ more events.
Besides a larger data set, this work also includes an improved efficiency
determination, improved likelihood analysis, and more thorough testing
of systematic errors, especially with respect to the treatment
of potential backgrounds to microlensing.
We note that an important source of background are supernovae in
galaxies behind the LMC.  

\end{abstract}
% \vspace{-10mm}

%-------------------- institutions and email ----------------------

\altaffiltext{1}{Lawrence Livermore National Laboratory, Livermore, CA 94550\\
    Email: {\tt alcock, dminniti, kcook, adrake, mgeha, cnelson, 
    popowski, stuart@igpp.llnl.gov}}

\altaffiltext{2}{Center for Particle Astrophysics,
    University of California, Berkeley, CA 94720}

\altaffiltext{3}{Supercomputing Facility, Australian National University,
    Canberra, ACT 0200, Australia \\
    Email: {\tt Robyn.Allsman@anu.edu.au}}

\altaffiltext{4}{Research School of Astronomy and Astrophysics, 
	Canberra, Weston Creek, ACT 2611, Australia\\
 Email: {\tt tsa, kcf, peterson@mso.anu.edu.au}}

\altaffiltext{5}{Department of Physics, University of California,
    San Diego, CA 92039\\
    Email: {\tt kgriest@ucsd.edu, endall@physics.ucsd.edu, vandehei@astrophys.ucsd.edu }}

\altaffiltext{6}{Departments of Astronomy and Physics,
    University of Washington, Seattle, WA 98195\\
    Email: {\tt becker, stubbs@astro.washington.edu}}

\altaffiltext{7}{Department of Physics, University of Oxford,
    Oxford OX1 3RH, U.K.\\
    Email: {\tt w.sutherland@physics.ox.ac.uk}}

\altaffiltext{8}{McMaster University, Hamilton, Ontario Canada L8S 4M1\\
    Email: {\tt welch@physics.mcmaster.ca}}

\altaffiltext{9}{Visiting Astronomer, Cerro Tololo Inter-American Observatory}

\altaffiltext{10}{Department of Physics, University of Notre Dame, IN 46556\\
    Email: {\tt bennett@bustard.phys.nd.edu}}

\altaffiltext{11}{Department of Physics, University of Sheffield, Sheffield S3 7RH, UK\\
    Email: {\tt m.lehner@sheffield.ac.uk}}

\altaffiltext{12}{Space Telescope Science Institute, 3700 San Martin Dr.,
    Baltimore, MD 21218\\
    Email: {\tt alves@stsci.edu}}

\altaffiltext{13}{Depto. de Astronomia, P. Universidad Catolica, Casilla 104, 
	Santiago 22, Chile\\
Email: {\tt dante@astro.puc.cl}}

\altaffiltext{14}{European Southern Observatory, Karl Schwarzchild Str.\ 2, 
	D-8574 8 G\"{a}rching bel M\"{u}nchen, Germany\\
Email: {\tt pjq@eso.org}}

\altaffiltext{15}{Department of Physics, University of California, Berkeley,
	CA 94720}

\setlength{\footnotesep}{\oldfootsep}

%%% 0000000 Marker for start of main text. 
\section{Introduction}
\label{sec-intro}

\def\yrone {A96} % abbreviation for LMC-year-1 main paper. 
\def\yrtwo {A97} % abbreviation for LMC-year-2 main paper. 

Following the suggestion of \pac\ (1986),
several groups are now engaged in searches for
dark matter in the form of massive compact halo 
objects (MACHOs) using gravitational microlensing, 
and many candidate microlensing events have been reported. 
Reviews of microlensing in this context are given by
\pac\ (1996) and Roulet \& Mollerach (1996).

Previously \cite{macho-lmc2}
we conducted an analysis of 2.1 years of photometry of 8.5 million
stars, and found 6--8 microlensing events, implying an optical depth 
towards the LMC of $2.9^{+1.4}_{-0.9} \ten{-7}$ for the 8-event sample
and $2.1^{+1.1}_{-0.7} \ten{-7}$ for the 6-event sample
(\cite{macho-lmc1,macho-lmc2}, hereafter \yrone\ \& \yrtwo, respectively). 
Interpreted as evidence for a MACHO contribution to the Milky Way dark
halo, this implied a MACHO mass out to 50 kpc of $2^{+1.2}_{-0.7} 
\ten{11}\msun$.  Depending on the halo model this meant
a MACHO halo fraction
of between 15\% and 100\%, and a typical MACHO mass of 0.1 to 1 $\msun$
\cite{macho-lmc2,gates}.  The EROS group has reported
2 candidates \cite{eros-nat,eros-followup}, consistent
with the above results.  Recently the EROS's updated and expanded survey,
EROSII, has reported two new events that they
interpret as limiting the amount of halo dark matter \cite{eros-taup99},
but are consistent with both \yrtwo\ and the results of this paper.
The OGLE \cite{ogleII} collaboration also reported one LMC microlensing 
event in 1999. 
All claimed LMC events have characteristic timescales between
$\that \sim 34$ and $230$ days,
while searches for short-timescale events with timescales
1 hour $\lesssim \that \lesssim$ 10 days have revealed no candidates to date
\cite{eros-ccd,macho-spike,macho-eros-spike}
allowing important limits to be set on
low mass dark matter.
In addition, two candidates have been observed towards
the SMC 
\cite{macho-smc1,macho-smc2,planet-smc2,eros-smc1,eros-smc2,
rhie-smc,ogle-smc,combined-smc2},
but the small number of events, location of the lenses,
and large expected SMC self lensing rate
reduce their usefulness as a probe of the dark halo.

Conclusions based on our previous work suffered due to Poisson 
error as a result of the small number of events. 
Increasing the time span monitored from 2.1 years to 5.7 years and increasing
the number of monitored fields from 22 to 30, 
gives 13 to 17 events (depending on the cuts used), as well as
greatly increases our sensitivity to long duration events and 
therefore to higher mass MACHOs.   It also increases
coverage over the face of the LMC, providing a useful tool 
to distinguish between various interpretations of the microlensing events.
In addition, while our previous analyses contained the most 
careful evaluations of microlensing detection efficiency ever done, 
we have made several important improvements, and have thoroughly
tested the robustness of our methods.  In addition we have more
fully and carefully investigated sources of potential background
to microlensing, in particular LMC variable star background
(i.e., bumpers \yrtwo) and supernovae in galaxies
behind the LMC.  For example, one candidate event classified in A97 as 
microlensing (LMC-10) is now removed as a probable background  
supernova.

The nature of microlensing implies that
many of our events will be low signal-to-noise, so to test
the robustness of our results and to estimate systematic
error due to our event selection methodology, 
we present two independently derived sets of selection criteria, 
with two corresponding sets of events and efficiency determinations.
One set is designed to select only high signal-to-noise events and is modeled
on the selection criteria used in \yrtwo.  The other is designed to be inclusive
of lower signal-to-noise events and also exotic microlensing events, 
and makes heavier use of several new statistics.
While the number of microlensing candidate events selected by the two
sets of cuts differ, the 
corresponding efficiencies compensate, 
and the resulting optical depth values, halo fractions,
etc. are essentially the same.  
This suggests that the systematic error in our optical depth, etc.
due to our choice of cuts is small.  
Finally, we implement an improved likelihood analysis that 
self-consistently 
incorporates currently available information on known stellar
backgrounds.

The increase in number of events, improved efficiency determination,
and more thorough investigation of systematic errors and backgrounds,
such as bumpers and supernovae, makes
the results of this paper the most accurate to date.
At this point,
uncertainties in the model of the Milky Way and the model of the LMC dominate
both the quantitative and interpretational aspects of microlensing 
as a probe of dark matter.

The plan of the paper is as follows:
in \S~\ref{sec-obs} we outline the 
observations and photometric reductions.  In \S~\ref{sec-det} we 
describe our microlensing event selection criteria, present the
resulting candidates, and discuss several sources of background 
to microlensing, including bumpers and supernovae.
In \S~\ref{sec-eff} we estimate
our detection efficiency, which has been improved in a number of ways. 
In \S~\ref{sec-dist} we show the distributions of the selected
events 
%in luminosity, 
in the 
color-magnitude diagram, location on the sky, 
%photometric centroid offsets, 
and impact parameter.  We compare with
predicted distributions, thereby testing the microlensing hypothesis.
In \S~\ref{sec-implic} we provide various analyses of the sample.
We calculate the optical depth, and discuss why it is a factor of two
smaller than in \yrtwo.   We perform a likelihood analysis that explicitly
includes models of the Milky Way and LMC stellar populations to
find new estimates of the MACHO contribution to the dark halo and 
the number of expected events from known stellar populations. 
We find new favorable mass ranges for the lenses if they are halo objects.  
We also
discuss various interpretations of our results, including the possibility
that no MACHOs exist and all the lensing is due to stellar lenses.
 
Note that many of the reduction and analysis procedures used here are
very similar to those in \yrone\ \& \yrtwo, to which we refer extensively. 
A more rigorous description of our detection efficiency, which is only
briefly outlined in this paper, may be found in the companion
paper, Alcock \etal\ (2000a), and in Vandehei (2000). 
The reader is encouraged to consult these papers to understand the details
of the experiment, but we will repeat the main points here for clarity.  
 
\section{Observations and Photometric Reductions} 
\label{sec-obs}

The MACHO Project has had full-time use of the $1.27$-meter telescope at 
Mount Stromlo Observatory, Australia, since 1992 July.  Observations
are scheduled to be completed at the end of 1999 December.
Details of the telescope system are given by Hart \etal (1996)
and of the camera system by Stubbs \etal (1993) and Marshall \etal (1994). 
Briefly, corrective optics and a dichroic are used to 
give simultaneous imaging of a $42 \times 42$ arcmin$^2$ field 
in two colors, using eight $2048^2$ pixel CCDs. 
As of 1998 March, over 70,000 exposures had been taken with the system, 
over 5 TBytes of raw image data.  About $55\%$ are of the LMC, 
the rest are of fields in the Galactic center and SMC. 

In this paper, we consider the first 5.7 years of data from 
30 well-sampled fields, located in the central $5^{\circ} \times 3^{\circ}$ 
of the LMC; field centers are listed in Table~\ref{tab-fields}, 
and shown in Figure~\ref{fig-lmc}. 

The observations described here comprise 21,570 images 
distributed over the 30 fields.  These include most of
our observations of these fields in the time span of 2067 days from
1992 September 18 to 1998 March 17 as well as a fraction of our observations
taken between 1992 July 22 and 1992 August 23 when our system was still in
an engineering phase.
The mean number of exposures per field is $21570/30 = 719$,  
with a range from 180 to 1338.  The sampling varies  
between fields (Table~\ref{tab-fields}), since the higher priority fields
were often observed twice per night with an average of about 4 hours
between exposures.

The photometric reduction procedure was very similar to 
that described in \yrone\ \& \yrtwo; briefly, a good-quality image of each 
field is chosen as a template and used to generate a list 
of stellar positions and magnitudes.  The templates are used to
``warm-start'' all subsequent photometric reductions, and for each
star we record information on the flux, an error estimate, 
the object type, the $\chi^2$ of the Point Spread Function (PSF) fit, 
a crowding parameter,
a local sky level, and the fraction of the star's flux rejected due
to bad pixels and cosmic rays.  
Details of the MACHO image data and photometry code (SoDOPHOT) are
provided in Alcock \etal\ (1999c).
The resulting data are reorganized into
lightcurves, and searched for variable stars and microlensing events. 
The LMC 5.7-year photometry database is about 200 Gbytes in size.

For the 22 fields reported on in \yrtwo, 
we have well calibrated photometry \cite{macho-cal},
but for the eight new fields our photometry has been only roughly
calibrated on a global basis (see \S~2 in \yrtwo).  Event
selection is generally based on this rough calibration, but as noted 
below, we 
report the well calibrated magnitudes and colors when possible.

We have corrected a minor complication in \yrtwo\ where, for
software-related reasons, we had used different templates 
for the first and second year's reductions of 6 of our
fields\footnote{Fields 1, 7, 9, 77, 78, \& 79}.
For these fields there was not a one-to-one correspondence between the set
of stars in the 2 distinct years used, and the first and second years
had to be analyzed separately.  All photometry in these 6 fields
have been re-run using the new generation of templates and
the lightcurves have now been merged onto a common photometric system.

%%%%%%% SSSSSSS
\section{Event Detection}
\label{sec-det}

The data set used here consists of about 256 billion individual photometric
measurements.  Discriminating genuine microlensing from stellar
variability, background, and systematic photometry errors 
is hard,  
and the significance of the results depends upon the event selection
criteria.

The selection criteria should accept `true' microlensing events,
and reject events due to intrinsic stellar variability and 
instrumental effects.
The determination of our event selection criteria could not be made
before looking in detail at the lightcurves.
We had to 
discover various background sources and learn how to perform
event selection from the data we gather itself,
making the selection criteria dependent
on the data.  As much as possible we have tried to base the selection
criteria upon our Monte Carlo artificial events (see \S~\ref{sec-eff}).
Although this allows us to place cuts along natural breaks in
parameter space (which lessens the sensitivity of the final results
on the exact placement of the cut) it did not allow us to fully explore
the background of variable stars.
This adds some subjectivity to our analysis, which we quantify below
by considering two limiting cases.
 
For each lightcurve, we compute a set of over 150
temporal variability statistics.
We use two levels of statistics:  level-1 statistics are calculated for
all stars, while 
level-2 statistics are calculated only for those stars that
pass the level-1 selection criteria.
We have developed selection criteria (``cuts") that use the 
level-2 statistics to distinguish microlensing from backgrounds such 
as variable stars and noise.
The selection criteria have evolved over the course of the experiment.
As the volume of data on a lightcurve increases, 
the meaning of some statistics change in subtle ways. 
Thus one must be careful not to blindly apply selection criteria from 
one data set to another.
For example, a fit $\chi^2$ to a constant flux star over 2.1 years of
data will not necessarily be the same when computed using 5.7 years of data,
due to changes in weather patterns (and thus seeing \& sky level) 
and the CCD camera over the course
of the experiment.  In addition, in order to increase our sensitivity to 
low signal-to-noise and exotic microlensing, our level 1 criteria (see below) 
have been loosened relative to those used in \yrtwo.
This means that our set of level-1 candidates contains more
variable stars and other noisy events, thus requiring changes to the
final level-2 selection criteria.
Because of the changes in the level-1 criteria, the
selection criteria used in \yrtwo\ are no longer appropriate for the
year 5.7 data.  About 45 lightcurves, 26 of which are clearly noise
or variable stars, would pass the \yrtwo\ criteria applied to the current
data.
%\machonote{Internal Note: 19 of the 45 were good microlensing candidates.  
%Of the 26 `bad' events that pass the A97 cuts, 7 are variables stars, 7 are
%seeing/crowding induced events, and 12 are the strange $\tmax < 310$ events
%who's cause is still more or less a mystery?  Tim thinks these are caused 
%by objects that lie near the edge of template/CCD (?) boundaries and have
%strange photometry performed on them during telescope slips.  Dave B. seems
%to disagree?}

We select the events using two different
and independently developed sets of level-2 selection criteria.  This allows
us to explore possible systematic error due to the choice of cuts.
While the goal of both sets of cuts is to select as many microlensing
lightcurves as possible, while rejecting as many non-microlensing lightcurves
as possible, the two sets of cuts were explicitly developed with 
complementary philosophies in mind.
The first selection criteria (hereafter referred to as criteria A)
was designed to be rather tight, only accepting events with a 
single highly significant bump in either passband, while requiring
the baseline to remain very flat, as expected in simple microlensing. 
These cuts resemble those in \yrtwo\ and for the most part statistics
similar to those described in \yrtwo\ were employed.  The second selection
criteria (criteria B) was designed to be rather loose, in an attempt to search
for exotic or low signal-to-noise microlensing candidates.  This second set
of cuts also looked for a single significant bump in either passband with
a flat baseline, but made use of some new statistics not
available in \yrtwo.  The new statistics (described below) better
characterize and filter out some variable stars and noisy events. 
To conservatively estimate the subjective
nature of our event selection, marginal events suspected of being 
supernova, etc. are preferentially rejected from set A, but
kept in set B.

Note that as long as the experiment's event
detection efficiency is calculated properly,
and the selection criteria
are sufficiently stringent to accept only real microlensing events,
changes in the selection criteria should be accounted for in 
the efficiency calculations, and 
the details should not greatly affect the final results
(in the limit of a large number of detected events).
This statement implicitly ignores exotic microlensing events such as
binary lens or parallax microlensing events which have lightcurves that
differ from those used in our efficiency determination. We have not
determined our efficiency for exotic events, but selection criteria B
is designed, in part, to be more sensitive to such events. 
Furthermore, a much
more sensitive search for exotic lensing events has been carried out, and
none were found.  So, we do expect that
the difference between selection criteria A \& B is a reasonable indication
of our selection criteria systematic error, and we find that this
difference is fairly small as
outlined here and discussed in \S~\ref{sec-implic}.

We have summarized the old (\yrtwo) and new selection criteria (A \& B) in 
Table~\ref{tab-cuts}.  In \S~\ref{sec-cuts} we briefly describe some of the 
analysis and statistics used by these selection criteria and in 
\S~\ref{sec-events} we present the set of events selected by criteria
A \& B.  In \S~\ref{sec-background} we identify two main sources of 
background to microlensing and discuss how they can be 
removed from the true microlensing pool.  Finally, in \S~\ref{sec-nevents}
we remove our identified background events and summarize the final sets of
microlensing candidates (the final set A \& B) used to compute the results
of this paper.

\subsection{Selection Criteria Statistics}
\label{sec-cuts}

Photometric
measurements with questionable
PSF fit, too much crowding, missing pixels, or cosmic rays are flagged as
suspect and removed from further consideration. 
The event detection then proceeds in two stages.  The first stage, 
defining a level-1 collection of candidate events, is similar to 
that described in \yrtwo; a set of matched filters
of timescales 7, 15, 45, and 100 days is run over each lightcurve.
If after convolution, a lightcurve shows a peak above a pre-defined
significance level in either color, it is defined as a `level-1 candidate'.
We also make use of a new filter that looks for bumps of any duration
and add these lightcurves to the level-1 pool of candidates.
(We found no additional candidates by this change, however.)
For level-1 candidates, a full 5-parameter fit to microlensing is made, 
and many level-2 statistics describing the significance
of the deviation, goodness of fit, etc. are calculated. 
We use the standard point-source, point-lens 
approximation \cite{refsdal,macho-lmc1}.
The 5 free parameters of the fit are
the baseline flux in red and blue passbands $f_{0R}$, $f_{0B}$, 
and the 3 parameters of the microlensing event: 
the minimum impact parameter in units of the Einstein radius, $\umin$, 
the Einstein diameter crossing time, $\that \equiv 2 r_E / v_{\perp}$,
and the time of maximum magnification, $\tmax$. 
Later, instead of $\umin$ we will often use
the fit maximum magnification $\Amax \equiv A(\umin)$, which is more
closely related to the observed light curve.

Lightcurves passing loose cuts on these statistics 
are defined as `level-1.5' candidates, and are output as individual files along
with their associated statistics. 
In the present analysis, there are approximately 150,000 level-1.5 candidates,
almost all of which are variable stars or noise.

Some of the important statistics used by both criteria A \& B 
are the chi-squares of various fits.  For example, a powerful
signal-to-noise statistic is $\Delta \chi^2 \equiv 
\chi^2_{\rm const} - \chi^2_{ml}$ where $\chi^2_{\rm const}$ and 
$\chi^2_{ml}$ are the $\chi^2$ values for the
constant flux and microlensing fits, respectively. 
$\Delta \chi^2$ is the effective `significance' of the event summed 
over all data points.  $\chi^2_{\rm peak}$ refers
to the $\chi^2$ of the microlensing fit in the ``peak" region where
$A_{\rm fit} > 1.1$.  A reduced $\chi^2$ of the microlensing fit outside
the interval $\tmax \pm 2\that$ is also computed, $\chi^2_{ml-out}/N_{dof}$.
Other useful statistics include the average values of crowding for a star,
the microlensing fit values, the magnitude and color, 
the number of 2-sigma high points in the peak region, 
the number of points on the rising and falling side of the peak region,
and the number of points outside the peak region.

Based upon our experience gained in \yrone\ \& \yrtwo\ we have 
developed a number of new statistics.  One such statistic is
the fraction of points in the peak that lie above the
lightcurve's median, $N_{\rm hi}/N_{\rm pk}$.  This new statistic is useful
for removing events with
spurious deviant points associated with crowding/seeing induced fluctuations, 
satellite/asteroid 
tracks and other causes.  To further help in removing similar spurious events
we also compute the fraction of points in the peak
rejected due to bad PSF measurements, $pkpsfrej$, and large crowding
values, $pkcrdrej$.

We have found it beneficial to concentrate on statistics
that help in rejecting variable star background.
One such statistic is the ratio
of powers in the two passbands, $bauto/rauto$.
The quantity $bauto$ is the sum of the absolute
values of the flux away from the median in the blue bandpass, 
and likewise for $rauto$ for the red bandpass.
Our variable
star background generally has more power in the red than in the blue,
while our Monte Carlo microlensing events tend to have equal power 
in both passbands, even when heavily blended (that this is the case is not
surprising given the fact that the color difference between any two
stars in the LMC is somewhat restricted).  A second new statistic on the 
ratio of powers in each filter, $pfwsr$ is also computed, but uses only
points measured simultaneously in both filters and is thus normalized 
differently than $bauto/rauto$.  To further help reject variable stars 
we compute a cross-correlation coefficient between the red \& blue filters,
$rbcrossout$, but use only data points that are outside the peak of
the event, $\pm\,1.0 \that$, to avoid the peak contributing.  A powerful
new statistic for rejecting variable stars is a robust reduced chi-square 
fit to a constant flux, $\chi^2_{\rm robust-out}/N_{\rm dof}$, also computed outside
the interval $\pm\,1.0 \that$.

Another new signal-to-noise statistic we compute is
$pfrdev$, which is a $\chi^2$ from the baseline counting only upward
excursions in the filter window and subtracting a penalty per measurement
in the bin so that 2-sigma points break even.  We subtract the largest 
single date contribution (1 or 2 points) to ensure robustness against
single strongly deviant observations.  We also compute a similar $\chi^2$ for
the second most significant non-overlapping filter trigger, $pfrdev2$.
The statistic $pfrdev$ is very similar to
$\Delta \chi^2/(\chi^2_{\rm ml}/N_{\rm dof})$,
and a comparison of the two on both real data and artificial data yields
similar results, even though
the latter is derived using the microlensing shape.  This gives us added
confidence that a cut on $\Delta \chi^2/(\chi^2_{\rm ml}/N_{\rm dof})$ is not
very `shape' dependent. 
We have opted to use $\Delta \chi^2/(\chi^2_{\rm ml}/N_{\rm dof})$ throughout,
but make use of the second peak significant $pfrdev2$ to ensure the 
uniqueness of the event.  This statistic is useful for eliminating 
variable stars, but must be used with caution to avoid missing
exotic lensing events such as binary lenses or sources that could exhibit
a second `bump'.  Similarly, cuts on $\chi^2_{\rm ml-out}/N_{\rm dof}$
and $\chi^2_{\rm robust-out}/N_{\rm dof}$ could potentially bias against
detecting widely separated binary microlenses.  We ran our
selection criteria with and without these cuts to ensure that no 
exotic lensing events were missed.  No additional candidates were found.  

Our set A selection criteria are designed to accept high 
quality microlensing candidates, while using mainly the statistics
described in \yrtwo.  Some of the cuts on these statistics have
been loosened as we have developed a better understanding of
our variable star background; for example the cut on magnification
was loosened from $\Amax > 1.75$ to its present value $\Amax > 1.49$,
because our main background of variable stars, a class of
blue variables called bumpers (see below), almost
never show fit magnifications larger than 1.5, and are well isolated
in the color magnitude diagram.  We also loosened our 
main significance cut $\Delta \chi^2/(\chi^2_{\rm ml}/N_{\rm dof}) < 400$ 
(from 500 in \yrtwo)
because a number of other statistics were tightened.  These include,
$\Amax > 3\overline{\sigma}$ (from $2\overline{\sigma}$ where $\bar \sigma$ is
the average red/blue error in magnitudes) which is a signal-to-noise cut,
$\chi^2_{\rm ml-out}/N_{\rm dof} < 1.8$ to reject variables (from $< 4.0$ in \yrtwo) 
and $\Delta\chi^2/(\chi^2_{\rm peak}/N_{\rm dof}) > 350$ (from $> 200$ in \yrtwo),
another signal-to-noise
cut.  This later cut is somewhat more reliant on the shape of the 5 parameter
microlensing fit and increases the likelihood of rejecting exotic 
microlensing; for this reason, criteria B below does not use this cut.

\begin{figure}
\plotone{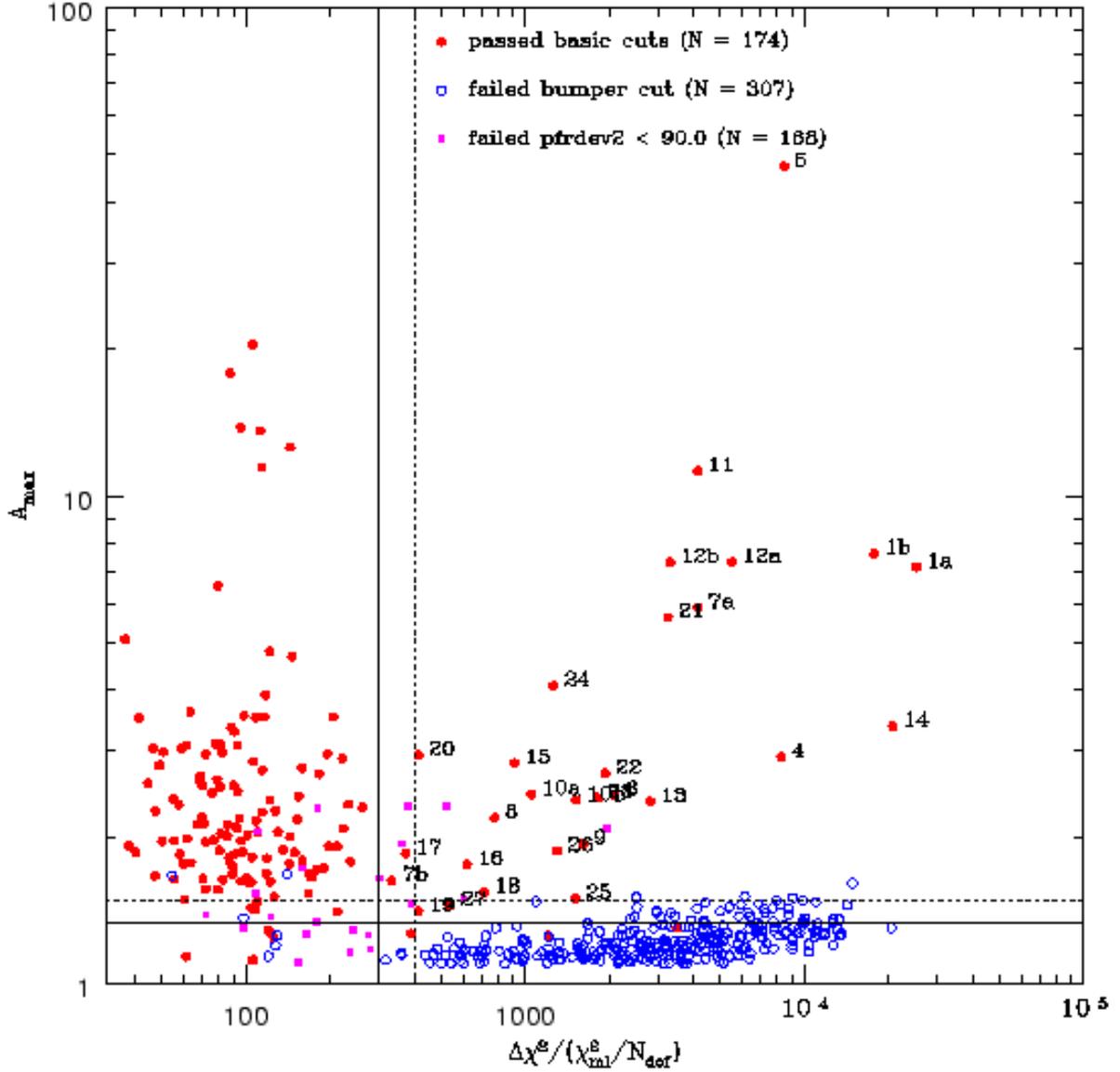}
\caption{Illustration of the cuts used to select microlensing candidates
for criteria B.  The x-axis is $\Delta\chi^2/(\chi^2_{\rm ml}/N_{\rm dof})$, where
$\Delta\chi^2 \equiv \chi^2_{\rm const} - \chi^2_{\rm ml}$ is the improvement
in $\chi^2$ between a constant brightness fit and a microlensing fit.
The y-axis is the fitted maximum magnification.  The 29 lightcurves are
shown as solid dots and are labeled.  The remaining symbols are 
explained in the figure and in detail in \S~\ref{sec-events}. 
The solid lines show the final cuts for criteria B.  The dotted line shows
the same for criteria A.  See text for details.
\label{fig-cuts} } 
\end{figure}

\begin{figure}
\plotone{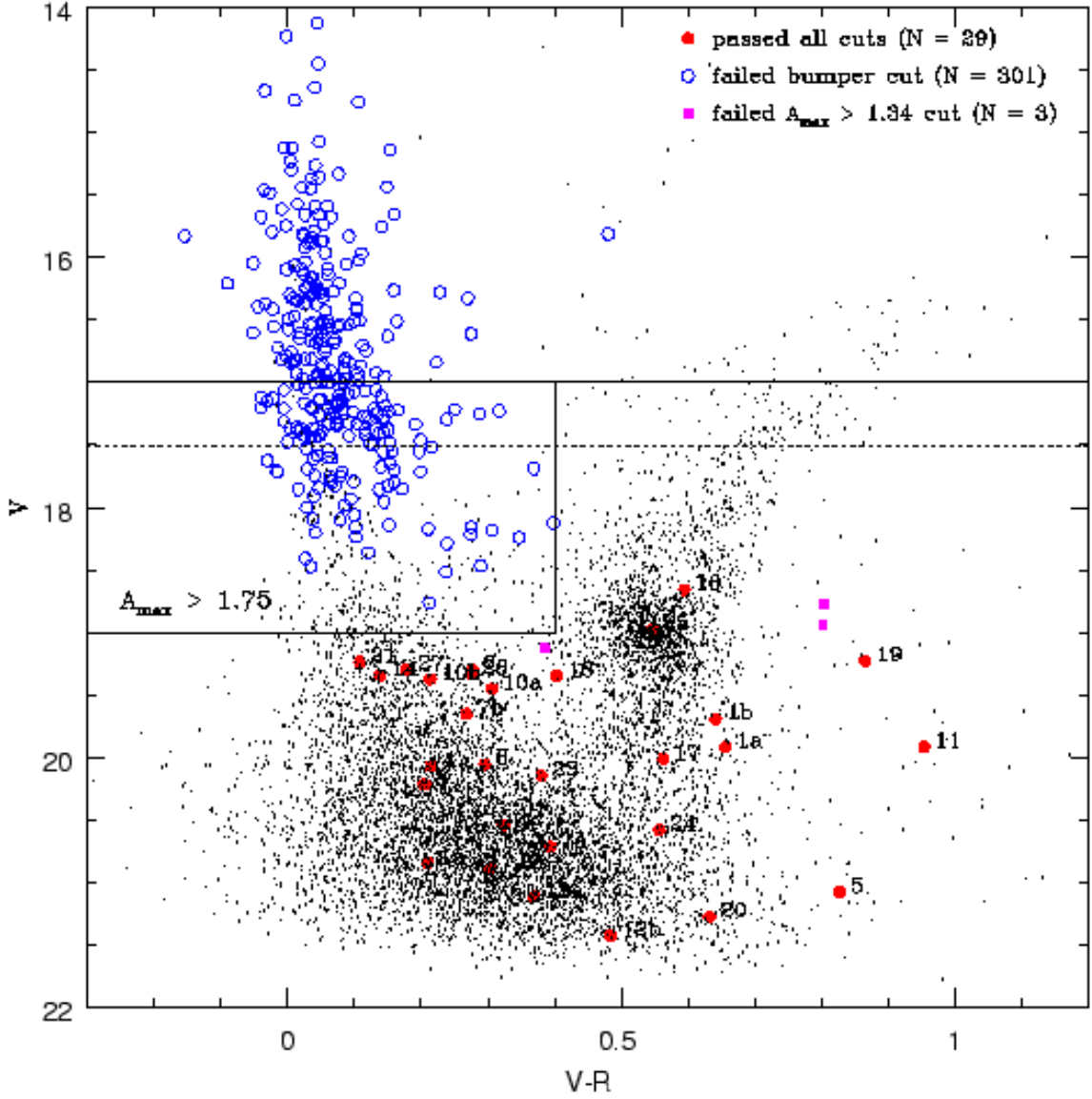}
\caption{Illustration of the cuts used to select microlensing candidates
for criteria B in the color--magnitude diagram.  The 29 lightcurves are
shown as solid dots and are labeled.  The `bumper' cut is outlined
with solid lines and labeled with `$\Amax > 1.75$' for criteria B.  
The dotted lines shows the same for criteria A.
The symbols are explained in the figure and in detail in \S~\ref{sec-bumpers}. 
The magnitudes displayed here use the rough global calibrations
\S~\ref{sec-obs}.
\label{fig-cmd-cuts} } 
\end{figure}

Our set B selection criteria are designed to accept any lightcurves
with a significant unique peak and a fairly flat baseline.
Our selection criteria B are summarized in Figure~\ref{fig-cuts} which 
illustrates our two most important signal-to-noise cuts: the cut
on magnification $\Amax$ and the cut on 
$\Delta\chi^2/(\chi^2_{\rm peak}/N_{\rm dof})$. Events that passed the
basic cuts (all cuts used by criteria B minus the ordinate \& abscissa
cuts of Figure~\ref{fig-cuts}) are shown as solid dots and labeled. 
The final cut on magnification
$\Amax$ and $\Delta\chi^2/(\chi^2_{\rm peak}/N_{\rm dof})$ are shown as solid
lines (for comparison dotted lines for criteria A are also illustrated).
Open circles indicate events which fail criteria B's bumper cut and
solid squares events which fail criteria B's uniqueness cut.
The new statistics on the number of points rejected in the peak region,
$pkpsfrej + pkcrdrej$, and the fraction of points above the baseline
in the peak, $N_{\rm hi}/N_{\rm pk}$ are useful for eliminating spurious
noise-induced events.  New statistics on uniqueness,
$pfrdev2$, the passband power ratio $pfwsr$, the red/blue cross-correlation
coefficient outside the peak $rbcrossout$, and the robust
$\chi^2_{\rm robust-out}/N_{\rm dof}$ baseline statistic are useful in removing
periodic and quasi-periodic variable stars.
We have also more carefully characterized our main source of
variable star background, the bumpers, in
a magnitude--color--magnification space (see Figure~\ref{fig-cmd-cuts}).
With the background of variable stars more effectively removed 
we can reduce both our reliance on any `shape' dependent
criteria %(such as $\Delta\chi^2/(\chi^2_{\rm peak}/N_{\rm dof})$)
and lower the significance level of a detection.  In 
\S~\ref{sec-eff} we demonstrate the relative looseness of 
criteria B over A as well as the decreased dependence
on shape.  However, one potential difficulty with
criteria B is its inability to discriminate against some types of
variable stars, such as CVs and SNs, that might exhibit strongly
asymmetric and/or chromatic lightcurves during the `event' but remain
constant for long periods of time.  Supernova removal is discussed separately
in more detail below.
See Table~\ref{tab-cuts} for a complete list of the
individual cuts used by selection criteria A and B.

\subsection{Microlensing Candidates} 
\label{sec-events}

\begin{figure}
\plotfiddle{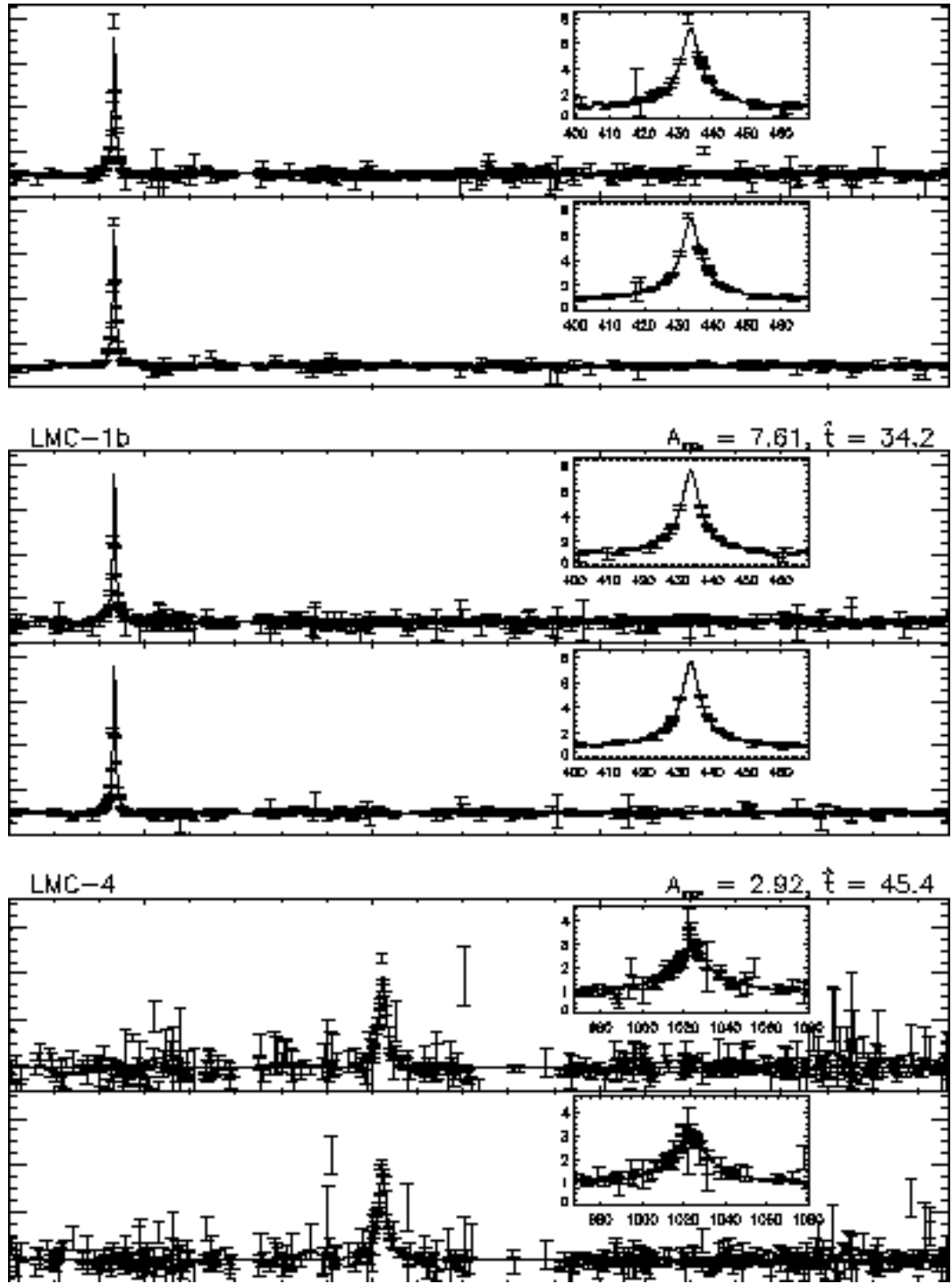}{8.0truein}{0}{100}{100}{-200}{50} 
%\plotone{fig_event_a.eps}
\caption{The lightcurves for the 29 candidates (25 stars) in 
\S~\protect\ref{sec-events}.
For each object, the upper and lower panels show red and blue passbands. 
Flux is in linear units with $1\sigma$ estimated errors, 
normalized to the fitted unlensed brightness.  
Full lightcurves are shown with 2 day binning, insets
of the event regions are unbinned.
The full lightcurves can be found on the World Wide Web at 
{\tt{ http://wwwmacho.anu.edu.au/}}
The thick line
is the fit to unblended microlensing (Table~\ref{tab-blend}), except
for probable supernovae where both the blended fit (solid line)
and type Ia fit (dashed line) are shown. 
%\machonote{Currently the fits displayed here are the standard fits.
%Perhaps we should show the blend fits instead?  This would be a more fair
%comparison with the SN fits.}
\label{fig-events} }
\end{figure}

We find 19 lightcurves that pass criteria A and 29 lightcurves that
pass criteria B (before applying the supernova cuts described below).  
All the lightcurves passing criteria A also pass criteria B.
The 29 lightcurves are shown in Figure~\ref{fig-events} and 
their microlensing fit parameters are listed in Table~\ref{tab-events}. 
Events that do not pass criteria A are marked with an asterisk.
Parameters for fits including the possibility of blending with an
unlensed star in the same seeing disk as the lensed star are given
in Table~\ref{tab-blend}.  The unblended fits 
are displayed as a thick
line in Figure~\ref{fig-events},   
%\machonote{We should probably display the blend fits in Marla's lc's so
%as to more fairly compare these against the SNIa fits.}
used for all
statistics except comparison with supernova. Note that our events here are numbered 
as in \yrtwo\ to avoid any possible ambiguity.  Thus the
first event described here that was not described in \yrtwo\ is event~13.
Finding charts for the events, as well as the full lightcurves can be
found on the internet at {\tt{http://wwwmacho.anu.edu.au/}}
\footnote{A mirror site exists at {\tt{http://wwwmacho.mcmaster.ca/}} and the
site for microlensing alerts is 
{\tt{http://darkstar.astro.washington.edu.}}}

Six of these 29 lightcurves (1a, 1b, 10a, 10b,
12a and 12b) actually correspond to only three stars that occur in field 
overlap regions; the two lightcurves for each star are based on independent
data and reductions.  Two lightcurves are also of the same star 
(7a and 7b) but were not in field overlaps.  Event~7 was bright enough
and in a locally crowded enough region that some of the flux from the primary
(7a) contaminated a secondary (7b) neighbor causing a spurious 
detection.  Event~7a passed both criteria A \& B, while event~7b passed only
criteria B due to its low signal-to-noise.  Thus before supernova removal
there are 16 unique events
found by the criteria A, and 25 unique events found by the criteria B.

\subsection{Background} 
\label{sec-background}

\subsubsection{Bumpers} 
\label{sec-bumpers}

As noted in \yrtwo\ a potential source of
background to microlensing is a class of bright blue variables which we
refer to as bumpers.  Although associated with Be stars, which are known
to show periodic outbursts in our Galaxy, the true nature of these
variable stars is still unknown.  However, it is possible to
eliminate bumpers as a serious source of background as they can be
well isolated in a multi-parameter space.  For example, microlensing
fits to bumpers seen
in our data almost never return magnifications larger than 1.5,
typically much less, as can be seen in Figure~\ref{fig-cuts} (open circles). 

One does not have to be confined to magnifications
above 1.5 as the bumpers are also well isolated in the color--magnitude
diagram (CMD) as illustrated in Figure~\ref{fig-cmd-cuts}.
Here a typical CMD of the LMC is shown with
a scattering of small dots.  The 29 lightcurves that pass selection
criteria B are shown as filled circles and labeled.  As in 
Figure~\ref{fig-cuts} open circles indicate events which fail 
criteria B's bumper cut.  Filled boxes indicate events which fail
criteria B's magnification cut $\Amax > 1.34$.
The final cut on brightness $V > 17$ and 
color--magnitude--magnification (the `bumper' cut) are marked 
as solid lines for criteria B (for comparison the dotted line 
illustrates the `bumper' cut for criteria A).
If a potential event falls within the boxed region labeled 
`$\Amax > 1.75$' then it must have a magnification greater then 
1.75 to be included in the criteria B set of events. 
We have visually inspected the $\sim\!300$ bumper candidate lightcurves
and confirm that the vast majority of them show the slight
asymmetry and other characteristics 
typical of the bumpers described in \yrone.
In fact,
due to the cut on uniqueness these bumpers have only single bumps
and evidently represent one-time only bumpers or bumpers with inter-bump
intervals longer than 5.7 years.  

\subsubsection{Supernovae} 
\label{sec-sn}

Another serious source of potential contamination in microlensing
surveys, that has not been given sufficient attention before,
are supernovae (SN) occurring in galaxies behind the LMC.  These
background SN are picked up in the crowded fields, and their host galaxies 
are not always easy to
identify in ground--based images.  The fact that they 
occur only once and show a flat baseline before and after the
`event' make SN interlopers a serious concern.

A first step in understanding this source of contamination
is to estimate the number of SN we might see during the course
of the experiment.  Because of the recent interest in SN type Ia
as standard candles, the rate of SN (both type I and II) occurring
in field galaxies is now fairly well known.  
We use a typical rate of 0.5 SN/year/$\Box^{\circ}$ with peak
magnitude brighter than $V\sim\!20$
\cite{woods}.  
The duration of the experiment
is 5.7 years and covers 13.5 $\Box^{\circ}$, which suggests we should have 
approximately 38 SN in our data set.  This does not include
our SN detection efficiency and so is an overestimate.
We expect our efficiency for
detecting SN to be on the order of 5-15\%, similar to that of detecting
microlensing events (see \S~\ref{sec-eff}), due to the similar shapes of
the corresponding lightcurves, 
implying we are likely to see $\sim\!2-6$ SN in the 
current data set.  
%This number is quite uncertain, and could be a 
%factor of 2-3 larger or smaller.

If the density of galaxies behind the LMC is average, then the probability
of finding a galaxy in proximity of a star should be low. 
Therefore, a robust way of eliminating potential SN 
interlopers would be a search for a background host galaxy in an image.
If such a galaxy were found within a pre-determined radius, the microlensing 
interpretation would be unlikely. Since the area covered
by nearby galaxies is small, the correction to our experimental efficiency
would be small.
Unfortunately, our ground--based
images do not have the resolution and low enough sky count to perform
such a search with confidence.  We do have HST observations of 8
of our events (see Table~\ref{tab-summary}) and these give us 
high confidence that 6 of them are not SN as there are no obvious
background galaxies anywhere near these events.  
Two events for which we have an 
HST image (events~10 and 12) do show a fairly bright spiral 
galaxy within 5 arcsecs.  In retrospect, the host galaxy for 
event~10 shows up in our ground--based images but could
not have been recognized as a galaxy without further data.  This
event was noticeably asymmetric 
in \yrtwo\ and was classified as a marginal microlensing candidate.
In calculating the optical depth, it was rejected from the 
6 event sample, but included in the 8 event sample.
Although the effect of its inclusion was small, increasing the reported
optical depth of the 8 event sample by only 7\%, its presence 
underscores the need to take the potential SN contamination seriously.

Unfortunately, with an incomplete sample of HST images we are unable to
follow the above prescription to reject SN interlopers completely.
We have requested HST time to observe the remaining events, and have examined
the best available ground based images for evidence of a background galaxy.
However, we can also use the shape of the 
lightcurve to help distinguish SN from microlensing.
Even though the lightcurves of SN type II are not well understood and
exhibit a wide range of behavior,
SN of Type Ia have been studied in detail.  Their lightcurves
are very similar once distance, reddening, and a shape parameter
are allowed for \cite{phillips-snIa,riess-snIa}.  
In addition, type Ia's
probably dominate the SN rate in flux limited 
samples \cite{woods}.
Using the SN type Ia templates of Riess, Press, \& Kirshner (1996) transformed
to the MACHO photometric system,
we have applied a
6 parameter SN type Ia fit to all 29 lightcurves.
The 6 free parameters of the fit are the baseline flux of the photometered
object in red and 
blue passbands $f_{0R}$, $f_{0V}$, and 4 parameters that describe the
SN type Ia lightcurve: the time of peak $t_{\rm peak}$, the distance 
modulus in red band $\mu_{R}$ and blue band $\mu_{V}$ (fit independently
to account for the possibility of reddening) and a shape parameter
$\delta$ which parameterizes how SN of type Ia become longer in duration
when intrinsically brighter.  
Emperically, we discovered that by allowing any acceptable value of $\delta$, 
in many cases our best fit SN shape was well outside the range of observed
SN (e.g. $\delta = 5$).  This was especially true for
high-quality microlensing events, where the SN template
provides a poor fit to the shape of the lightcurve.  We
therefore limit our range in $\delta$ to be between -0.5 and
0.75, and note that for most of the events with poor SN fits,
we find $\delta$ pegged at one of these values.
%These SN type Ia fits are displayed
%as a thin line in Figure~\ref{fig-events} for the 29 events and the fit
%parameters are given in Table~\ref{tab-sn}.  
The SN type Ia fits are displayed 
as a dashed line for the events we categorize (see below) as SN
in Figure~\ref{fig-events} and the fit parameters for all 29 
lightcurves are given in Table~\ref{tab-sn}.  
%We note that in many cases the
%SN type Ia fits do not look very physical for typical Ia's.  This
%possible source of confusion is the result of trying to fit a SN 
%type Ia template to something that is not a SN type Ia lightcurve.  In
%these cases the shape parameter $\delta$ runs into one of the walls
%that physically constrains it ($-0.5\,\leq\,\delta\,\leq\,0.75$
%in our passbands) to lie in the region where real SN type Ia lightcurves
%are actually observed to reside in.}

We summarize all the relevant available
information for the 29 events in Table~\ref{tab-summary}.  Columns 
2-4 compare the blended microlensing fits $\chi^2_{\rm ML}/N_{\rm dof}$
with the SN type Ia fits $\chi^2_{\rm SN}/N_{\rm dof}$.
A positive value of 
$\Delta\chi^2_{\rm SN-ML}$ indicates a better fit to blended microlensing,
while a negative value a better fit to SN type Ia.  Inspection
of Table~\ref{tab-summary} reveals 10 lightcurves (8 events) that are
better fit by SN type Ia.  As a consistency check column 5 indicates
the presence or absence of an obvious background galaxy within $\sim\!10-15$
arcsecs as determined using the best available image of the event (image
source given).  This background galaxy identification is
subjective.  Part of the problem is the severe crowding of the 
ground-based images used (MACHO or CTIO images).
In every case, 
where a galaxy is probably present the
fit to a SN type Ia template is preferred over the blended fit, giving
us some assurances of the overall correctness of the procedure.  We 
have no spectra to confirm the hypotheses that
any of these 8 events are, in fact, SN of type Ia.  But given
the fact that we should see 2--6 SN in our survey we feel it is nevertheless
prudent to eliminate these 8 events as potential
interlopers.  We thus implement as our final cut (on both selection
criteria A \& B) the requirement that the blended microlensing fit be
preferred over the SN type Ia fit, or $\Delta\chi^2_{\rm SN-ML} > 0.0$.
This eliminates events~10 and 12 from set A and events~10, 11, 12, 16,
17, 19, 24, and 26 from set B.  
Event 22, clearly not a type Ia, 
is a special case and will be discussed later.
The advantage of using this simple cut
is that we may quantify the effect it has on the detection 
efficiency (see \S~\ref{sec-eff}).  The effect is 
negligible since less then 0.7\% of artificial standard
microlensing events are falsely rejected by this cut.

There are several important caveats to using the SN Ia fits to 
reject potential SN interlopers.
First, both microlensing
and SN come in a variety of flavors.  Although type Ia have well
defined lightcurves, other types of SN are not so well behaved
(types Ib, Ic, IIp, IIl, etc.) and can come in a variety of durations
and asymmetries.  We note here that two of our SN interlopers 
(events~11 and 24) are likely to be SN type IIp as judged by the plateau
seen in both passbands about 25-50 days after maximum 
(never seen in both passbands for type Ia).
However, even these lightcurves are usually better fit
by SN type Ia than blended microlensing, mostly due to the asymmetry
that SN lightcurves typically exhibit.

Second, exotic microlensing
such as found in binary lenses or parallax events, could mimic the
asymmetry of type Ia SN.
There are several reasons why we believe that this is not a major problem:
1) event 9, a binary lens event, is
better fit by a single lens microlensing lightcurve than a SN Ia lightcurve,
2) out of many microlensing events towards the Galactic
bulge less than 10\% are of clearly exotic type
\cite{macho-binary}\footnote{However, the lens populations towards the LMC may
be different than that towards the bulge.},
3) exotic microlensing should show the ``wrong'' sign of asymmetry
50\% of the time, and we have no examples of this among our events.

However, two events
do stand out as potentially worrisome.  Event 26
is better fit by SN type Ia, but does not show a clear background
galaxy in our deepest CTIO 0.9m images. It could be an example of
exotic lensing.  For purposes of this paper,
we reject this as a potential SN interloper,
but must await a better image before making any definitive
conclusions on this event.  
Event~22 is our longest duration event and is clearly asymmetric to the eye.  
However, even with the clear asymmetry this event is better fit
with blended microlensing than with a SN type Ia, mainly because type
Ia's are not observed to last this long.  
The asymmetry of this event is well fit by microlensing
parallax, which would be a natural
explanation given the very long duration of this event.
There is also no strong evidence of a galaxy in ground-based images,
though the object is slightly extended.   
%Thus while microlensing
% parallax is a good explanation for this event, it is
Microlensing parallax is
a good explanation for this event, and if this interpretation is
correct, it appears that the lens is probably a white dwarf in a
flattened halo or thick disk population\rlap.\cite{macho-lmc22}
On the other hand, it is
also possible that event~22
belongs to a class of SN (``slow" type IIn) similar to
SN 1988Z \cite{snIIn,schlegel,stathakis}.
Therefore, in the spirit of our selection criteria, we reject event 22
from the exclusive set A and keep it in the inclusive set B.
Since event~22 is our longest duration event and therefore contributes
maximally to the optical depth, this is also the conservative
approach to exploring the sensitivity of our results to
the selection criteria.

\subsection{How Many Events?}
\label{sec-nevents}

Here we briefly summarize the events that will constitute
set A \& set B.  The main results of this paper 
rest upon these two sets of candidate microlensing events.

Criteria A selected 19 lightcurves corresponding
to 16 unique events (events~1, 10, \& 12 in field overlaps).  
Of these 16 unique events two (events ~10 \& 12) are rejected
as SN interlopers due to the final SN cut $\Delta\chi^2_{\rm SN-ML} > 0.0$
and an unexplained event~22 is rejected 
in the spirit of criteria A being exclusive.  This leaves
set A containing 13 events: ~1, 4-8, 13-15, 18, 21, 23,
\& 25. 

Criteria B selected 29 lightcurves corresponding
to 25 unique events (events~1, 10, \& 12 in field overlaps
and event~7 duplicated via contamination).  
Of these 25 unique events 8 (events~10, 11, 12, 16, 17, 19, 24, \& 26)
are rejected as SN interlopers. Here, we do not reject 
event~22 but leave it in the set as a
potential exotic lensing event.
As a result, set B contains 17 
events: ~1, 4-9, 13-15, 18, 20-23, 25, \& 27.

%%%% SSSSSSSS
\section{ Detection Efficiency } 
\label{sec-eff}

The detection probability for individual events 
depends on many factors, e.g. the 3 event parameters $\Amax$, $\that$, 
$\tmax$, and the unlensed stellar magnitude, as well as our
observing strategy and weather conditions.  Such a complicated
dependence is most naturally found with a Monte Carlo technique.
We may simplify the dependence to some extent by averaging over the known 
distributions in $\Amax$, $\tmax$, the stellar magnitudes, 
and the known time-sampling and weather conditions, to derive our 
efficiency as a function only of event timescale, $\eff(\that)$. 

We have computed our detection efficiency using a method similar
to that outlined in \yrtwo, but with a number of improvements.
A full discussion of the method, with detailed results, is given
in Alcock \etal\ (2000a) and Vandehei (2000).  Briefly, we generate simulated 
microlensing events with $\that$ logarithmically distributed
in the range 1--2000 days over the slightly wider time interval, $({\rm
JD}-2,448,623.5) = 190.0$ to 2277.0, and add these simulated events
into the extended time span of observations.  A large database of 
artificial star tests is used to simulate the effects of blending.
A number of systematic photometric effects,
including the response of flux, error bars, and the photometric
flags outlined in \S~\ref{sec-obs} are also included.  The Monte-Carlo procedure
takes into account the actual spacing and error bars of the
observations, so any variations in sampling frequency, weather, seeing,
etc.~between the \yrtwo\ data set and the current data set are automatically
accounted for.

One of the primary shortcomings of the efficiency analysis presented 
in \yrtwo\ \footnote{We consider here only the `photometric' efficiency
defined in \yrtwo.  The `sampling' efficiency, also described in \yrtwo\,
is of little value for the discussion of this paper.
`Sampling' efficiencies
for the present analysis may be found in Alcock \etal\ (2000a).} 
was a lack of faint `stars' in the artificial star
tests, which are used to add simulated events.  
In the current analysis, a large 
number of faint stars down to $V\sim\,24.5$ are used.  This is 2.5
magnitudes fainter than in \yrtwo. 
It is also 2.5 magnitudes fainter 
than our faintest detected objects.  The present analysis also makes
use of a much larger database of artificial stars ($\times5$) sampled
over a larger ($\times15$) and more fairly distributed set of observing 
conditions (stellar density, seeing, \& sky).  
Another major improvement,
not fully recognized as a major source of uncertainty
in \yrtwo, is a normalization of our fields to the underlying
luminosity function.  A main issue in efficiency determination is the
distinction between ``objects" and stars.  Objects are flux concentrations
identified by the photometry code as stellar-like objects.  Each object
is typically a blend of many underlying LMC stars, any one of which can
undergo microlensing, and it is important to identify the correct density of
underlying LMC stars in each of our fields.

The present efficiency is based on all stars in our fields, even
those not uniquely identified because of signal-to-noise or crowding effects.
These are accounted for by integrating the detection efficiency per 
star over the true underlying luminosity function (LF) in the LMC. 
Our LF in the LMC is derived from a combination of ground-based
MACHO photometry (for stars with $V < 20$) and multiple HST WFPC2
fields in the LMC bar (for stars with $V > 20$) and is described
in Alcock \etal\ (2000a) and Alcock \etal\ (1999c).  
The shape of the LF appears universal in
most of our fields and is well constrained for $V < 22$. But, any 
reasonable deviation from the adopted shape down to $V\sim\,24$ 
has little effect on our efficiency \cite{macho-eff}. 

An important and yet uncertain factor is the normalization of the
LF in each of our fields, which determines our effective sensitivity
or exposure in star-years.  Because our exposure in {\it object-years}
is well known and the efficiency should converge at some magnitude
(as fainter stars contribute less and less) we have chosen to factor
this normalization into the efficiency.  Thus our efficiency is properly
defined as $\eff(\that)\,=\,(S/O)(V<24)\eff_{\rm stars}(V<24,\that)$, where
$\eff_{\rm stars}(V<24,\that)$ is the efficiency per star to $V = 24$
and $(S/O)(V<24)$ is the normalization, or the number of stars per object
(defined as the ratio of the true number of stars with $V<24$ to the
number of SoDoPhot objects).
The value of $(S/O)(V<24)$ is  
$10.84 \pm 2.4$ stars per object and represents a weighted average
over all 30 fields.  The limiting magnitude $V\,=\,24$ was chosen because
the efficiency for durations $\that < 300\,$days converges.
The efficiency for longer durations does not
converge by $V\,=\,24$, and thus these are likely underestimated,
with the underestimation becoming worse for longer duration events.
The exact point and speed of convergence is sensitive to the assumed
shape of the LF and the cuts used, with
criteria A converging more rapidly (see \cite{macho-eff}).  Since none of
our 13-17~events have durations longer than 300 days our efficiency 
determination 
for them is sound.  The uncertainty in the efficiency is dominated by the
uncertainty in the normalization, which we estimate to be 
$\approx\,20\%$ (see \cite{macho-eff} for a more detailed error budget).

\begin{figure}
\epsscale{0.9}  % 0.9 can often fit 4 to a page
\plotone{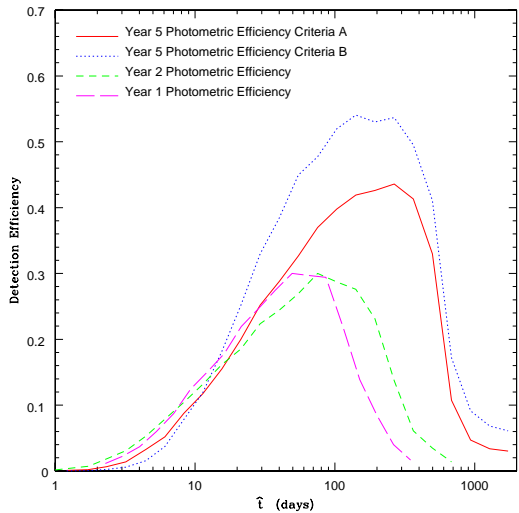}
\caption{ Microlensing detection efficiency (normalized to $\umin < 1$)
for the year 5.7 MACHO data, as a function of event timescale $\that$.
The \textit{solid line} shows the `photometric' efficiency computed for cut A,
and the \textit{dotted line} for cut B as described in \S~\protect\ref{sec-eff}.
For comparison the corresponding curves from year-1 (\yrone) and
year-2 (\yrtwo) are also shown.
\label{fig-eff} } 
\end{figure}

Efficiency results are shown in Figure~\ref{fig-eff}. 
Selection criteria A is shown as a {\it solid} line and criteria
B as a {\it dotted} line.  Also shown for comparison are the 
efficiencies presented in \yrone\ (\textit{long dash}) and
\yrtwo\ (\textit{short dash}).  Note that the efficiencies presented
here have been scaled by a normalization term that accounts
for the increased exposure due to all stars in our fields, down
to $V\,=\,24$ (as described above).  Strictly speaking, the efficiency
defined in this manner is not constrained to lie below one, though in practice
it always does.
This efficiency is defined relative to an `exposure' of 
$E = 6.12 \ten{7}$ object-years, which arises as follows: 
there are 11.9 million lightcurves in our total sample,
and 20\% occur in field overlaps.  The relevant time span 
is the 2087-day interval over which we add the simulated events; 
thus the exposure is $10.7 \ten{6}$ objects $\times 2087 \
{\rm days} = 6.12 \ten{7}$ object-years.  This exposure
is 3.4 times larger than in \yrtwo.  Note that number of stars in
field overlaps has increased from 12\% in \yrtwo, due to additional
fields and a more careful calculation of the field overlap size.

The most striking difference between the previous two data sets 
(\yrone\ \& \yrtwo) and the current set is the much higher efficiency
at long durations.  
Most of this difference is a reflection of having 5.7 years of data
instead of 2.1 years.  Explicit cuts in A97 were made that removed any
events with $\that > 300$ days, while the current cuts both use
$\that > 600$ days.
We also made a slight modification to our observing strategy 
intended to increase our sensitivity to long duration events.
However, some of the difference also lies in a quirk of the year 2 data set.  
In A97, six of the densest fields
had their lightcurves cut in half, roughly, due to an early generation of
templates used to reduce the photometry for these fields.  As described 
in \S~\ref{sec-obs} the photometry for these six fields has been re-run
with the new generation of templates and the lightcurves merged onto a
common photometric system.  
The effect of this was a lowered
efficiency in \yrtwo\ for events with durations longer than
$\that\sim\,100\,$days
(mainly due to the required 40 baseline points
in the `halved' lightcurves).  The problem did not exist in
the year 1 data, thus the rather similar behavior of year 1
and year 2 at large $\that$, even though the later had twice the coverage.

The relative looseness of selection criteria B over criteria A discussed 
in \S~\ref{sec-cuts} (i.e. 17 vs. 13 events) is well illustrated in 
Figure~\ref{fig-eff}.  Only for short durations, $\that < 10\,$days,
is criteria A more efficient, due to criteria B's larger number 
of required high points ($\geq 10$) as compared with criteria A ($\geq 7$).
Less than half of the difference in efficiency between criteria A \& B
is explained by the different $\Amax$ cuts; given $\Amax$ cuts
of 1.49 \& 1.34 for criteria A \& B respectively, we naively expect 
criteria B to recover 17\% more events.  In fact, only event~27 (7\%)
would have been missed had criteria B's $\Amax$ cut been increased
to 1.49.  The remainder of the difference lies primarily
in the effect of two cuts.  Many of the events that failed criteria A
did so because they failed either the cut in
$\Delta\chi^2/(\chi^2_{\rm peak}/N_{\rm dof})$ or the
cut in $\chi^2_{\rm ml-out}/N_{\rm dof}$.  Both of these cuts have been tightened
from their year 2 values.  From our Monte Carlo events we find that
tightening these two cuts has the following two
effects: (1) lessens our sensitivity to moderately or strongly
blended events, and (2) tightening the
$\Delta\chi^2/(\chi^2_{\rm peak}/N_{\rm dof})$
cut also decreases sensitivity to exotic microlensing and 
other asymmetric lightcurves, such as supernovae.
For example, the slightly asymmetric event~26 did not pass 
criteria A because of the cut $\Delta\chi^2/(\chi^2_{\rm peak}/N_{\rm dof}) > 350$.
As a result, set A has fewer events removed by the supernova cut.

In Alcock \etal (2000a) we describe in detail a robust way of statistically 
correcting for the $\that$ bias induced by blending.  Briefly,
this method is an integration over the LF of the median $\that$ bias
induced by blending in our sample of Monte Carlo events.  As a
check that this method gives a truly unbiased optical depth estimate
we ran a series of secondary Monte Carlo simulations that make use of this 
correction and a number of Galactic halo models.  The statistical
correction, although it blurs the individual events together,
does a satisfactory job of reproducing an unbiased optical
depth.

The primary shortcoming of the present efficiency analysis
is that all simulated events are assumed to be ``normal" microlensing
events with a single lens, a point source, and constant velocities.
This assumption is used in the present analysis for simplicity and 
because of the highly uncertain characterization of exotic lensing
events.  A careful study using selection criteria much looser than
criteria A \& B has
convinced us that it is unlikely that we have missed any
exotic lensing events in the present data set.  The primary concern
is what effect the addition of exotic lensing might have on our
detection efficiency (in particular because of binary lensing events).
Although this is as yet uncertain it is probably a small effect
due to the small number of exotic lensing events seen so far and the
fact that criteria B does find some exotic lensing events.

\section{ Event Distributions } 
\label{sec-dist}

There are a number of statistical 
tests that can be performed on microlensing event
distributions to test the hypothesis 
that events are gravitational microlensing, or to test
hypotheses regarding the lens population.
As the sample of events becomes larger these tests become
important discrimination tools.  

\subsection{ Impact Parameters }
\label{sec-impact}

\begin{figure}
%this rotates the figure and controls the dimensions
%\plotfiddle{fig_umin.ps}{3.0truein}{-90}{35}{35}{-150}{250} 
\plotone{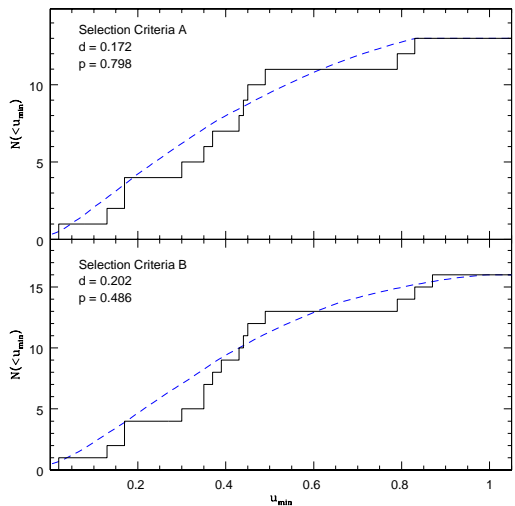}
\caption{The upper (lower) panel shows the cumulative distribution 
in $\umin$ for set A (set B; excluding the binary event~9). The
expected distribution is shown
as a \textit{dashed line}: a uniform distribution modified by our efficiency.
The results of K-S tests comparing the observed and expected distributions
are also shown.
\label{fig-umin}} 
\end{figure}

An important model-independent test of the
hypothesis that we have observed gravitational microlensing
is to compare the distribution of peak magnifications to the
theoretical prediction.
It is convenient to switch variables from the
maximum magnification ($A_{\rm max}$) to the minimum distance of approach
between the MACHO and the line of sight, in units of the
Einstein radius, $\umin = b / r_E$.
Events should be uniformly distributed in $\umin$; this distribution is
then modified by the experimental detection probability
which is typically higher for small $\umin$ \cite{macho-eff}.
The observed and predicted distributions for
our LMC events for both selection criteria are shown in
Figure~\ref{fig-umin}.  A K-S test shows a probability of $79.8\%$ of
getting a K-S deviation worse than the observed value 0.172 for 
criteria A by chance, and a probability of 
$48.6\%$ of being worse than 0.202 for 
criteria B.  The binary event~9 is excluded from this comparison.  
We conclude that the
distribution of events in $\umin$ is consistent with the microlensing
interpretation.

%\machonote{If we drop event~22 from set B the probability rises
%to 66.5\% chance of getting a K-S deviation worse then 0.180.  \ie\
%It's better.}

The $\umin$ distribution and the high magnification events
may be used to lend support to the microlensing
interpretation of our lower magnification events.
Our high magnification events are striking,
and are clearly separated from the background in Figure~\ref{fig-cuts}.
If these high magnification events are accepted as microlensing then 
there must exist many more microlensing events with smaller peak magnification.
Figure~\ref{fig-umin} shows that we find these in just the right proportion.

\subsection{The Color-Magnitude Distribution}
\label{sec-cmd}

The gravitational microlens does not distinguish between types
of stars, so naively one expects microlensing to occur uniformly on every
type of source star.  However, both selection criteria A \& B employ 
various signal-to-noise cuts that bias us against detecting 
microlensing events on faint stars.  In addition, the measured 
baseline flux of an event may be significantly larger because of 
the blending of non-lensed flux, and it is not always
possible to accurately determine the amount of blending.
That is, the blending fits in Table~\ref{tab-blend}
may not be reliable, since there is considerable fit parameter
degeneracy between $A_{\rm max}$ and the blend fraction.
Thus, detected microlensing candidates, while occurring on many types
of source stars, may not follow the observed 
distribution of stars (or rather objects) exactly.  
%Our 
%Monte Carlo efficiency analysis, however, predicts the 
%luminosity distribution of detected events (see \S~\ref{sec-eff} and
%especially \cite{macho-eff}).  In Figure~\ref{fig-lumin} we show
%the distribution of apparent magnitudes $V_{\rm obj}$ for 
%set A \& B along with the 
%expected distribution of $V_{\rm obj}$ from our
%Monte Carlo events.  The match is marginal with a K-S
%probability of $9.4\%$ of getting a K-S deviation worse than
%the observed value 0.329 for criteria A and a probability of
%$10.3\%$ of being worse than 0.285 for criteria B.

\begin{figure}
\plotone{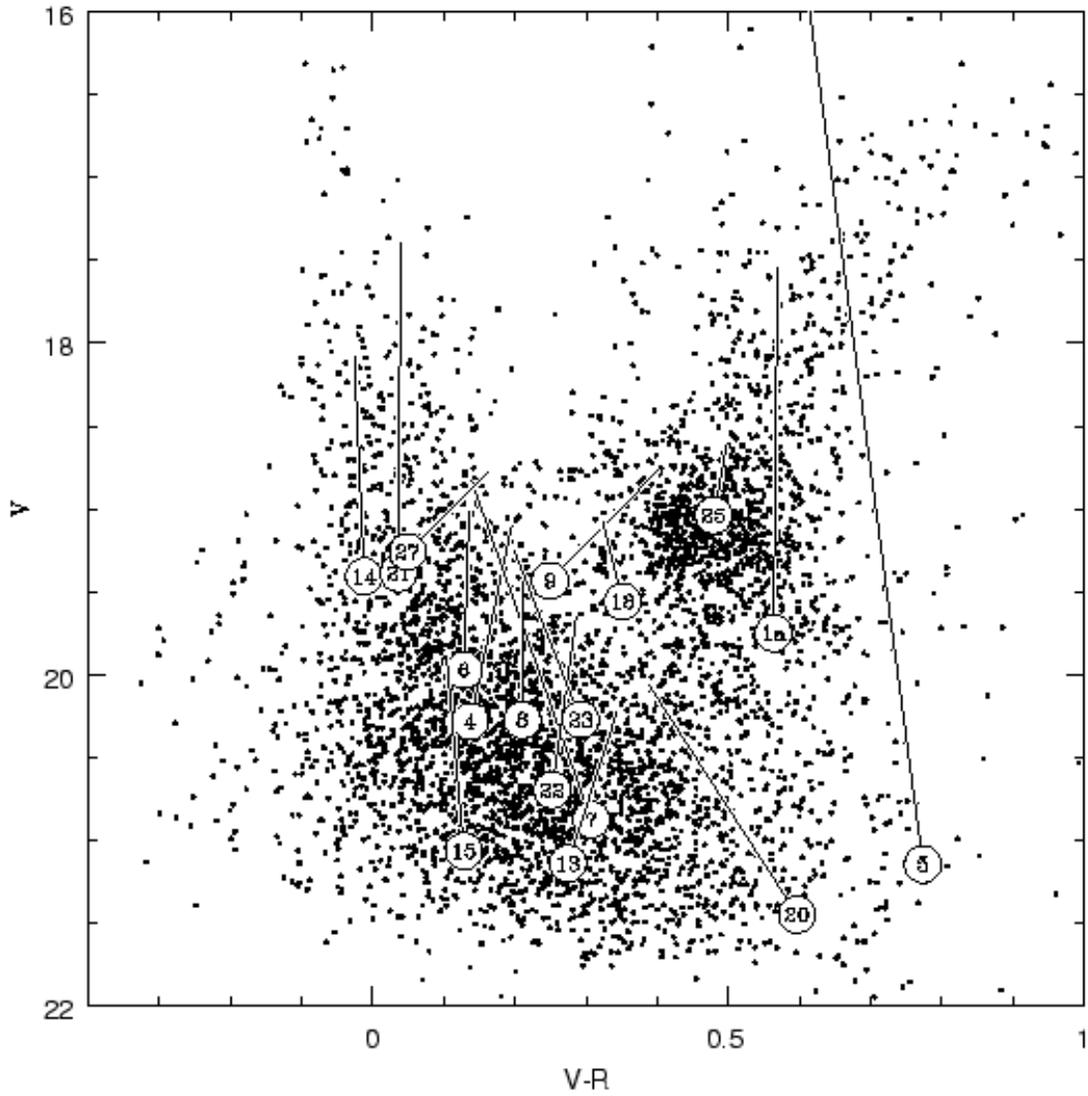}
\caption{Candidate microlensing events.  The open circles indicate
the position in the CMD of the blended MACHO object.  The lines extend
to the peak brightness of the event.  The points correspond to the
200 MACHO objects nearest each event.  The best available calibrations
for each field described in \S~\ref{sec-obs} have been used for this figure.
\label{fig-cmd} }
\end{figure}

Figure~\ref{fig-cmd} shows a CMD with each 
of the 17 microlensing candidates, along with all the closest
200 stars around each candidate.
Most of the events lie along the faint main sequence
where most of the observed LMC stars reside.  Event~5 is quite
red for its brightness and could represent a foreground
population of bright M--dwarf lenses as noted in \yrtwo, but
as a whole the distribution in the CMD is
quite representative. 
The distribution of events 
is not clustered in luminosity or the color magnitude diagram
and is consistent with the microlensing interpretation.

\subsection{Spatial Distribution}
\label{sec-spatial}

For microlensing by MACHOs smoothly distributed in the
Galactic halo, or stellar lensing by stars in the Milky way thin or thick 
disk, we expect the detected events to be distributed across
our fields in proportion to the local exposure.  
An extended LMC halo population could also form a smooth distribution.
In contrast, models in which LMC disk and bar stars dominate the lensing
population predict that the lensing events will be concentrated
within the LMC (\cite{macho-lmc2,aubourg-lmc,salati,gyuk-lmc},
but also see Alves \& Nelson 1999 where a flared disk could 
widen this distribution to some extent).

Figure~\ref{fig-lmc} indicates that the detected
events are apparently spread evenly across our 30 fields. 
To quantify this impression we perform two simple tests.
For criteria A \& B we computed a concentration
parameter, $\tilde{\theta}$, as described in Gyuk \etal\ (1999).
This parameter is a mean spatial separation between all
combinations of events.  For our two selection criteria we
find $\tilde{\theta}_A = 1.94^{\circ}\pm0.29$ and 
$\tilde{\theta}_B = 1.86^{\circ}\pm0.23$, where the error bars have
been estimated using the observed number of events and the models
of Gyuk \etal (1999) (private communication).
These numbers should be compared with predictions from the various
models of LMC self--lensing.  Gyuk \etal\ (1999) find 
$\tilde{\theta} = 1.3^{\circ} \pm 0.2$ over the MACHO 30 fields for all 
their LMC disk+bar self--lensing models, and 
$\tilde{\theta} = 1.85^{\circ} \pm 0.15$ for LMC disk+bar+halo and Galactic
halo models.  Thus by this measure our event distribution is 
inconsistent with their most favored LMC disk+bar self--lensing at 
the $\sim\,2\,\sigma$ level, but is consistent with an extended 
lens population such as is expected for a Galactic or LMC halo 
population.

\begin{figure}
\plotone{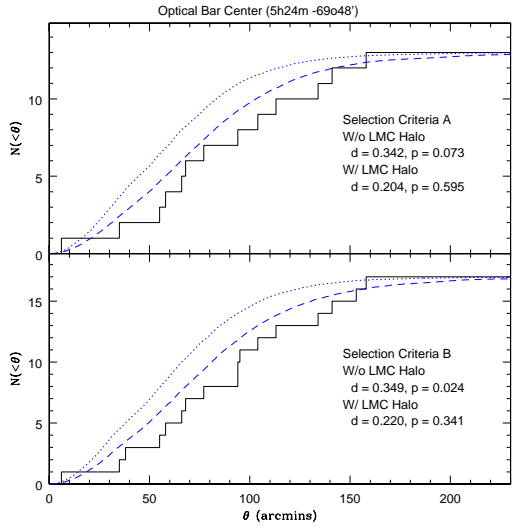} 
\caption{
The upper (lower) panel shows the cumulative spatial distribution on
the sky as measured from the optical center of the bar ($\alpha\!=\!5h24m$,
$\delta\!=\!-69^\circ 48'$) for set A (set B).  The predictions of 
Gyuk \etal\ (1999) for the case of stellar LMC disk+bar self--lensing
only (\textit{dotted line}) and LMC disk+bar+halo self--lensing (
\textit{dashed line})
are also shown.  The models of Gyuk \etal\ (1999) have been
folded into our efficiency per field for each of the 30 fields reported
on here.  The results of K-S tests comparing the observed and expected
distributions are also shown.
\label{fig-lmc_space} }
\end{figure}

Our second test is illustrated in Figure~\ref{fig-lmc_space}
where the cumulative distribution in spatial distances on the
sky, as measured from the optical center of the bar ($\alpha\!=\!5h24m$,
$\delta\!=\!-69^{\circ}48'$), is plotted.
Also shown are two predictions based on the models of
Gyuk, \etal\ (1999): the dashed line is the 
predicted distribution for uniform lensing (LMC disk+bar+halo) over
the face of the LMC, convolved with our detection efficiency per field
\cite{macho-eff}, and the dotted line is the favored LMC 
disk+bar self--lensing model, also convolved with
our detection efficiency per field. 
For criteria A, a K-S test yields a probability of $7.3\%$ of getting
a K-S deviation worse than the observed value 0.342 for the disk+bar
model.  For the disk+bar+halo model there is a probability of $59.5\%$
of getting a K-S deviation worse than the observed value 0.204.
For criteria B, a K-S test yields a probability of $2.4\%$ of getting
a K-S deviation worse than the observed value 0.349 for the disk+bar
model and for the disk+bar+halo model there is a probability of $34.1\%$
of getting a K-S deviation worse than the observed value 0.220.  We
note that these results are dependent on a direct comparison to the
30 fields used in this analysis, and that the LMC disk+bar self--lensing 
predictions are inconsistent with the data at the 93\% C.L.

One difference between a relatively smooth distribution of events over
the face of the LMC caused by a Galactic population and that caused by a LMC
halo is a slight east-west asymmetry due to the $\sim\,30^o$ tilt of
LMC's disk.  This tilt would induce a slightly higher optical depth
on the western side of the LMC, due to the longer path length through
any LMC halo.  No such east-west asymmetry should exist if the
microlenses are due to a Galactic population.  A casual inspection
of Figure~\ref{fig-lmc} hints at the possibility of such an east-west
asymmetry.  However, a simple Monte Carlo shows that the asymmetry is not 
statistically significant given the number of events in our sample.
\section {Implications} 
\label{sec-implic}

We start with the implied
microlensing optical depth, which is compared with the optical
depth expected from known populations of stars along the line
of sight to the LMC.  We then discuss our likelihood estimate 
of microlensing rate, MACHO masses, and optical depth for both
the dark halo and known stellar populations.

\subsection{ Optical Depth Estimates }
\label{sec-tau}

The simplest measurable quantity in a gravitational microlensing experiment is
the microlensing optical depth, $\tau$, which is defined to be the instantaneous
probability that a random star is magnified by a lens by more than
a factor of 1.34.  This probability depends only on the density profile of
lenses, not on their masses or velocities.  Experimentally, one can
obtain an estimate of the optical depth as 

\begin{equation}
\label{eq-taumeas}
\tau_{\rm meas} = {1 \over E} {\pi\over 4} 
\sum_i {\hat t_i \over \eff(\that_i)} \ ,
\end{equation}

\noindent
where $E=6.12 \ten{7}$ object-years is the total exposure, $\that_i$ is
the Einstein ring diameter crossing time of the $i$th event, 
and $\eff({\that_i})$ is its detection efficiency. 
Here, and below, we use the statistically corrected values of 
the blended durations $\that_{st}$ (Table~\ref{tab-blthat}).  These take into account the
fact that our typical star is blended, and so the fit $\that$ is
typically underestimated.  This statistical correction depends upon
the selection criteria used and is described more fully in 
\S~\ref{sec-eff}.  
It is also convenient to define the function 

\begin{equation}
\label{eq-tau1}
\tau_1(\that) = {1 \over E}{\pi\over 4}{\hat t \over \eff(\that)}\ , 
\end{equation}

\noindent
which is the contribution to $\tau_{\rm meas}$ 
from a single observed event with timescale $\that$. $\tau_1$ values 
for each of our events are also listed in Table~\ref{tab-blthat} for
both selection criteria.  

Using the criteria A set of 13 events, we find (Tables \ref{tab-tau} and
\ref{tab-taucl}) an optical depth 
for events with durations $2$ days $< \that < 400$ days of
$\tau_2^{400} = 1.1 {+0.4 \atop -0.3} \ten{-7}$. 
With the criteria B set of 17 events, we find 
$\tau_2^{400} = 1.3 {+0.4 \atop -0.3} \ten{-7}$. 
This is to be compared with $\tau = 4.7 \ten{-7}$ for a typical dark
halo consisting entirely of MACHOs and with predicted
$\tau_{\rm stars} = 0.24 \ten{-7}$ to $ 0.36 \ten{-7}$ from known stellar
populations (from Table~\ref{tab-stars} below).
Subtracting the stellar lensing background from our observed optical depth, we
find that the observed excess is about 15\%-25\%
of the predicted microlensing optical depth for a typical
all-MACHO halo of equation~(\ref{eq-stdhalo}) below.

\begin{figure}
\plotone{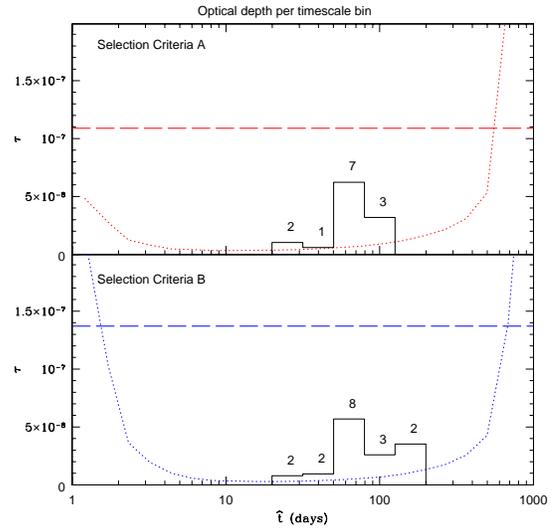}
\caption{
The contribution to the optical depth of eq.~\protect\ref{eq-taumeas}
from events binned in timescale for selection criteria A (upper panel)
and selection criteria B (lower panel).  The \textit{solid line} shows 
the observed 
values from the samples, with the number of events shown in each bin.
The \textit{dotted curve} shows the contribution to $\tau$ which would
arise from a single observed event with timescale
$\that$.  The \textit{dashed line} shows the total optical depth. 
\label{fig-tau} } 
\end{figure}

This optical depth estimate has the virtue of simplicity; 
however, since the events are ``weighted'' $\propto \tau_1$, 
it is hard to assign meaningful confidence intervals 
to $\tau$ without assuming some particular $\that$ distribution. 
This is illustrated in Figure~\ref{fig-tau}, which shows the 
contribution to the sum in equation~(\ref{eq-taumeas}) 
from events in various bins of $\that$. 
In comparison with A97, we note
that while the contributions become large at large $\that$, they
are substantially smaller in the 100-300 day range due
to the increased baseline and looser cut $\that < 600$ days.
The large contribution at long $\that$ implies that the overall uncertainty
in $\tau$ is greater than simple Poisson statistics based on 13
or 17 events. However, this uncertainty will continue to decrease as
the experiment progresses.  For example, in A97 we estimated that
should we expect to have observed on average 1 additional event 
with $\that \sim 300$ days (but happened to observe no such event),
the real $\tau$ would have been a factor
of two higher, and we were not able to exclude such a possibility
with any confidence.  The equivalent situation with the current data set is 
less dangerous, as such a missed event would result in a real $\tau$ being 
only 
about 20\% higher.  However, our optical depth estimate is valid only for
a specific mass or timescale interval.
The likelihood method of \S~\ref{sec-likely}
gives another way of estimating the optical depth and the errors
on the optical depth.

\subsubsection{Optical Depth Cut Dependence}
\label{sec-cutdepend}

\begin{figure}
\plotone{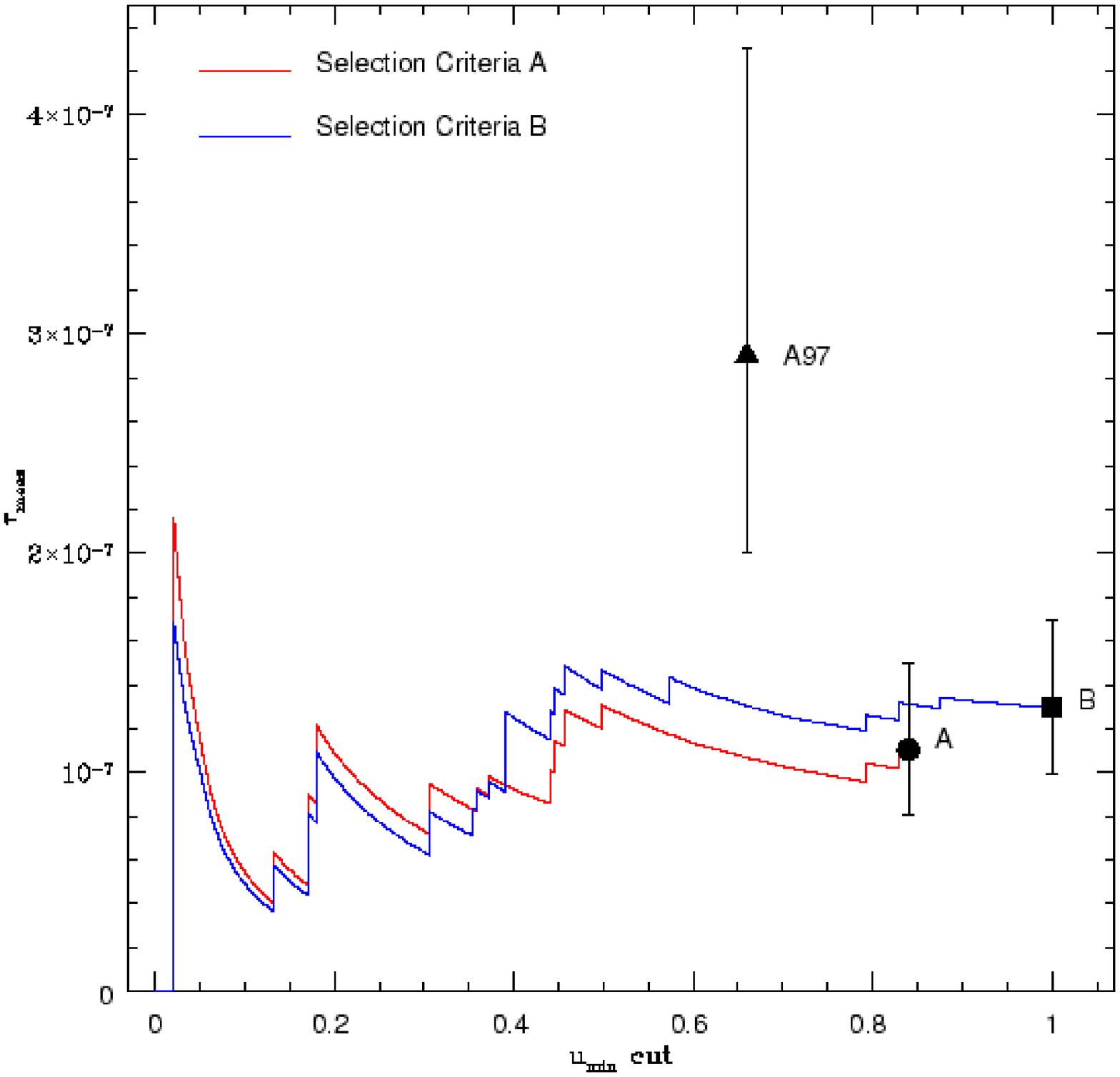}
\caption{Measured microlensing optical depth is plotted as a function of the
$u_{\rm min}$ cut for selection criteria A (\textit{thick line}) and selection
criteria B (\textit{thin line}).  Our selected cuts for criteria A \& B
are marked with large dots plus $1\,\sigma$ error bars.  The optical depth
reported in \yrtwo\ is also shown.
\label{fig-tau-vs-umin} } 
\end{figure}

\begin{figure}
\plotone{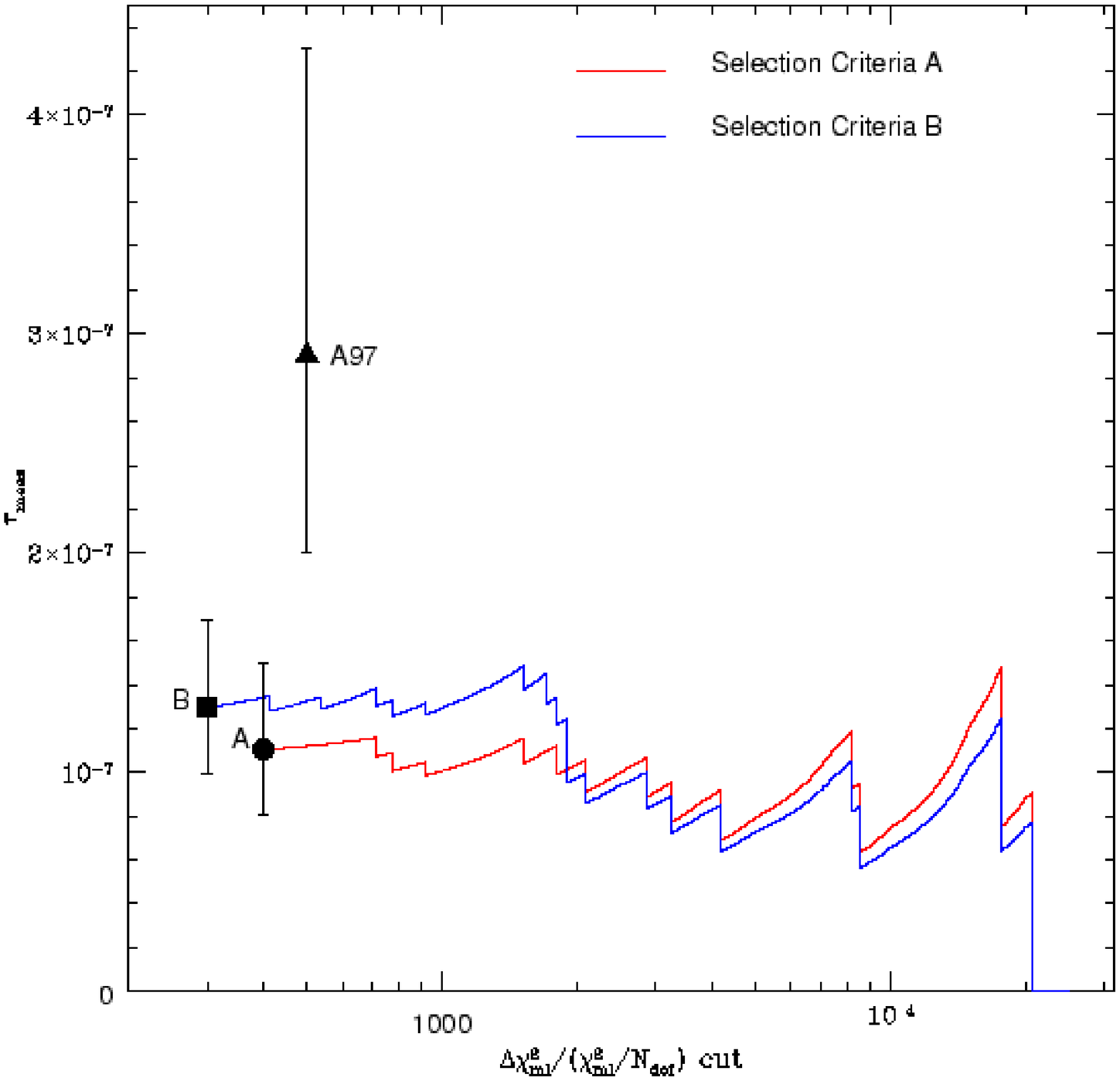}
\caption{Measured microlensing optical depth is plotted as a function of the
$\Delta\chi^2/(\chi^2_{\rm ml}/N_{\rm dof})$ cut for selection criteria A 
(\textit{thick line}) and selection criteria B (\textit{thin line}).
Our selected cuts for criteria A \& B are marked with large dots plus 
$1\,\sigma$ error bars.  The optical depth reported in \yrtwo\ is also shown.
\label{fig-tau-vs-delc2} } 
\end{figure}

Figures~\ref{fig-tau-vs-umin} and \ref{fig-tau-vs-delc2} show the
dependence of the the measured optical depth on the $u_{\rm min}$
and $\Delta\chi^2/(\chi^2_{\rm ml}/N_{\rm dof})$ cuts.  The heavy curves
indicate $\tau_{\rm meas}$ for set A  while the light curves
give $\tau_{\rm meas}$ for set B.  For the binary event we 
have assigned a $u_{\rm min}$ value of 0.573, which is the value obtained
for the single-lens fit.

Figures~\ref{fig-tau-vs-umin} and \ref{fig-tau-vs-delc2} clearly indicate
that our optical depth results are not very sensitive to our
cut values.  The $\tau_{\rm meas}$ values generally do not vary by more then
the $1\sigma$ statistical error bars for $u_{\rm min}$ and
$\Delta\chi^2/(\chi^2_{\rm ml}/N_{\rm dof})$ cuts in the ranges $0.2 \leq u_{\rm min}
\leq 1.0$ and $300 \leq \Delta\chi^2/(\chi^2_{\rm ml}/N_{\rm dof}) \leq 5000$.
We note the largest single contribution is from event 22, which was
included in set B and excluded in set A.

\subsubsection{Comparison of \yrtwo\ with the Present Analysis}
\label{sec-why}

Why is our new value of the optical depth 
a factor of two smaller than the value reported in {\yrtwo}?  The reasons
are manyfold and somewhat difficult to separate out completely.
By far the largest effect can be classified as
Poisson in nature.  We list the causes in order of decreasing effect
on the optical depth.

Inspection of Figure~\ref{fig-events} reveals that
a disproportionate number
of our `high' quality events were observed in the first 2.1 years
of the data set.  Events~1, 5, 7, and 9 are all of high quality.
Only event~14, 21, 23, in the following 3.6 years stand out as having
comparable quality.  This `qualitative' feel for the events is
backed-up by the fact that in the first 2.1 years, 6-7 events were
observed: a rate of 2.9-3.3 events/year.  And for the last 3.6 years
only 7-10 events were observed: a rate of 1.9-2.8 events/year.
Since the efficiency has not changed drastically over this interval,
we conclude that while the exposure increased by a factor of 3.4, the
number of events did not.  That is, we got ``lucky" during the 2.1 years
of \yrtwo\ and detected more microlensing than average.  This possibility
was reflected in the large Poisson errors quoted in A97.

%A reason the optical depth reported in \yrtwo\ is 
%large compared to this work
%is related to the splitting of the top 6 fields for analysis
%described in \S~\ref{sec-obs}.  These 6 fields represented $\sim 27\%$
%of our exposure in object-years.  Because they were split into
%two sets of lightcurves of approximately one year duration
%affected efficiencies for durations $\simgt 100$ days.
%At least 4 of the events in the 8 event sample of \yrtwo\ (events~5,
%6, 7, 9) had larger contributions to the optical depth 
%than they would have had if the top 6 fields had not been split for analysis. 
%This in itself was not in error, for in principle one would have
%expected that one or two events would have been missed because of the
%splitting of the top 6 fields and these `missed' events would have been
%counter-balanced the decreased efficiency.  By chance,
%no such events were missed, as is evident in the current analysis
%where the top 6 fields have been merged and analyzed as full 
%lightcurves.  We estimate that this ``small numbers" effect
%increased the optical depth reported in \yrtwo\ by
%$\sim 25\%$.

Another reason the optical depth reported in \yrtwo\ is
large compared to this work
is related to the splitting of the top 6 fields for analysis
described in \S~\ref{sec-obs}.  These 6 fields represented $\sim 27\%$
of our exposure in object-years.  Because they were split into
two sets of lightcurves of approximately one year duration
the efficiency was lowered for durations $\simgt 100$ days.
At least 4 of the events in the 8 event sample of \yrtwo\ (events~5,
6, 7, 9) contributed somewhat more to the optical depth
then they would have had the top 6 fields not been split for analysis.
This in itself was not in error, for one would have
expected that one or two events would have been missed because of the
splitting of the top 6 fields and these `missed' events would have
counter-balanced the decreased efficiency.  By chance,
no such events were missed, as is evident in the current analysis
where the top 6 fields have been merged and analyzed as full
lightcurves.  We estimate that this ``small numbers" (Poisson) effect
increased the optical depth reported in \yrtwo\ by
$\sim 25\%$.

For reasons discussed in more detail in \S~\ref{sec-eff} (and
Alcock, \etal 2000a), our detection efficiency is somewhat
higher than in \yrtwo.  Briefly, we neglected a 
contribution from faint stars with 
$V > 21.5$ and thus our efficiency in \yrtwo\ had not yet
converged for durations $\simgt 100$ days.  In addition, our new
normalization, which has been more carefully determined 
using HST data, and has more realistically estimated
errors leads to a somewhat increased sensitivity.
Together, these effects
spuriously increased the optical depth in \yrtwo\
by $\sim 10\%$ with respect to the current results.
This is within our estimated uncertainty.

As mentioned in \S~\ref{sec-sn}, one of the events
used in \yrtwo\ (event~10) is most likely a SN interloper.
This interpretation is supported by the presence of an obvious spiral galaxy 
in our HST frames of
this event and the fact that it is quite reasonably fit by
a type Ia SN lightcurve.
The effect of this interloper spuriously increased
the optical depth by $\sim\,7\%$ in the 8 event sample 
and had no effect on the 6 event sample (which rejected it). 

\subsubsection{Errors in Present Analysis}
\label{sec-errors}

While the current analysis is the most careful yet performed and a
substantial improvement over earlier efforts, there are still a number
of errors or potential errors that exist in our results.
The errors due to small number statistics are included in the
error bars we report (about 30\% uncertainty). 
The errors due to model dependency are explored
by considering a range of models.  We believe these are the
largest errors in our results.  In this short section 
and in Table~\ref{tab-tauerrorbudget}
we list some other sources of error that might be worth considering in 
more detail in the future.

{\it Normalization of star to object ratio:}
The ratio of actual LMC stars to SodoPhot objects
varies across our fields, and it is very difficult to accurately estimate.
We have HST images for three areas, and
attempt to tie together MACHO object based photometry to the HST
star-based photometry to create a unified luminosity function (LF).
We estimate a 20\% uncertainty in our final results because of this.
Issues include the underlying luminosity function, the magnitude calibration
of our objects, blending effects in matching the object LF to the stellar
LF, and the unknown effects of crowding, seeing, and sky in the template images,
among several others.

{\it Selection Criteria:}
Since we do not have a complete understanding of the background, and because
we examine our data before deciding upon the selection criteria, it is
possible that we differentiate background and microlensing in an imperfect
and/or biased way.
We implemented two independent sets of selection criteria as a test of
our sensitivity to this bias.  We estimate about a 20\% uncertainty due
to our selection criteria.

{\it Correction to $\that$:}
Blending causes the durations of events to shift from their naive fit values.
We chose to correct for this in a statistical manner,  and estimate
about a 3\% uncertainty in this correction. 
This is due to several factors, but mainly
the uncertainty in the true $\that$ distribution which is needed to make the
correction.

{\it Binary source stars:}  Locally most stars reside in binary or multiple
star systems, and it is expected that this is also true of LMC stars.
Our LMC luminosity function does not include a correction for this.
This is a complicated correction which will be uncertain since
the binary fraction for LMC stars is not known.  We did not make an estimate
of the size of this effect, but will consider it in a later paper.

{\it Exotic microlensing:}  We do not explicitly add binary-lens microlensing
or other exotic lensing into our artifical lightcurves.  
Thus our efficiencies for these are unknown. 
We did try to explore two sets of selection criteria to detect any gross
sensitivity of our results to this effect, 
but a proper calculation should be done in the future.

{\it Others:}  There are several other sources of systematic error that
have been considered.
For example, in the artifical star
Monte Carlo we assume that all flux added goes to the nearest SoDoPhot object.
Direct tests show that this is not true in about 3\% of the cases.  The
effect of this mis-identification is not known, but could be around 3\%.
Some other small errors are discussed in Alcock, \etal\ (2000a)
and Vandehei (2000), and a summary is given in Table~\ref{tab-tauerrorbudget}.

%Finally we note that because systematic errors can have substantial biases,
%it is not straightforward to combine them with each other or with
%statistical errors.  
%Systematic error biases can also be strongly asymmetric.
Finally we note that due to the complex nature of systematic errors, it is
not straightforward to combine them with each other or with statistical
errors. For example, systematic errors can be strongly asymmetric.

We estimate our total systematic error to
be in the range of 20\% -30\%, though even this range is uncertain.

\subsection{Likelihood Analysis and Dark Matter}
\label{sec-likely}

We compare the number of detected events and the distribution of observed
timescales, $\that$, with predictions from models of various lens populations.
Microlensing can occur when any compact object travels in front of a monitored
star, so we expect microlensing events from any population of 
stars, remnants, or dark compact objects that lie between us and the LMC.
Luckily, much is known about the density and velocity distribution of stars
and remnants in the Milky Way and LMC.
Less is known about the dark halo of the Milky
Way (and even less about the dark halo of the LMC), but we can leave
the fraction, $f$, of dark objects that can lens, as well as the masses, $m$ of
these objects, as free parameters which we determine using a maximum
likelihood analysis.
This analysis was done in A97, with 6 or 8 observed microlensing events.
We found values of $f$ between 0.15 and 1.0 at 90\% confidence level (CL), 
and MACHO masses $m$ between 0.1 and 1.0 $\msun$. 
The large uncertainty in these results came largely from
small number statistics, but also from uncertainty in the models.
In the current analysis, we have a larger number of events and an improved
efficiency determination, so we can reduce the Poisson and some systematic
errors.  Here we also make improvements to our likelihood analysis, most notably
the inclusion of realistic estimates of the stellar lensing background within
the likelihood function.  We note that the results of this section still depend
heavily upon the models of Milky Way and LMC.  We will come back to this point
in \S~\ref{sec-discuss}.

\subsubsection{Microlensing Rate}
\label{sec-rate}

The microlensing event rate $\Gamma$ is more model-dependent than 
the optical depth $\tau$. Rate $\Gamma$ depends on the event timescales 
via the mass function of MACHOs and 
their velocity distribution, but the uncertainties in 
$\Gamma$ are given purely by 
Poisson statistics. Thus $\Gamma$ is useful in quantifying 
the errors on any measurement, 
once a halo model is specified. 

The number of observed events 
is given by a Poisson distribution with a mean of 

\begin{equation} 
\label{eq-nexp} 
\Nexp = E \int_0^{\infty} \, {d\Gamma \over d\that} \,
\eff(\that) \, d\that
\end{equation} 

\noindent
where $ E = 6.12 \ten{7}$ object-years is our total `exposure',
and  ${d\Gamma \over d\that}$ is the total differential microlensing
rate,
$$
{d\Gamma \over d\that} = 
f \dgamma{MWhalo} + \dgamma{thindisk} + \dgamma{LMC} + \cdots.
$$
For a typical dark matter halo (equation~\ref{eq-stdhalo})
consisting of 100\% MACHOs, the total rate of microlensing events with
$A_{\rm max}>1.34$
is given by equation~(A2) of \yrone:
$\Gamma \approx 1.6 \ten{-6} (m/\msun)^{-0.5}$ events/star/yr.  
Thus if all MACHOs had the same mass and our efficiency were 100\%, 
we would expect
about $100 (m/\msun)^{-0.5}$ events in the present data set.
The average timescale of an event scales oppositely,
$\avethat \approx 140 (m/\msun)^{1/2}$, since the product of the two
gives the optical depth which is independent of the MACHO masses,
$\tau = {\pi\over 4} \Gamma \avethat$.
Thus, although the optical depth is independent of MACHO mass, 
in a real microlensing experiment 
statements on the MACHO content of the halo  
will be dependent on the MACHO mass.
However, the masses of the lenses are constrained,
since we measure the distribution of event timescales.

\subsubsection{Milky Way and LMC Models}
\label{sec-models}

We consider four stellar components and a dark component of the Milky Way,
and a stellar and dark component of the LMC.  
Given our exposure and efficiency, and a model of the density, velocity 
distribution, and mass function
of a lens population, we can calculate the expected microlensing optical 
depth, microlensing rate, distribution of event durations, 
and the number of expected events detected in our experiment.
A summary of the results are shown in Tables~\ref{tab-stars} and 
\ref{tab-like}.
For the LMC self-lensing model, we note that the rate depends strongly
on the position on the sky, so the values we report depend on
the 30 specific fields we monitor.

We model the density of the
Milky Way and LMC thin and thick disks as double exponentials
$$
\rho_d =  {M_{\rm disk} \over 4 \pi z_d R_d^2} \exp{ \left( -\left| {R\over R_d} \right| - 
\left| {z\over z_d} \right|\right)},
$$
where  $z$ and $R$ are cylindrical coordinates,
$M_{\rm disk}$ is the total mass of the disk, $z_d$ is the scale height,
and $R_d$ is the scale length.   Instead of specifying the total mass,
Milky Way disks are often specified
by the column density, $\Sigma_0$, at the solar circle, $R=R_0\sim 8.5$ kpc,
and the relation is
$$
\Sigma_0 = {M_{\rm disk}\over 2 \pi R_d^2} e^{-R_0/R_d}.
$$
For simplicity, we characterize the velocity distribution of a disk as 
a constant rotation velocity, $v_c$, with some isotropic dispersion,
$\sigma_v$, in addition. 

For the normal Milky Way thin disk we use parameters: 
$R_d=4$ kpc, $z_d=0.3$ kpc,
$\Sigma_0 = 50 \msunpcth$, $R_0=8.5$ kpc, $v_c=220$ km/s, and
$\sigma_v = 31$ km/s.  This gives a total thin disk mass of 
$M_{\rm disk} = 4.2 \ten{10} \msun.$
Later, we will discuss models (e.g., model F) with a maximal thin disk, and a 
smaller dark halo.  In that case we use the above parameters except
with $\Sigma_0 = 80 \msunpcth$, which gives  $M_{\rm disk} = 6.7 \ten{10} \msun.$ 

For the Milky Way thick disk we use parameters: 
$R_d=4$ kpc, $z_d=1.0$ kpc,
$\Sigma_0 = 4 \msunpcth$, $R_0=8.5$ kpc, $v_c=220$ km/s, and
$\sigma_v = 49$ km/s, for a total mass of $M_{\rm disk} = 3.4 \ten{9}\msun$. 

For the LMC disk we use the preferred parameters from Gyuk \etal\ (1999),
$R_d=1.57$ kpc, $z_d=0.3$ kpc, $v_c = 70 $ km/s, $\sigma_v=25$ km/s, and
$M_{\rm disk} = 3.0 \ten{9} \msun$.  The LMC disk self-lensing also depends upon
its distance, $L=50$ kpc, inclination, $i=30^\circ$, and position
angle, $\phi=170^\circ$.  These are the parameters we use when considering
the LMC disk plus LMC dark halo model.  For the pure disk (no LMC halo)
case, we conservatively
increase $M_{\rm disk}$ to $5 \ten{9}\msun$, in good agreement with recent
analysis of the LMC rotation curve ($M_{\rm disk} = 5.3 \pm 1.0 \msun$
\cite{alves-flared-lmc},
corresponding to central surface densities of 190 $\msunpctwo$ and 
320 $\msunpctwo$ respectively.  We do not consider a separate bar component,
since the bar mass is strongly limited by the HI
kinematics \cite{kim,gyuk-lmc}.

The Milky Way spheroid density is modeled as \cite{guidice,macho-lmc1}
$$
\rho_{\rm spher} =  1.18 \ten{-4} (r/R_0)^{-3.5} \msunpcth,
$$
with no rotation, and an isotropic velocity dispersion of $\sigma_v = 120$ km/s.

For the LMC we consider two main cases: 1) pure disk/disk self-lensing, and
2) disk/disk self-lensing plus an LMC dark halo consisting of fraction $f$
of MACHOs, with $f$ being the same fraction used for the Milky Way dark halo.
Later, we also consider the possibility of an all stellar LMC halo.

For the Milky Way dark halo we consider three models: S, B, F, which
were used in A96 and A97.
The density of Model S is given by

\begin{equation}
\label{eq-stdhalo}
\rho_H(r) = \rho_0 { R_0^2 + a^2 \over r^2 + a^2 } 
\end{equation} 

\noindent
where $\rho_H$ is the halo density, 
$\rho_0 = 0.0079 \msun \, \pc^{-3} $ is the 
local dark matter density, $r$ is Galactocentric radius, 
$R_0 = 8.5 \kpc$ is the Galactocentric
radius of the Sun, and $a = 5$ kpc is the halo core radius
\footnote{Analysis of carbon star kinematics on the periphery of the LMC disk
support a pseudo-isothermal density profile for the Galactic dark
halo \cite{alves-flared-lmc}.}.  
With the standard thin disk, this model has a total rotation speed at
50 kpc of 200 km/s, with 190 km/s coming from the halo.
We assume an isotropic Maxwellian distribution of velocities
with a 1-D rms velocity of $155 \kms$,
and assume a $\delta$-function MACHO mass function
of arbitrary mass $m$.  
\footnote{
As discussed in A97, a delta-function mass distribution is a reasonable 
fit to the observed distribution, and more complicated forms are
difficult to distinguish with such a small number of events.}
Note that we always multiply the above density
by the MACHO halo fraction $f$, implicitly assuming that the remaining
$1-f$ fraction of the halo is filled with some exotic particle dark matter
or other non-lensing matter.
Dark halo models B and F are power-law models \cite{evans93,explore}, 
and are discussed in detail in A96, and A97.  Model B has a very large 
dark halo and a standard thin disk,
giving a rising rotation curve that reaches 258 km/s at 50 kpc.
Model F has an extremely low mass halo, somewhat inconsistent
with the known Galactic rotation curve, and a very massive disk.  At 50 kpc,
the model F
halo contributes 134 km/s towards a total of 160 km/s rotation speed.

Finally, for the LMC halo we follow Gyuk \etal\ (1999) and
use the same density distribution as model S above, but with
a central density of $0.0223 \msunpcth$, $a=2$ kpc, $v_c = 70$ km/s, and
a tidal truncation radius at 11 kpc.  
In this model the mass of the LMC halo in the
inner 8 kpc is $ 6 \ten{9}\msun$ and the total mass of the halo is
$9.2 \ten{9}\msun$.  This is a somewhat extreme model, probably
larger than allowed by the LMC rotation curve.  Like the Milky
Way halo, the LMC halo is assumed to consist of a fraction $f$ of MACHOs all
of mass $m$.  It is conceivable that the LMC MACHO halo fraction and make-up
is different than the MACHO fraction in the Milky Way halo, 
and we consider this possibility in the next section.
We note that no substantial stellar component of
an LMC halo has yet been observed.

For the stellar lensing populations we integrate the microlensing rate over a
mass function.  There have been several recent determinations
of the present day mass function (PDMF), but it is not clear that the
mass function determined locally is valid for all the stellar populations
we model.  However, for simplicity we will use the PDMF of 
Gould, Bahcall, \& Flynn (1997) 
for all the stellar populations. 
Table~\ref{tab-stars} shows some properties of the stellar population
calculated from the models above.  It also shows the expected number
of microlensing events from each population found by the likelihood method,
and thus including the effect of our efficiency calculation and
selection criteria.  These results differ to some degree from those presented
in A97 for several reasons.  First, the models we use are different in
some cases, and we are using a different PDMF.  Second, and most
importantly, we explicitly do not count
lenses that are too bright to be detected as microlensing in our experiment.
We have an explicit cut at around $V=17.5$ magnitudes, and so stars brighter
than this cannot be found as lenses.  Our Monte Carlo shows that
this cuts the expected number of thin disk stellar lensing events 
by more than half, with smaller effects for thick disk, spheroid, etc.
and almost no effect for LMC disk lensing.
The results displayed in Table~\ref{tab-stars},
use the full Monte Carlo for all stellar distributions.

\subsubsection{MACHO Halo Fraction and Mass}
\label{sec-like}

\begin{figure}
%this rotates the figure and controls the dimensions
\plotfiddle{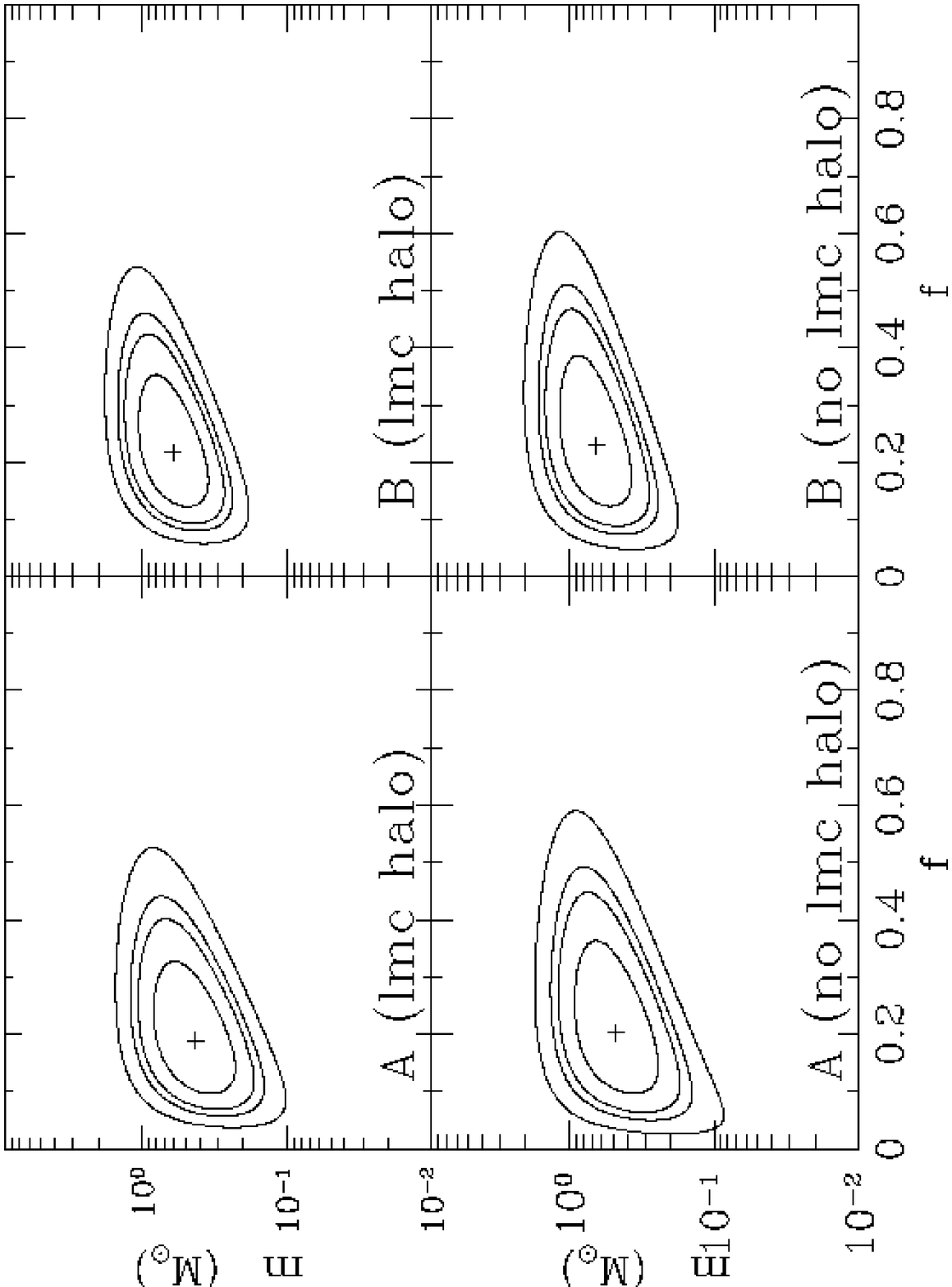}{3.0truein}{-90}{35}{35}{-150}{250} 
%\plotone{fig_mf_cow.eps}
\caption{ 
Likelihood contours for MACHO mass $m$ and halo fraction $f$ for Model
S which has a typical size halo.  See A96 for details of the model.
The plus sign shows the maximum likelihood estimate and the contours enclose
regions of 68\%, 90\%, 95\%, and 99\% probability.  
The panels are labeled according to which set of selection criteria
(A or B) is used, and whether or not an LMC halo with MACHO fraction $f$
is included.
\label{fig-like-cow} } 
\end{figure}

\begin{figure}
\plotfiddle{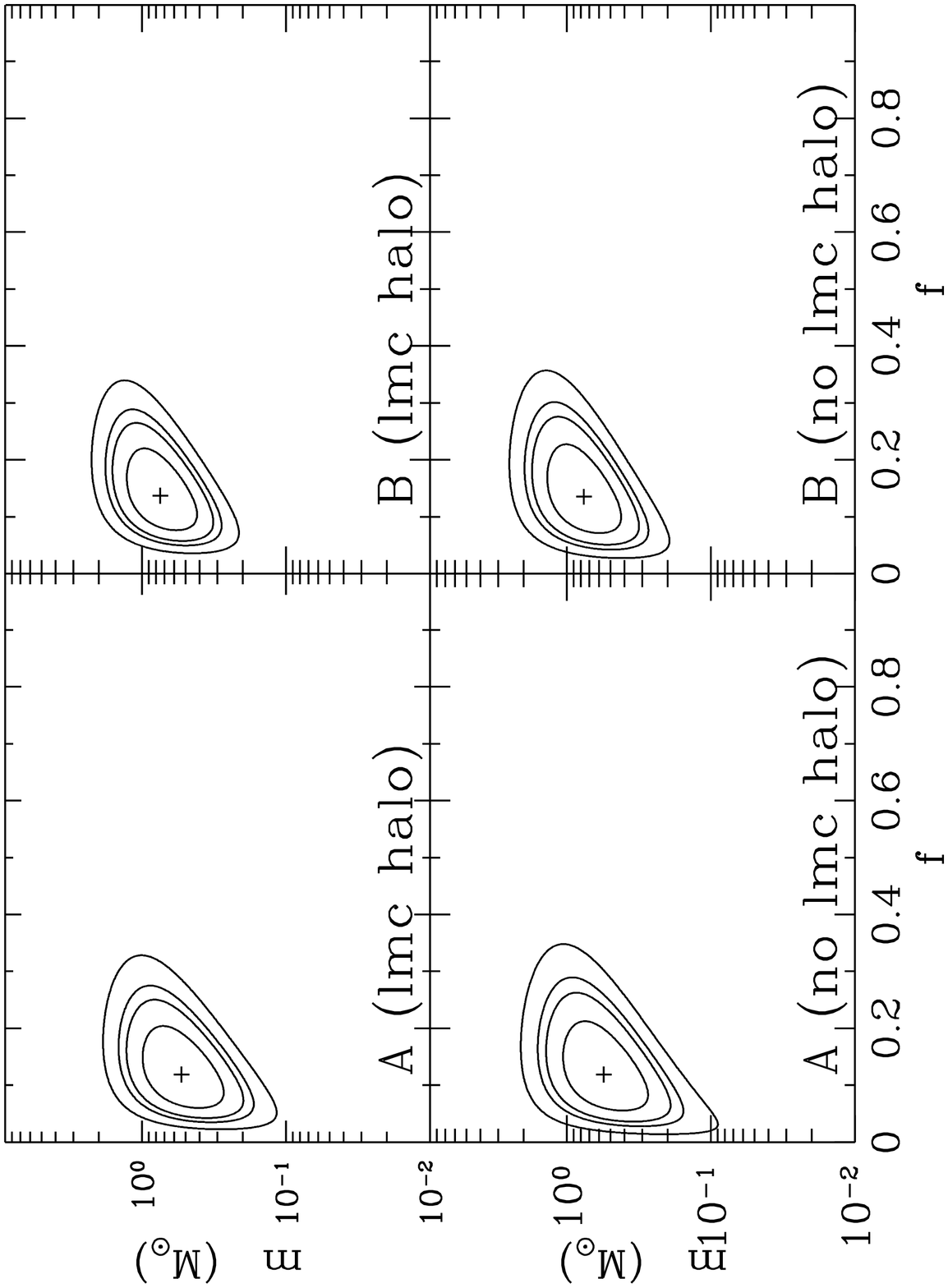}{3.0truein}{-90}{35}{35}{-150}{250} 
%\plotone{fig_mf_b.eps}
\caption{
Likelihood contours for MACHO mass $m$ and halo fraction $f$ for Model
B which has a very large dark halo.  See A96 for details of the model.
The plus sign shows the maximum likelihood estimate and the contours enclose
regions of 68\%, 90\%, 95\%, and 99\% probability.  
The panels are labeled according to which set of selection criteria
(A or B) is used, and whether or not an LMC halo with MACHO fraction $f$
is included.
\label{fig-like-b} } 
\end{figure}

\begin{figure}
\plotfiddle{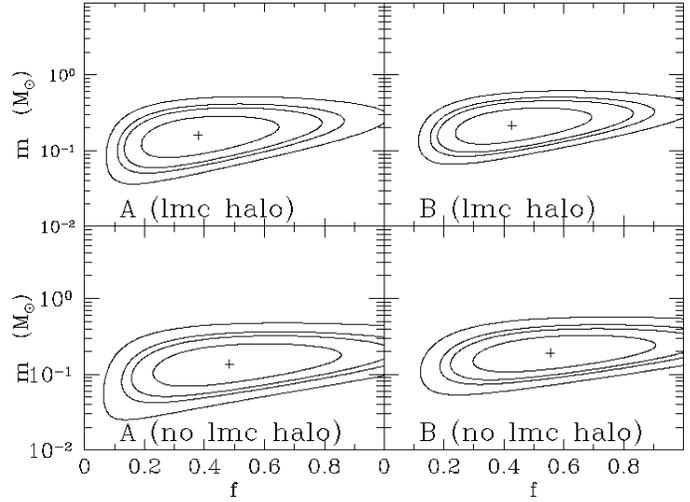}{3.0truein}{-90}{35}{35}{-150}{250} 
%\plotone{fig_mf_f.eps}
\caption{ 
Likelihood contours for MACHO mass $m$ and halo fraction $f$ for Model
F which has a very small dark halo, and a nearly maximal
thin disk.  See A96 for details of the model.
The plus sign shows the maximum likelihood estimate and the contours enclose
regions of 68\%, 90\%, 95\%, and 99\% probability.  
The panels are labeled according to which set of selection criteria
(A or B) is used, and whether or not an LMC halo with MACHO fraction $f$
is included.
\label{fig-like-f} } 
\end{figure}

We find the most likely values of halo fraction $f$ and unique  
MACHO mass $m$ using our set A (13 events) and set B (17 events), 
and their corresponding
efficiencies.  The likelihood function is

\begin{equation}
\label{eq-like}
L(m,f) = \exp(-N_{\rm exp}) \prod_{i=1}^{\Nobs} \left( E \eff(\that_i) 
{d\Gamma \over d\that}(\that_i;m) \right), 
\end{equation}

\noindent
where,
$
\dgamma{} = f \dgamma{\rm MWhalo} + \dgamma{\rm thindisk} \break
+ \dgamma{\rm thickdisk} +
\dgamma{\rm spheroid} + \dgamma{\rm bulge} + \break
\dgamma{\rm LMCdisk} 
+ f\dgamma{\rm LMChalo},
$
and each $\dgamma{j}$ is
the theoretical rate of microlensing derived from
model $j$.  The distributions $\dgamma{i}$ for stellar populations
are integrated over a mass function \cite{gouldpdmf} as described above,
and are calculated using code described in Gyuk \etal\ (1999).
\placefigure{fig-like-cow}
\placefigure{fig-like-b}
\placefigure{fig-like-f}
The results are dependent on the model so 
we explore a range of possible halos,
including a standard halo (model S from A96 and A97).
We also use two other dark halo models.
We choose model B from A96 and A97, because it is about as large a halo as the
data will allow.  We also choose model F from A96 and A97 because
it has a nearly maximal disk and a very low mass halo, 
and therefore is as small a halo as the data allows.  
These models are described in detail in A96.  We do not show models
A, C, D, or G from A96 and A97, or other possible halo models
since they are in general intermediate between the
extremes of models B and F.  
Model S is a common pseudo-isothermal sphere \cite{griest91}
with an asymptotic rotation velocity
of 220 km/s,  while models B and F are power-law Evans models \cite{evans93}.
Table~\ref{tab-like} shows the results for all the models, and 
Figures~\ref{fig-like-cow}, \ref{fig-like-b}, and \ref{fig-like-f} 
show the corresponding likelihood contours.

For model S, 
the resulting likelihood contours, assuming a $\delta$-function halo mass
function, are shown in Figure~\ref{fig-like-cow}.  The probabilities
are computed using a Bayesian method with a prior uniform in 
$f$ and $\log m$.  We show likelihood contours
for both our 13 event sample and our 17 event sample, and
with and without the LMC halo.  The best fit values and errors are given
in Table~\ref{tab-like}.
The errors shown in the table are one-sigma errors.
The peak of the likelihood contours gives the most probable mass and halo
fraction for a given model and for set A with an LMC halo
we find $m_{\rm 2D} = 0.48 \msun$, $f_{\rm 2D} = 0.20$.  For the 
corresponding set B,  $m_{\rm 2D} = 0.67 \msun$, and $f_{\rm 2D} = 0.23$.

We calculate the one-dimensional likelihood function by integrating over
the other parameter and find (for set A without an LMC dark halo)
a most likely MACHO mass $m_{\rm ML}= 0.60^{+0.28}_{-0.20} \msun$,
and most likely halo fraction $f_{\rm ML} = 0.21^{+0.10}_{-0.07}$.
The errors given are 68\% CL.
The values for set B are 
$m_{\rm ML}= 0.79^{+0.32}_{-0.24} \msun$ and
$f_{\rm ML} = 0.24^{+0.09}_{-0.08}$.
For model S, the 95\% CL contour includes halo fractions from about
8\% to about 50\%,
and MACHOs masses from about 0.12 $\msun$ to 1.1 $\msun$, depending upon
the selection criteria and LMC model used.
The likelihood method gives an optical depth for the halo population
of $1.1^{+0.5}_{-0.4}\ten{-7}$ almost independent of the section criteria,
the LMC model, and the Galactic model.

There are several important comments to be made.  
First, sets A and B
give results that are remarkably similar, implying
that the systematic error introduced by our selection
criteria methodology is small. The important parameters of estimated
MACHO halo fraction are nearly identical using the two different
sets of events and efficiency determinations. 
The estimated typical MACHO mass does vary between the two sets of events, but
the values lie within one sigma of each other.  This difference in lens
mass comes partially from the rejection of event 22 from set A. 

Second, consistent with our optical depth estimates, the values of the halo
fraction are approximately a factor of two lower than we found in A97.
As discussed in \S~\ref{sec-why},
this is mainly a result of finding more events per unit exposure during the
first two years, but is also due to changes in efficiency, etc.
We note that the optical depths reported in 
Table~\ref{tab-like} are the estimated MACHO contribution, and do not
include the background of stellar microlensing.
The contributions from stellar background are shown in Table~\ref{tab-stars}.
The values found here are quite similar to those found directly in
\S~\ref{sec-tau}.  

Third, our new confidence intervals are substantially smaller
than those of A97 due to the larger number of events.
Even though the central values have changed, there is reasonable
overlap of our new contours with the A97 contours, 
and our new most likely values lie within the A97 90\% confidence region.

Next, for Model S with a large LMC disk, but no LMC dark halo, 
and set A, 
we expect a total of 3.0 events from stellar
background sources, with the majority coming from LMC self-lensing.
For the same model and set B, the number of 
expected background events is 3.9.  
In both cases the predicted number of background events
is substantially below the number of detected events.
Thus, if these models are correct, the microlensing events are
very unlikely to come from the known stellar populations.

For the case of an LMC halo plus LMC disk, LMC disk self-lensing must be
smaller since part of the LMC rotation curve is supported by the halo.
In this case some of the lensing can come from the dark halo.
This changes the predictions of MACHO halo fraction, since the
LMC halo contributes very little to the total mass of the Milky Way,
but relatively more to the microlensing.  As shown in
Table~\ref{tab-like}, for Model S we 
find 1.1 events from the LMC halo, and 2.1 background events
using the set A.  For set B
we find 1.4 LMC halo events, with 2.7 background stellar events.
Again, the expected number of background events is significantly
smaller than the number of observed events.
When an LMC dark halo is included the events from the LMC halo count towards
dark matter that is not uniformly spread across the sky.
The predicted values change from $f=0.21$ to $f=0.20$
and from $m=0.60 \msun$ to $m=0.54\msun$ 
for set A,
and from $f=0.24$ to $f=0.22$, and from $m=0.79\msun$ to $m=0.72\msun$
for set B.  The change in MACHO fraction is small because
our LMC halo has an optical depth of $0.79 \ten{-7}$, substantially
smaller than the $4.7\ten{-7}$ contributed by Model S, and the LMC disk
contributes more background when no halo is present.
The most likely total mass in MACHOs in the Milky
Way dark halo (within 50 kpc) shows an expected drop of about 10\%,
(from $8.5 \ten{10} \msun$ to $7.9 \ten{10}\msun$) 
when an LMC dark halo is included.
As we discuss below, with Model F, which has a very small Milky Way halo,
the change in $f$ and the change in total MACHO mass is much more substantial.

Finally we note that with a typical halo model like S,
the likelihood contours in all cases
rule out a 100\% MACHO halo at high significance.  
This was not true in A97, or in any previous microlensing data set.
(But see the very recent EROS II reports \cite{eros-taup99}).
We note that our sensitivity to events longer
than a thousand days is small, so we cannot rule out dark matter objects
with masses of tens of solar masses.

\subsection{Interpretation}

\placetable{tab-like} 

Several interesting features can be seen in Table~\ref{tab-like}
and Figures~\ref{fig-like-cow}, \ref{fig-like-b}, and \ref{fig-like-f}.

Examination of the likelihood contours show that with our new data set,
the uncertain nature of the Milky Halo dominates
over Poisson error and the systematic error caused by our selection criteria.
For each model, the most likely values and confidence limits are
nearly the same, while between models there are significant differences.
This is an improvement over A96 and A97 where small number statistics
dominated the errors.
The values found are typically within 1.5 sigma
of those reported in A97, but a factor of two smaller 
for reasons given above.  

As noted in A96 and A97, the most likely halo fraction, $f$, depends strongly
on the halo model, with massive halos such as Model B giving small
MACHO fraction ($f \sim 13$\%), medium halos such as Model S giving
medium values ($f \sim 22$\%), and very low-mass halos such as Model F
giving large fractions ($f \sim 40$\% - 60\%).
However, there are some model independent conclusions that can be drawn. 
The total predicted mass in MACHOs within 50 kpc (column 5) is about
$9 \ten{10}\msun$ for all models.  This is again a factor of two smaller
than reported in A97 for the same reasons.  However, for very small halos,
such as model F, the total MACHO mass is somewhat dependent on the model of
the LMC halo.  With no dark LMC halo, masses up to $10 \ten{10}\msun$
are found, while with a large dark LMC halo, the prediction drops
to $8 \ten{10}\msun$.  This is because this MW halo has an optical
depth of only $1.9 \ten{-7}$, compared to $0.79\ten{-7}$ for our
LMC halo.
So with the large MACHO fraction caused
by the very small MW halo, the LMC halo can contribute substantially
to microlensing without contributing much to the mass within 50 kpc.
Note, however, that when one sums the {\it total } MACHO contribution
to microlensing optical depth (MW halo plus LMC halo), the result is 
$\tau_{\rm ML} = 1.1 \pm 0.4 \ten{-7}$ almost completely
independent of the MW and LMC halo models.
This is shown in column~6 of Table~\ref{tab-like}.
These values and
their confidence intervals are simple to interpret statistically, since
each model provides a distribution of event durations.  Thus the subtleties
discussed in Section~\ref{sec-tau} are absent.  We note that these values
include only the LMC and MW halo contributions, and
are close to the values we obtained in our direct estimates of optical
depth.

Interestingly, for models S and B, halos consisting 100\%
of MACHOs are strongly ruled-out.  Even for the rather extreme
model F, and no LMC halo, a 100\% MACHO halo is ruled-out.
The only way this data is consistent with a 
100\% MACHO halo is if there is an extremely small MW halo coupled with
a very small LMC halo.
This is the strongest limit to date on an all MACHO halo and 
is a major result of this work.  We note that we do not set 
strong limits on dark matter objects with masses in 
the tens of solar mass range.
However, the $9 \ten{10}\msun$ in MACHOs found in this work
still represents
several times the mass of all known stellar components of the Milky
Way. If the bulk of the lenses are located in the halo, then they
represent the dominant identified 
component of our Galaxy, and a major portion of the dark matter.

While $\tau_{\rm MACHO}$ and the total mass are fairly model independent,
as discussed in A96 and A97,
the typical mass of a MACHO, $m$, is not.  
Lighter halo models such as F 
have a smaller implied MACHO mass $m_{\rm ML} \sim 0.2\msun$, 
while heavier halos such as B have $m_{\rm ML} \sim 0.8\msun$, and
medium halos such as S giving $m_{\rm ML} \sim 0.6\msun$.
Sets A and B also differ slightly in this parameter.
We conclude that our estimate of $m$ is not very robust, but that
masses below the brown dwarf limit of $0.08\msun$ are quite unlikely.
Therefore, the nature of the lenses remains unclear.

Finally, using Table~\ref{tab-like}, and comparing the
number of expected events from the MW halo with the number of
expected events from known stellar populations,
one can ask how strong is the case for any MACHO contribution
to the dark matter.
For set A, the known stellar background ranges from 
2.1 to 3.2 events
depending upon the LMC and MW halo model.  For set B,
the range is 2.7 to 4.2 stellar events.  
If our models of the MW thin disk,
thick disk, spheroid, and LMC disk are adequate then it is very unlikely
to find 13 (or 17) events when expecting no more than 3.2 (or 4.2)
(probability less than $10^{-5}$ in both cases).
We note that the LMC disk we use in the pure-disk model is conservatively
large ($M_{\rm disk} = 5 \ten{9} \msun$).

The question remains whether an LMC halo could supply the observed 
microlensing. The halo we use is a good fit to the available data,
($M_{\rm halo} = 6 \ten{9} \msun$ within 8 kpc), 
and the rotation curve for this halo
plus LMC disk is a good fit to the data \cite{gyuk-lmc}.
Now, due to the lack
of stellar tracers found with velocity dispersion of the $\sim 50$ km/s
expected for a halo population,
we have assumed that this LMC halo is dark.  
A large dark halo is, of course, typical of dwarf spiral galaxies such as
the LMC.  So assuming the halo consists of a fraction $f$ of MACHOs,
is a reasonable first approximation.
Table~\ref{tab-like} shows almost no difference in predicted
background or other quantities for Models S and B that have medium
to large dark halos;  however, for the very small halo Model F,
nearly half the expected events come from stars or the LMC halo.
The likely final halo fraction, is still nearly 50\%, but with substantially
larger error bars.  Figure~\ref{fig-like-f} shows, however,
that even in this case, a no-MACHO halo is quite unlikely.

Recently, several groups \cite{aubourg-lmc,weinberg,gyuk-lmc} have
considered lensing by an extended stellar population around the LMC, and have
obtained different estimates of the optical depth contribution.
Interestingly, Graff, \etal\ (1999) claim tentative ($\sim 2\sigma$)
detection of a kinematically distinct population.  
We therefore explore the effect this
would have on our maximum likelihood analysis, by considering the effect
of an LMC halo consisting of MACHO fraction $f$
and a Milky Way halo with no MACHO population (thus presumably consisting
entirely of exotic elementary particles).  
The likelihood maximum for such a model
gives $f=1.35^{+0.6}_{-0.4}$, and $m=0.2^{+0.1}_{-0.08}\msun$ for
set A, and 
gives $f=1.52^{+0.6}_{-0.4}$, and $m=0.3^{+0.11}_{-0.09}\msun$ for
set B.
Thus a possible non-dark-matter explanation
for our results is an LMC halo of mass $\sim 9 \ten{9}\msun$, consisting
of stellar-like objects that have not yet been convincingly observed.
This value of LMC halo mass is somewhat extreme, 
though there are published models with masses this high.
Clearly
it is important to discover, or convincingly rule-out, the possibility
of a large LMC stellar halo.

\section{Summary and Discussion} 
\label{sec-discuss}

We have detected between 13 and 17 microlensing events towards the LMC.  
The implied optical depth,
microlensing rate, and MACHO halo fraction are a factor of two smaller than
found in our previous work, but are consistent with previous
results within the errors of small number statistics.  
The larger number of events allows us to reduce the Poisson
error considerably, which along with our improved efficiency analysis
and study of sources of systematic error, means that the interpretation 
of the microlensing events is now dominated by uncertainties in the
models of the Milky Way and LMC.
We find that the number of events is not consistent with known lens
sources and our measured optical depth, $\tau_2^{400} = 1.2 ^{+0.4}_{-0.3}
\ten{-7}$, is significantly larger
than allowed by known Galactic and LMC stellar populations.  The total 
implied mass in MACHOs within 50 kpc is $\sim 9 \ten{10} \msun$,
quite independent of the dark halo model. 
This is substantially larger than all known stellar components of the 
Galaxy.  However, one of our most important conclusions is that a 100\% all
MACHO Milky Way halo is ruled out at the 95\% C.L. for a wide
range of reasonable models.

One explanation of our results is a Milky Way halo consisting
of about 20\% MACHOs.
Another possibility is an 
LMC halo that dominates the microlensing, and no MACHOs in the Milky
Way halo.
The spatial distribution of events
makes lensing entirely by a stellar population in the LMC disk or bar
less likely, but given the highly
uncertain nature of an LMC halo, a previously 
unknown stellar component or an LMC halo consisting of MACHO dark matter,
could explain the measured optical depth, the
number of observed events and their spatial distribution on the sky. 
We note, however, that no known LMC stellar population exhibits kinematics
of this nature.  It would help to constrain the kinematics of old populations
in the LMC and to look for new populations that could represent the
lenses.
A direct measurement of the distance to some LMC lenses would be 
especially useful in distinguishing the two possibilities above.

There is intriguing evidence for a population of white dwarfs 
(Ibata \etal 1999;
M\'{e}ndez \& Minniti 1999) in the Hubble Deep Fields (North and South),
consistent with the Milky Way halo hypothesis.
These results were spawned primarily by the recent work of Hansen (1999) 
on low temperature white dwarf cooling curves.  Although the 
identification of these faint-blue objects as white dwarfs remains
to be confirmed and the small sample size restricts 
an accurate estimate, the suggestion that these white dwarfs
could contribute 1/3 to 1/2 of the dark matter in the Milky
Way is certainly stimulating
in light of the present work.  A third epoch HDF image
to check the proper motion
of these objects should help confirm or rule-out this hypothesis.
A Galactic halo composed of $\sim\,20$\% by mass of white dwarfs
would seem to be a natural explanation of both the 
microlensing data and this newly observed population,
but the formation of such objects and the chemical enrichment
they would cause trigger serious concern \cite{fields}.

\acknowledgments
\section*{Acknowledgments}

We are very grateful for the skilled support given our project 
by the technical staffs at the Mt.~Stromlo and CTIO Observatories,
and in particular we would like to thank Simon Chan,  Glen Thorpe,
Susannah Sabine, and Michael McDonald, for their invaluable assistance in 
obtaining the data.  
We would like to thank Dave Reiss for suppling the SN type Ia templates and
help with the SN fits.  We thank Geza Gyuk for help in modeling and 
many useful discussions.  We thank the NOAO for making nightly use
of the CTIO 0.9~m telescope possible.

Work performed at LLNL is supported by the DOE under contract W7405-ENG-48.
Work performed by the Center for Particle Astrophysics personnel 
is supported in part by the Office of Science and Technology Centers of
NSF under cooperative agreement AST-8809616.
Work performed at MSSSO is supported by the Bilateral Science 
and Technology Program of the Australian Department of Industry, Technology
and Regional Development. 
DM is also supported by Fondecyt 1990440.  CWS thanks
the Packard Foundation for their generous support.  WJS is supported
by a PPARC Advanced Fellowship. CAN was supported in part by an NPSC
Fellowship.
ND and KG were supported in part by the DOE under grant DEF03-90-ER 40546.
TV was supported in part by an IGPP grant.

%
%-------- RRRRRRRRR  references --------------
%

%

\onecolumn %----- switch back to single column mode

%%%%% TTTTTTTTTTTT  Tables here. 
%% WARNING - don't have blank lines between the \enddata and
%%  the \end{deluxetable}, or centering gets screwed. 

\clearpage

\begin{deluxetable}{rccc}  % 4 columns. 
\tablecaption{ Field Centers \label{tab-fields} }
\tablewidth{0pt}
\tablehead{
%% Use a footnote to explain numbering. 
\colhead{Field No.} & 
  \colhead{ Center: RA } & \colhead{Dec (2000)}  & \colhead{Observations} 
}  % end header. 

\startdata
  1 ...... & 05 05 23 & -69 05 24 & 1017 \nl
  2 ...... & 05 12 47 & -68 30 21 &  860 \nl
  3 ...... & 05 22 24 & -68 28 01 &  720 \nl
  5 ...... & 05 11 17 & -69 40 18 &  839 \nl
  6 ...... & 05 20 00 & -70 17 10 &  856 \nl
  7 ...... & 05 28 54 & -70 27 31 & 1027 \nl
  9 ...... & 05 10 57 & -70 23 40 &  811 \nl
 10 ...... & 05 04 34 & -69 52 19 &  665 \nl
 11 ...... & 05 36 56 & -70 31 34 &  930 \nl
 12 ...... & 05 45 36 & -70 35 16 &  772 \nl
 13 ...... & 05 19 39 & -70 51 40 &  752 \nl
 14 ...... & 05 35 53 & -71 09 22 &  741 \nl
 15 ...... & 05 45 34 & -71 14 36 &  718 \nl
 17 ...... & 04 57 04 & -69 43 11 &  345 \nl
 18 ...... & 04 57 55 & -68 56 08 &  594 \nl
 19 ...... & 05 06 09 & -68 21 03 &  672 \nl
 22 ...... & 05 11 18 & -71 00 19 &  398 \nl
 23 ...... & 05 02 51 & -70 35 39 &  339 \nl
 24 ...... & 05 00 39 & -67 56 45 &  297 \nl
 47 ...... & 04 53 05 & -68 01 26 &  610 \nl
 53 ...... & 05 02 09 & -66 40 19 &  180 \nl
 55 ...... & 05 02 15 & -65 58 50 &  241 \nl
 57 ...... & 05 09 07 & -65 46 07 &  189 \nl
 76 ...... & 05 44 13 & -69 49 41 &  386 \nl
 77 ...... & 05 27 24 & -69 45 24 & 1338 \nl
 78 ...... & 05 19 26 & -69 42 27 & 1312 \nl
 79 ...... & 05 12 59 & -69 05 43 & 1226 \nl
 80 ...... & 05 22 44 & -69 05 18 & 1186 \nl
 81 ...... & 05 35 56 & -69 49 34 &  792 \nl
 82 ...... & 05 32 50 & -69 03 18 &  757 \nl
\enddata
\tablenotetext{} { This table lists the 30 well-sampled 
 fields used in the current analysis, the 8 new fields
 are 17, 22, 23, 24, 53, 55, 57, \& 76 and typically have
 less than half the observations of the top 22 fields. We
 observe 82 LMC fields in total, but the remaining 52
 were observed less often ($\sim 120$ observations each).
 } 
\end{deluxetable} 

\clearpage
\begin{deluxetable}{cccc}  % 5 columns. 
\tablecaption{ Selection Criteria \label{tab-cuts} }
\tablewidth{0pt}
\tablehead{
%% Use a footnote to explain numbering. 
 \colhead{Description} & 
 \colhead{Year 2 (A97)} & \colhead{Criteria A} &
 \colhead{Criteria B}
}  % end header. 
\startdata
Min. coverage & $bmrN >= 7$  & $rN > 0$ \& $bN > 0$ & $>65$ simul. \nl
              & $>40$ baseline points & $>45$ baseline points & baseline points \nl
			  & $\that < 300$ & $\that < 600$ & $\that < 600$ \nl   
			  &               & $\tmax > 310$ & $\tmax > 310$ \nl\nl   
SN87A echo  & $10'\times10'$ square excl. & $10'\times10'$ square excl. &
$10'\times10'$ square excl. \nl\nl
Crowd \& PSF & 
$f_{CRD} < \left(\Delta\chi^2 /(\chi^2/N_{\rm dof})\right)^{10/9}/520$ & None &
$pkcrdrej\,+\,$ \nl
& \& $crdrej < 0.05$ & & $pkpsfrej < 0.2$ \nl\nl
Bumper cut  & $V > 17.5$ & $V > 17.5$ & $V > 17$ \nl
            & \& $V-R < 0.9$ & \& $V-R < 0.9$ & \& ($\Amax > 1.75$ or    \nl
            &                &                & $V > 19$ or $V-R > 0.4$) \nl\nl
Variable cut  & None & $bauto/rauto > 0.75$ & $pfwsr > 0.6$ \nl
              &      &                      & \& $rbcrossout < 0.75$ \nl\nl
High points & 6 pts. $>2\sigma$ & 7 pts. $>2\sigma$ & 10 pts. $>2\sigma$ \nl
            & \& $\geq 1$ pt.~on rise \& fall &  & \& $N_{\rm hi}/N_{\rm pk} > 0.9$ \nl\nl
Baseline fit & $\chi^2_{\rm ml-out}/N_{\rm dof} < 4$ & $\chi^2_{\rm ml-out}/N_{\rm dof} < 1.8$ &
               $\chi^2_{\rm ml-out}/N_{\rm dof} < 4$ \nl 
             & & & \& $\chi^2_{\rm robust-out}/N_{\rm dof} < 1.5$ \nl\nl
2nd S/N  & $\Delta\chi^2 / (\chi^2_{\rm peak}/N_{\rm dof}) > 200$ & 
                   $\Delta\chi^2 / (\chi^2_{\rm peak}/N_{\rm dof}) > 350$ & None \nl\nl
Main S/N & $\Delta\chi^2 /(\chi^2_{\rm ml}/N_{\rm dof}) >500$ &
           $\Delta\chi^2 /(\chi^2_{\rm ml}/N_{\rm dof}) > 400$ &
           $\Delta\chi^2 /(\chi^2_{\rm ml}/N_{\rm dof}) > 300$ \nl\nl
Magnification & $\Amax > {\rm max}(1.75, 1+2\overline{\sigma})$ & 
                $\Amax > {\rm max}(1.49, 1+3.5\overline{\sigma})$ & 
                $\Amax > {\rm max}(1.34, 1+4\overline{\sigma})$ \nl\nl
2nd peak      & None & None & $pfrdev2 < 90$ \nl\nl
Supernova cut & By eye  &  $\Delta\chi^2_{\rm SN-ML}>0$ \& not~event~22 & 
$\Delta\chi^2_{\rm SN-ML}>0$ \nl 
\enddata
\end{deluxetable} 

\clearpage
\begin{deluxetable}{rccccccccc}  % 10 columns. 
\tablecaption{ Candidate Microlensing Events
\label{tab-events} }
\tablewidth{0pt}
\tablehead{
%% Use a footnote to explain numbering. 
\colhead{Event \tablenotemark{a}} & \colhead{ID} &
\colhead{RA (2000)} & \colhead{Dec (2000)}  & \colhead{V} & 
\colhead{V-R}  & \colhead{$\tmax$} & \colhead{$\that$} & 
\colhead{$\Amax$} & 
\colhead{$\chi^2/N_{\rm dof}$}
}  % end header. 

\startdata
  1a...... &  2.5628.5917  & 05 14 44 & -68 48 01 &  19.75 &  0.56 &
  433.6  &   34.4  &   7.15  &  1.083 \nl

  1b...... & 79.5628.1547  & 05 14 44 & -68 48 00 &  19.73 &  0.55 &
  433.8  &   34.2  &   7.62  &  1.051 \nl

  4....... & 13.5961.1386  & 05 17 14 & -70 46 58 &  20.25 &  0.14 &
  1023.4  &   45.4  &   2.92  &  1.415 \nl

  5....... &  6.5845.1091  & 05 16 41 & -70 29 18 &  21.15 &  0.76 &
  400.4  &   75.6  &  47.28  &  1.512 \nl

  6....... &  7.7420.2571  & 05 26 13 & -70 21 14 &  19.97 &  0.12 &
  573.6  &   91.6  &   2.43  &  0.763 \nl

  7a...... & 10.3802.872   & 05 04 03 & -69 33 18 &  20.87 &  0.32 &
  840.0  &  102.9  &   5.91  &  1.398 \nl

  *7b..... & 10.3802.494   & 05 04 04 & -69 33 18 &  19.85 &  0.18 &
  832.9  &   43.9  &   1.63  &  1.046 \nl

  8....... & 77.7307.4800  & 05 25 09 & -69 47 53 &  20.31 &  0.21 &
  764.3  &   66.4  &   2.19  &  1.733 \nl

  *9\tablenotemark{b}...... &
             80.6468.2746  & 05 20 20 & -69 15 11 &  19.58 &  0.27 &
  976.4  &  179.2  &   1.95  &  6.579 \nl

 10a\tablenotemark{c}.....  &  1.3324.122   & 05 01 15 & -69 07 33 &  19.60 &  0.23 &
 582.2  &   41.8  &   2.45  &  1.455 \nl

 10b\tablenotemark{c}.....  & 18.3324.1765  & 05 01 16 & -69 07 33 &  19.47 &  0.14 &
 581.8  &   42.7  &   2.38  &  1.455 \nl

 *11\tablenotemark{c}...... & 11.8746.130   & 05 34 21 & -70 41 07 &  20.04 &  0.85 &
 368.5  &  280.7  &  11.23  &  2.365 \nl

 12a\tablenotemark{c}.....  & 11.8622.1257  & 05 33 51 & -70 50 57 &  21.34 &  0.25 &
 366.6  &  137.4  &   7.32  &  1.238 \nl

 12b\tablenotemark{c}.....  & 14.8622.4762  & 05 33 51 & -70 50 59 &  21.51 &  0.39 &
 365.7  &  170.5  &   7.31  &  1.237 \nl

 13......  & 80.7080.5384  & 05 24 03 & -68 49 12 &  21.02 &  0.26 &
 1510.5  &  100.1  &   2.36  &  1.170 \nl

 14......  & 11.8871.2108  & 05 34 44 & -70 25 07 &  19.37 &  0.01 &
 1767.8  &  100.1  &   3.37  &  0.751 \nl

 15......  & 10.4162.3555  & 05 05 46 & -69 43 51 &  21.03 &  0.14 &
 1848.9  &   36.8  &   2.83  &  0.905 \nl

 *16\tablenotemark{c}...... & 79.4655.4621  & 05 09 16 & -69 08 15 &  18.71 &  0.56 &
 1934.1  &   27.0  &   1.76  &  0.984 \nl

 *17\tablenotemark{c}...... &  9.5362.408   & 05 13 35 & -70 24 43 &  20.05 &  0.48 &
 2199.5  &   24.1  &   1.85  &  1.059 \nl

 18......  & 15.10554.465  & 05 45 21 & -71 09 11 &  19.55 &  0.35 &
 1159.3  &   74.2  &   1.54  &  1.217 \nl

 *19\tablenotemark{c}...... & 15.10669.178  & 05 46 18 & -71 31 48 &  19.44 &  0.82 &
 687.5  &   16.3  &   1.41  &  0.734 \nl

 *20...... & 17.2221.1574  & 04 54 19 & -70 02 15 &  21.35 &  0.57 &
 2151.2  &   72.7  &   2.95  &  1.298 \nl

 21......  & 17.2714.1058  & 04 57 14 & -69 27 48 &  19.37 &  0.03 &
 589.7  &   93.2  &   5.64  &  1.620 \nl

 22......  & 22.5472.1126  & 05 14 32 & -71 09 12 &  20.64 &  0.25 &
 1333.2  &  229.3  &   2.70  &  1.277 \nl

 23......  & 23.4143.256   & 05 06 17 & -70 58 47 &  20.27 &  0.31 &
 1138.8  &   85.2  &   2.41  &  1.492 \nl

 *24\tablenotemark{c}...... & 24.2862.1187  & 04 57 46 & -67 41 08 &  20.63 &  0.49 &
 2201.4  &  186.5  &   4.09  &  3.927 \nl

 25......  & 24.3583.2286  & 05 02 16 & -68 00 52 &  19.04 &  0.48 &
 1110.8  &   85.2  &   1.50  &  0.733 \nl

 *26\tablenotemark{c}...... & 47.1764.464   & 04 51 11 & -68 16 41 &  19.46 &  0.21 &
 2170.6  &   44.8  &   1.88  &  1.305 \nl

 *27...... & 1.4289.1748   & 05 06 35 & -69 20 48 &  19.24 &  0.07 &
 890.6  &   50.5  &   1.45  &  1.239 \nl

\enddata
\tablenotetext{} {
The magnitudes and colors are fit baselines using the best available
calibrations for each field as described in \S~\ref{sec-obs}. 
Time of peak magnification
$\tmax$ is in JD - 2,448,623.5 (January 2, 1992).}
\tablenotetext{a} {Events 1-12 appeared in \yrtwo.
We number the current sample 1, 4 \ldots 27 to avoid any ambiguity with
the previously published events.} 
\tablenotetext{b} {Event 9 is the binary microlensing event; the parameters
here are those resulting from a single-lens fit, and are not strictly
appropriate.} 
\tablenotetext{c}{Probable supernova.}
\tablenotetext{*} {Events marked with ``*" do not pass selection criteria A.}
\end{deluxetable} 

\clearpage
\begin{deluxetable}{rcccccccc}  % 9 columns. 
\tablecaption{ Microlensing Fits with Blending
\label{tab-blend} }
\tablewidth{0pt}
\tablehead{
%% Use a footnote to explain numbering. 
\colhead{Event} & \colhead{$\tmax$} & 
\colhead{$\that$} & \colhead{$\Amax$} &
\colhead{$f_{0V}$} & \colhead{$f_{0R}$} & 
\colhead{$f_{V}$} & \colhead{$f_{R}$} & 
\colhead{$\chi^2/N_{dof}$}
}  % end header. 

\startdata

%event    t0        that     Amax  Vflux    Rflux     fV     fR        chi2/dof
%--------------------------------------------------------------------------------
  1a...... &  433.6 &   34.5 &   7.19 &  58.60 &  93.30 & 0.984 & 1.000 &  1.083 \nl   
  1b...... &  433.7 &   34.7 &   7.83 &  47.23 &  77.38 & 0.972 & 0.982 &  1.052 \nl   
   4...... & 1023.0 &   83.3 &   6.98 &  40.85 &  35.00 & 0.322 & 0.365 &  1.380 \nl   
   5...... &  400.4 &  109.8 &  1.2e7 &  15.98 &  34.35 & 1.000 & 0.457 &  0.843 \nl   
   6...... &  573.6 &   92.0 &   2.45 &  42.63 &  41.33 & 0.981 & 1.000 &  0.764 \nl   
  7a...... &  840.1 &  112.6 &   6.87 &  23.12 &  25.51 & 1.000 & 0.748 &  1.328 \nl   
 *7b...... &  833.2 &   46.5 &   1.71 &  61.37 &  56.25 & 1.000 & 0.761 &  1.042 \nl   
   8...... &  764.3 &   66.4 &   2.19 &  38.45 &  32.52 & 1.000 & 1.000 &  1.735 \nl   
% *9\tablenotemark{a}......
%    &  977.3 &  201.8 &   2.22 &  82.67 &  73.87 & 0.637 & 0.999 &  6.269 \nl   
  *9\tablenotemark{a}......
     &  979.5 &  143.4 &   .... &  79.37 &  83.30 & 0.260 & 0.170 &  1.755 \nl   
 10a\tablenotemark{b}...... &  582.1 &   43.6 &   2.56 &  74.38 &  71.49 & 1.000 & 0.871 &  1.451 \nl   
 10b\tablenotemark{b}...... &  582.0 &  128.8 &  10.12 &  79.99 &  67.05 & 0.163 & 0.160 &  1.447 \nl   
 *11\tablenotemark{b}...... &  367.7 &  436.9 &  20.96 &  31.24 &  93.80 & 0.998 & 0.366 &  1.965 \nl   
 12a\tablenotemark{b}...... &  367.1 &  213.5 &  13.41 &  13.09 &  15.59 & 0.604 & 0.440 &  1.184 \nl   
 12b\tablenotemark{b}...... &  367.0 & 1002.0 &  65.94 &   9.15 &  12.45 & 0.140 & 0.091 &  1.138 \nl   
  13...... & 1510.0 &  222.7 &   6.95 &  17.79 &  17.64 & 0.219 & 0.260 &  1.158 \nl   
  14...... & 1768.0 &  106.5 &   3.67 &  81.16 &  59.87 & 0.901 & 0.874 &  0.750 \nl   
  15...... & 1849.0 &   41.9 &   3.48 &  20.11 &  16.88 & 0.776 & 0.735 &  0.906 \nl   
 *16\tablenotemark{b}...... & 1934.0 &   27.0 &   1.76 & 148.00 & 226.70 & 1.000 & 1.000 &  0.985 \nl   
 *17\tablenotemark{b}...... & 2200.0 &   24.3 &   1.88 &  42.44 &  61.96 & 1.000 & 0.960 &  1.060 \nl   
  18...... & 1159.0 &   75.8 &   1.58 &  80.26 &  91.15 & 1.000 & 0.892 &  1.217 \nl   
 *19\tablenotemark{b}...... &  687.5 &   18.6 &   1.62 &  86.28 & 194.90 & 1.000 & 0.465 &  0.702 \nl   
 *20...... & 2152.0 &   99.4 &   5.42 &  12.30 &  20.21 & 0.628 & 0.407 &  1.253 \nl   
  21...... &  589.4 &  141.5 &  11.59 &  91.01 &  67.39 & 0.499 & 0.502 &  1.592 \nl   
  22...... & 1333.0 &  233.9 &   2.73 &  26.04 &  26.31 & 0.927 & 1.000 &  1.278 \nl   
  23...... & 1139.0 &   88.9 &   2.61 &  39.25 &  41.98 & 1.000 & 0.801 &  1.452 \nl   
  *24\tablenotemark{b}...... & 2201.0 &  186.3 &   4.09 &  22.58 &  32.85 & 1.000 & 1.000 &  3.941 \nl   
  25...... & 1111.0 &   85.3 &   1.51 & 112.50 & 156.40 & 0.924 & 1.000 &  0.734 \nl   
 *26\tablenotemark{b}...... & 2169.0 &  260.0 &  21.34 &  82.81 &  77.29 & 0.056 & 0.057 &  1.277 \nl   
 *27...... &  895.0 & 3247.0 & 258.90 &  85.17 &  67.56 & 0.002 & 0.003 &  1.218 \nl

\enddata
\tablenotetext{} {
 Time of peak magnification $\tmax$ is in JD - 2,448,623.5 (January 2, 1992).}
\tablenotetext{a} {For the binary microlensing event (9) the fit
 parameters are given for the binary lens fit
 \cite{macho-binary}.
Not all of the these parameters are appropriate for this fit.}
\tablenotetext{b}{Probable supernova.}
\tablenotetext{*} {Events marked with ``*" do not pass selection criteria A.}
\end{deluxetable} 

\clearpage
\begin{deluxetable}{rccccccc}  % 8 columns. 
\tablecaption{ Supernova Type Ia Fits
\label{tab-sn} }
\tablewidth{0pt}
\tablehead{
%% Use a footnote to explain numbering. 
\colhead{Event} & 
\colhead{$\tmax$} & \colhead{$\mu_{V}$} & 
\colhead{$\mu_{R}$} & \colhead{$\delta$} &
\colhead{$f_{0V}$} & \colhead{$f_{0R}$} & 
\colhead{$\chi^2/N_{dof}$}
}  % end header. 

\startdata

%event     t0       muV       muR   delta      Vflux      Rflux    chi2/dof
%------------------------------------------------------------------------------
  1a...... &    431.1 &  37.44 &  36.82 &  0.75 &    57.86 &    92.39 &    3.946 \nl
  1b...... &    431.1 &  37.68 &  37.10 &  0.75 &    46.74 &    76.76 &    3.283 \nl
   4...... &   1020.0 &  38.11 &  37.98 &  0.75 &    40.75 &    34.88 &    1.617 \nl
   5...... &    397.1 &  36.94 &  36.76 &  0.75 &    15.21 &    34.57 &    4.187 \nl
   6...... &    566.4 &  39.49 &  38.89 & -0.50 &    42.99 &    41.66 &    0.865 \nl
  7a...... &    837.1 &  37.44 &  37.50 &  0.70 &    23.14 &    25.60 &    1.385 \nl
 *7b...... &    830.1 &  38.50 &  38.69 &  0.75 &    61.39 &    56.26 &    1.058 \nl
   8...... &    758.1 &  39.63 &  39.19 & -0.22 &    38.51 &    32.24 &    1.750 \nl
  *9...... &    952.1 &  39.02 &  38.10 & -0.50 &    84.93 &    77.97 &    7.030 \nl
 10a\tablenotemark{a}...... &    577.2 &  37.72 &  37.68 &  0.59 &    74.15 &    71.18 &    1.286 \nl
 10b\tablenotemark{a}...... &    576.3 &  38.02 &  37.95 &  0.33 &    80.06 &    67.04 &    1.252 \nl
 *11\tablenotemark{a}...... &    362.5 &  36.97 &  36.26 & -0.42 &    36.47 &    98.99 &    1.871 \nl
 12a\tablenotemark{a}...... &    363.2 &  38.45 &  38.24 &  0.14 &    13.33 &    15.74 &    1.101 \nl
 12b\tablenotemark{a}...... &    363.9 &  38.66 &  38.45 &  0.11 &     9.59 &    12.77 &    1.128 \nl
  13...... &   1497.0 &  40.52 &  39.56 & -0.50 &    18.41 &    18.15 &    1.287 \nl
  14...... &   1758.0 &  38.31 &  38.06 & -0.50 &    82.35 &    60.47 &    1.754 \nl
  15...... &   1845.0 &  39.04 &  39.10 &  0.75 &    20.30 &    16.97 &    0.948 \nl
 *16\tablenotemark{a}...... &   1930.0 &  37.78 &  36.91 &  0.75 &   148.00 &   226.60 &    0.976 \nl
 *17\tablenotemark{a}...... &   2196.0 &  39.13 &  38.71 &  0.75 &    42.26 &    61.82 &    1.036 \nl
  18...... &   1147.0 &  39.49 &  38.92 & -0.50 &    80.91 &    91.76 &    1.406 \nl
 *19\tablenotemark{a}...... &    684.7 &  38.52 &  38.30 &  0.75 &    86.21 &   194.80 &    0.699 \nl
 *20...... &   2150.0 &  38.89 &  38.71 &  0.72 &    12.42 &    20.49 &    1.305 \nl
  21...... &    586.1 &  36.10 &  36.25 &  0.75 &    91.31 &    67.57 &    2.287 \nl
  22...... &   1318.0 &  39.40 &  38.89 & -0.50 &    26.83 &    27.32 &    1.797 \nl
  23...... &   1128.0 &  39.76 &  39.22 & -0.50 &    39.98 &    42.12 &    1.855 \nl
  *24\tablenotemark{a}...... &   2183.0 &  38.86 &  37.96 & -0.50 &    23.56 &    34.24 &    3.544 \nl
  25...... &   1103.0 &  39.49 &  38.49 & -0.50 &   113.00 &   157.80 &    1.243 \nl
 *26\tablenotemark{a}...... &   2165.0 &  38.16 &  38.06 &  0.41 &    83.10 &    77.45 &    1.222 \nl
 *27...... &    892.9 &  38.64 &  38.62 &  0.75 &    86.24 &    68.44 &    1.302 \nl

\enddata
\tablenotetext{} {
Time of peak magnification $\tmax$ is in JD - 2,448,623.5.
Events marked with ``*" do not pass selection criteria A.}
\tablenotetext{a}{Probable supernova.}
\end{deluxetable} 

\clearpage
%\begin{landscape}
\begin{deluxetable}{rccclcl}  % 7 columns. 
\tablecaption{ Supernova and Event Summary Table: \label{tab-summary} }
\tablewidth{0pt}
\tablehead{
%% Use a footnote to explain numbering. 
\colhead{Event} & 
\colhead{$\chi^2_{\rm ML}/N_{\rm dof}$}\tablenotemark{b} &
\colhead{$\chi^2_{\rm SN}/N_{\rm dof}$} &
\colhead{$\Delta\chi^2_{\rm SN-ML}$} & 
\colhead{Galaxy?} & 
\colhead{Follow-Up} &
\colhead{Notes}
}  % end header. 
\startdata
   1a...... & 1.083 &  3.946 &  6444   & No (HST)    &  ....        
      &  \nl 
   1b...... & 1.052 &  3.283 &  3685   & No (HST)    &  ....         
      &  \nl 
   4......  & 1.380 &  1.617 &   306   & No (HST)    &  CTIO+spec. 
      &  \nl 
   5...... & 0.843 &  4.187 &  4021   & No (HST)    &  ....     
      &  \nl 
   6......  & 0.764 &  0.865 &   173   & No (CTIO)   &  ....      
      &  \nl 
   7a...... & 1.328 &  1.385 &   71.0  & No (HST)    &  ....      
      &  \nl 
  *7b...... & 1.043 &  1.058 &   20.0  & No (HST)    &  ....      
      &  \nl 
   8......  & 1.735 &  1.751 &   28.0  & No (HST)    &  ....    
      &  \nl 
  *9......  & 6.269 &  7.029 &  1047   & No (HST)    &  ....    
      &  Binary ML  \nl 
  10a\tablenotemark{a}...... & 1.451 &  1.286 &  -163   & Yes (HST)   &  ....    
      &  \nl 
  10b\tablenotemark{a}...... & 1.447 &  1.252 &  -226   & Yes (HST)   &  ....        
      &  \nl 
 *11\tablenotemark{a}......  & 1.966 &  1.870 &  -117   & Yes (CTIO)  &  ....     
      &  \nl 
  12a\tablenotemark{a}...... & 1.184 &  1.101 &  -113   & Yes (HST)  &  ....    
      &  \nl 
  12b\tablenotemark{a}...... & 1.138 &  1.128 &  -12.0  & Yes (HST)  &  ....   
      & \nl 
  13......  & 1.158 &  1.287 &  276    & No (CTIO)   &  CTIO    
      &  \nl 
  14......  & 0.750 &  1.755 &  1724   & No (HST)    &  CTIO   
      &  \nl 
  15......  & 0.906 &  0.948 &   45.6  & No (CTIO)   &  CTIO   
      &  \nl 
 *16\tablenotemark{a}......  & 0.985 &  0.976 &  -13.0  & Yes (CTIO)  &  CTIO    
      & \nl 
 *17\tablenotemark{a}......  & 1.060 &  1.036 &  -31.0  & Yes (CTIO)  &  CTIO    
      &  \nl 
  18......  & 1.217 &  1.406 &  274    & No (CTIO)   &  ....   
      &  \nl 
 *19\tablenotemark{a}......  & 0.701 &  0.699 &   -3.1  & Yes (CTIO)  &  ....   
      &  \nl 
 *20......  & 1.253 &  1.305 &   23.9  & No (CTIO)   &  ....    
      &  \nl 
  21......  & 1.592 &  2.287 &  428    & No (CTIO)   &  ....   
      &  \nl 
  22......  & 1.278 &  1.797 &  352    & ? (CTIO 4m)   &  .... 
      &  ML parallax or rare SN?  \nl 
  23......  & 1.452 &  1.855 &  263    & No (MACHO)  &  ....   
      &  \nl 
 *24\tablenotemark{a}......  & 3.941 &  3.544 &  -218   & Yes (MACHO) &  ....   
      &  \nl 
  25......  & 0.734 &  1.243 &  232    & No (CTIO)   &  ....  
      &  \nl 
 *26\tablenotemark{a}......  & 1.278 &  1.222 &  -64.0  & No (CTIO)   &  ....   
      &  SN Type Ia or exotic ML? \nl 
 *27......  & 1.218 &  1.302 &  160    & No (CTIO)   &  ....   
      &  \nl 
\enddata
\tablenotetext{a} {Blended microlensing fits.}
\tablenotetext{b} {Probably supernova.}
\tablenotetext{} {
This table summarizes the available information for each event,
and a subjective `yes-or-no' determination of whether a background galaxy
is present, along with the source of the image used in making the determination.
Events marked with ``*" do not pass selection criteria A.}
\end{deluxetable} 
%\end{landscape}

\clearpage
\begin{deluxetable}{rccccc}  % 6 columns. 
\tablecaption{Microlensing Events Used, Efficiency Corrected $\that$,
and Single Event Optical Depths\label{tab-blthat} }
\tablewidth{0pt}
\tablehead{
%% Use a footnote to explain numbering. 
%% The qquads widen the table. 
\colhead{\qquad Event \qquad} & 
\colhead{\qquad $\that$ \qquad } &
\colhead{\qquad $\that_{\rm st}$(A) \qquad } &
\colhead{\qquad $\that_{\rm st}$(B) \qquad } &
\colhead{\qquad $\tau_1/10^{-9}$(A) \qquad } &
\colhead{\qquad $\tau_1/10^{-9}$(B) \qquad } 
}  % end header. 
\startdata

  1......  &   34.2 &  41.9 &  44.5 &  5.0  &  3.8 \nl
  4......  &   45.4 &  55.5 &  59.0 &  5.9  &  4.5 \nl
  5......  &   75.6 &  92.4 &  98.1 &  8.3  &  6.7 \nl
  6......  &   91.6 & 112.0 & 118.9 &  9.7  &  7.9 \nl
  7......  &  102.9 & 125.8 & 133.6 & 10.7  &  8.7 \nl
  8......  &   66.4 &  81.1 &  86.2 &  7.5  &  6.1 \nl
 *9......  &  143.4 & ..... & 143.4 & ....  &  9.3 \nl
 13......  &  100.1 & 122.4 & 130.0 & 10.5  &  8.5 \nl
 14......  &  100.1 & 122.4 & 130.0 & 10.5  &  8.5 \nl
 15......  &   36.8 &  45.0 &  47.7 &  5.2  &  4.0 \nl
 18......  &   74.2 &  90.7 &  96.4 &  8.2  &  6.6 \nl
*20......  &   72.7 & ..... &  94.3 & ....  &  6.5 \nl
 21......  &   93.2 & 113.9 & 121.0 &  9.9  &  8.0 \nl
*22......  &  229.3 & ..... & 297.8 & ....  & 20.0 \nl
 23......  &   85.2 & 104.2 & 110.7 &  9.1  &  7.4 \nl
 25......  &   85.2 & 104.2 & 110.7 &  9.1  &  7.4 \nl
*27......  &   50.5 & ..... &  65.6 & ....  &  4.9 \nl

\enddata
\tablenotetext{} {The quantity $\that_{\rm st}$ is the average actual event 
timescale for events in our Monte Carlo calculations which are detected 
with an unblended fit timescale of $\that$.  For the binary event 9,
the blended binary fit $\that$ value is used.  The quantity $\tau_1$ is the
contribution of each event to the total microlensing optical depth,
computed using equation~(\ref{eq-tau1}).  Columns are marked (A) or
(B) to indicate which selection criteria was used in the efficiencies.}
\end{deluxetable} 

\clearpage
\begin{deluxetable}{lccc} 
\tablecaption{Total Model Independent Optical Depths \label{tab-tau} }
\tablewidth{0pt}
\tablehead{
%% Use a footnote to explain numbering. 
\colhead{Criteria \& $\that$} &
\colhead{\# of events} & \colhead{measured $\tau(10^7)$} &
\colhead{$\tilde{\theta}(degrees)$}
}  % end header. 
\startdata

A $\that$-stat......................   & 13 & $1.10 ^{+0.4}_{-0.3}$ &
                                              $1.94\,\pm\,0.29 $ \nl
A $\that$-fit.........................  & 13 & $1.14 ^{+0.4}_{-0.3}$ & .... \nl
A no $\that$-correction........        & 13 & $0.94 $ & .... \nl\nl

B $\that$-stat......................   & 17 & $1.29 ^{+0.4}_{-0.3}$ &
                                              $1.86\,\pm\,0.23 $ \nl
B $\that$-fit.........................  & 17 & $1.24 ^{+0.4}_{-0.3}$ & .... \nl
B no $\that$-correction........        & 17 & $1.08 $ & .... \nl\nl

\enddata
\tablenotetext{} {The table entries 
show the microlensing optical depth $\tau$ in units of $10^{-7}$, 
for the two selection criteria A \& B using
different $\that$ corrections (due to blending).
The quoted errors are $1\sigma$ standard errors computed as
described in A97.
The statistical $\that$ correction is preferred because it is unbiased.
The `no $\that$ correction' values are for comparison only.}
\end{deluxetable} 
\clearpage

\begin{deluxetable}{ccccccccc} 
\tablecaption{Optical Depth Confidence Intervals \label{tab-taucl} }
\tablewidth{0pt}
\tablehead{
%% Use a footnote to explain numbering. 
\colhead{Selection Criteria} & \colhead{\# of events} & 
  \multicolumn{7}{c}{$\tau (10^{-7})$ for confidence level:} \nl
\multicolumn{2}{c}{} & 
                       \colhead{0.025} & \colhead{0.05} & \colhead{0.16} & 
 \colhead{measured} & \colhead{0.84} & \colhead{0.95} & \colhead{0.975}
}  % end header. 

\startdata

A       &  13 &  0.60 & 0.67 & 0.83 & 1.10 & 1.47 & 1.73 & 1.86\nl
B      & 17 & 0.73 & 0.81 & 0.99 & 1.29 & 1.69 & 1.97 & 2.10\nl
{\bf A B average}  & {\bf -} & {\bf 0.67} & {\bf 0.74} & {\bf 0.91} &
	{\bf 1.20} & {\bf 1.58} & {\bf 1.85} & {\bf 1.98}\nl
A likelihood&  13 &  0.40 & 0.46 & 0.62 & 0.92 & 1.32 & 1.61 & 1.76\nl
B likelihood& 17 & 0.52 & 0.58 & 0.74 & 1.05 & 1.44 & 1.73 & 1.87\nl
\enddata
\tablenotetext{} { The table entries 
show limits at various confidence levels on 
the microlensing optical depth $\tau$ in units of $10^{-7}$ for different
choices of selection criteria and different calculational methods.  
Rows marked A and B are model independent values found using the Monte Carlo 
method described in A97.  
The boldface row is the average of the model independent A and B calculations.
The rows marked ``likelihood" are background subtracted 
(that is, for {\bf halo} microlensing only) and depend upon
the model of the Galaxy and LMC used, in this case model ``S" with
a dark LMC halo included, as described in \S~\ref{sec-like}. 
}
\end{deluxetable}

\begin{deluxetable}{cll}  % 3 columns.
\tablecaption{Optical Depth Error Budget \label{tab-tauerrorbudget} }
\tablewidth{0pt}
\tablehead{
%% Use a footnote to explain numbering.
\colhead{Cause} &
\colhead{Size ($\tau/ 10^{-7}$}) &
\colhead{Relative Size} 
}  % end header.
\startdata

Poisson         	&  0.4  &  30\%  \nl
Exposure/Normalization of S/O &  0.25     &  20\%    \nl
Selection Criteria 	& 0.25      &  20\% 	\nl
$\that$ bias      	&   0.05   &   3\%       \nl
Binary star sources    	&   ?       &   ?       \nl
Exotic Lensing  	&   ?       &   ?       \nl
\enddata
\end{deluxetable}

\begin{deluxetable}{lcccccc}  % 7 columns. 
\tablecaption{ Microlensing by Stars 
\label{tab-stars} 
    \tablenotemark{} }
\tablewidth{0pt}
\tablehead{
%% Use a footnote to explain numbering. 
\colhead{Population } & 
  \colhead{$\tau (10^{-8}) $} & 
  \colhead{$ \VEV{\that}$ (days) } & 
  \colhead{$ \VEV{l} (\kpc) $}  & 
  \colhead{$ \Gamma (10^{-8} {\rm yr^{-1}}) $}  &
  \colhead{$ \Nexp $ (A)} &
  \colhead{$ \Nexp $ (B)}
}  % end header. 
\startdata
Thin disk ................... & 0.36 & 101 & 1.3 & 1.7 & 0.38 & 0.49  \nl
Big thin disk (F) ....... & 0.59 & 101 & 1.3     & 2.7 & 0.60 & 0.79 \nl
Thick disk .................. & 0.20 & 104 & 3.6 & 0.90 & 0.20 & 0.26 \nl
Spheroid ..................... & 0.20 & 129 & 8.8 & 0.90 & 0.19 & 0.25 \nl
%Bulge ................ & $2.3\ten{-4}$ & 38  & 1.8 & $3.6\ten{-4}$ & 
%	$5\ten{-5}$ & $7 \ten{-5}$ \nl
LMC disk(w/ halo) ..... & 1.6 & 120 & 50 & 5.8 & 1.3 & 1.7 \nl
LMC disk(w/o halo).... & 2.6 & 120 & 50 & 9.8 & 2.2 & 2.9 \nl

\enddata
\tablenotetext{} {This table shows microlensing quantities for
 various lens populations, with the density and velocity distributions
 and PDMF described in the text.
 $\tau$ is the optical depth, $\VEV{l}$ is the mean lens distance, and
 $\Gamma$ is the total theoretical microlensing rate, but in all cases 
 excluding bright lenses (see text).
 The expected number of events $\Nexp$ includes our detection
 efficiency averaged over the $\that$ distribution. 
 The LMC values are averaged over the locations of our 30 fields.
 $\Nexp$ is the number of expected events using either selection
 criteria A (13 events), or criteria B (17 events). Two models of the LMC
 are considered, one with a dark halo and one without.  Lensing
 from the LMC stellar disk only is shown in this table, lensing from the
 dark LMC halo is discussed elsewhere.
 }
\end{deluxetable} 

\clearpage

%%% This table is `flipped' 90^\circ relative to year-1. 
%%%  Its easiest to have a separate column for 6/8 events, 
%%%  so that it all ends up on one page. 
%% Doublespace doesn't seem to work, so I've stuck in 2 \nl's
%% between each line to avoid the +-'s colliding. 

\begin{deluxetable}{ccccccccc}  % 9 columns. 
\tablecaption{ Maximum Likelihood Fits\tablenotemark{}
\label{tab-like} }
\tablewidth{0pt}
\tablehead{
%% Use a footnote to explain numbering. 
\colhead{Model/} & 
\colhead{Comment} &
  \colhead{$ m_{\rm ML}$ } & 
   \colhead{$f_{\rm ML} $} & 
   \colhead{$ f_{\rm ML} M_{\rm H} $ }  & 
   \colhead{$\tau_{\rm ML}  $} &
   \colhead{$\Nexp$ } &
   \colhead{$\Nexp$ } &
   \colhead{$\Nexp$ }
   \nl
\colhead{Events } & 
\colhead{} &
  \colhead{$(\msun)$ } & 
   \colhead{ } & 
   \colhead{$(10^{10}\msun)$ }  & 
   \colhead{$(10^{-8}) $} &
   \colhead{MW} &
   \colhead{LMC halo} &
   \colhead{stars}
}  % end tablehead. 

\startdata
S/13 & standard &0.60\pms{0.28}{0.20}	& 0.21\pms{0.10}{0.07}	
	& 8.5\pms{4}{3}	& 10\pms{5}{3} & 9.6 & 0 & 3.0
\nl\nl
S/13 & standard & 0.54\pms{0.26}{0.18}	& 0.20\pms{0.08}{0.06}
	& 7.9\pms{3.4}{2.6}	& 11\pms{5}{4} & 9.4 & 1.1 & 2.1
\nl\nl
S/17 & standard & 0.79\pms{0.32}{0.24} & 0.24\pms{0.09}{0.08}	
	& 10\pms{4}{3}	& 11\pms{4}{4} & 12.7 & 0 & 3.9
\nl\nl
S/17 & standard & 0.72\pms{0.30}{0.20}	& 0.22\pms{0.08}{0.07}	
	& 9.1\pms{3}{3}	& 12\pms{5}{4} & 12.4 & 1.4 & 2.7
\nl\hline\nl
B/13 & big halo & 0.68\pms{0.35}{0.22} & 0.12\pms{0.06}{0.04}	
	& 8.8\pms{4}{3}	& 10\pms{5}{4} & 9.7 & 0 & 3.0 
\nl\nl
B/13 & big halo & 0.66\pms{0.30}{0.22}	& 0.12\pms{0.05}{0.04}	
	& 8.8\pms{4}{3}	& 11\pms{5}{4} & 9.8 & 0.62 & 2.1
\nl\nl
B/17 & big halo & 0.92\pms{0.40}{0.28} & 0.14\pms{0.06}{0.04}	
	& 10\pms{4}{3} 	& 11\pms{5}{4} & 12.5 & 0 & 3.9
\nl\nl
B/17 & big halo & 0.87\pms{0.35}{0.26}	& 0.14\pms{0.05}{0.04}	
	& 10\pms{4}{3} 	& 12\pms{5}{4} & 12.9 & 0.78 & 2.7
\nl\hline\nl
F/13 & small halo & 0.16\pms{0.08}{0.05} & 0.50\pms{0.22}{0.18}
	& 10\pms{4}{4} & 10\pms{4}{3} & 9.5 & 0 & 3.2
\nl\nl
F/13 & small halo & 0.19\pms{0.09}{0.06} & 0.39\pms{0.17}{0.13}
	& 8.0\pms{3}{3} & 11\pms{4}{3} & 7.0 & 3.3 & 2.3
\nl\nl
F/17 & small halo & 0.22\pms{0.09}{0.06} & 0.57\pms{0.21}{0.17}
	& 11\pms{4}{4} & 11\pms{4}{3} & 12.5 & 0 & 4.2
\nl\nl
F/17 & small halo & 0.25\pms{0.10}{0.07} & 0.44\pms{0.16}{0.13}
	& 9.0\pms{3}{3} & 11\pms{4}{4} & 9.2 & 4.3 & 3.0
\nl\nl

\enddata
\tablenotetext{} { The first column shows the model as defined
in \yrone\ and \yrtwo, and the number of 
 microlensing candidates used; either 13 from selection criteria A, 
 or 17 from criteria B.
 Model S is given by eq.~\ref{eq-stdhalo} and has a typical size
 halo, Model B has a halo as large as possible, and Model F has a halo as
 small as possible with a large thin disk.
 Columns 3 \& 4 show the maximum likelihood MACHO mass
 and halo fraction from Section~\ref{sec-like}.  
 Columns 5 and 6 show the implied total mass of MACHOs 
 within $50 \kpc$ of the Galactic center, 
 and the resulting halo optical depth.  For models with dark LMC halos,
 the sum of LMC and Milky Way MACHO optical depth is shown.
 Column 7 shows the number of expected events from the Milky Way Halo,
 column 8 shows the number of expected events from the LMC halo,
 and column 9 shows the expected number of events from stars 
 (from Table~\ref{tab-stars}). 
 Every Milky Way model is shown twice, once with a dark LMC halo, and once with
 the dark LMC halo set to zero.  See the text for more explanation.  }
\end{deluxetable}

\clearpage

%%%%%%%%%  FFFFFFFFFFFFFFF   Figures.... 
%% NB: due to a Latex bug, put a \protect before \ref{xxx} 
%%  within \caption{}

%% See below for use of \plotone, \plotfiddle etc. 
%% This is currently using dummy files blankl.ps, blankp.ps 
%%  (landscape & portrait) to check the page layout without generating
%%  megabytes of junk every time. 

\twocolumn  % Do 2-col mode to get more figs / page. 

\onecolumn
\vspace{-2.5cm}
\begin{figure} 
%\plotone{fig_lmc.ps}
\caption{An R-band image of the LMC, $8.2$ degrees on a side
(G. Bothun, private communication), showing the locations of the
30 MACHO fields used here.  Also shown are the locations of the 17
microlensing candidates discussed in the text.
\label{fig-lmc} } 
\end{figure} 

\onecolumn

\begin{figure}
\plotfiddle{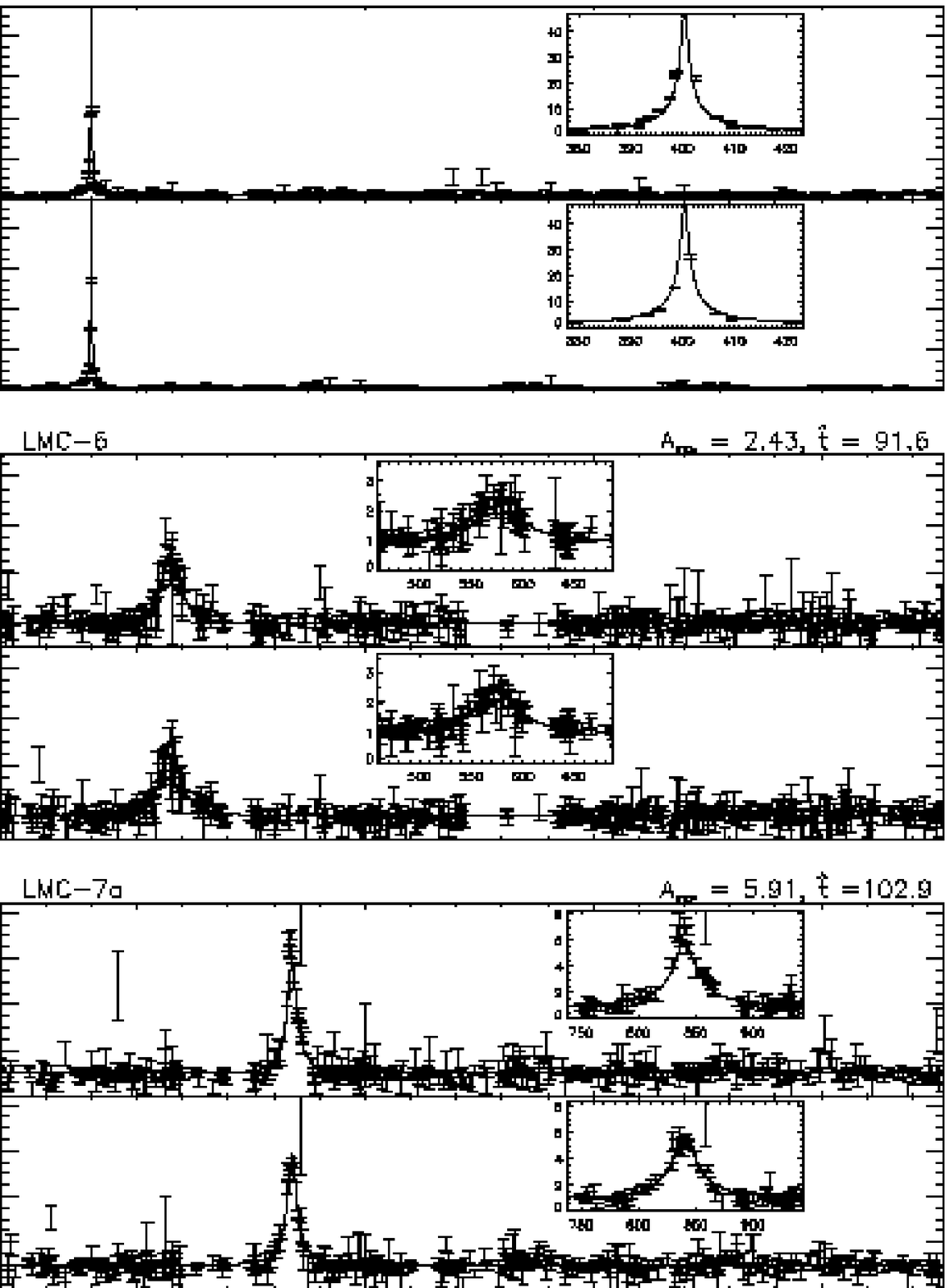}{8.0truein}{0}{100}{100}{-150}{50} 
%\plotone{fig_event_b.eps}
%% Dont use \caption as it advances numbers. 
%%  Blank line above to center.
Figure~\protect\ref{fig-events} (continued)
\end{figure}

\begin{figure}
\plotfiddle{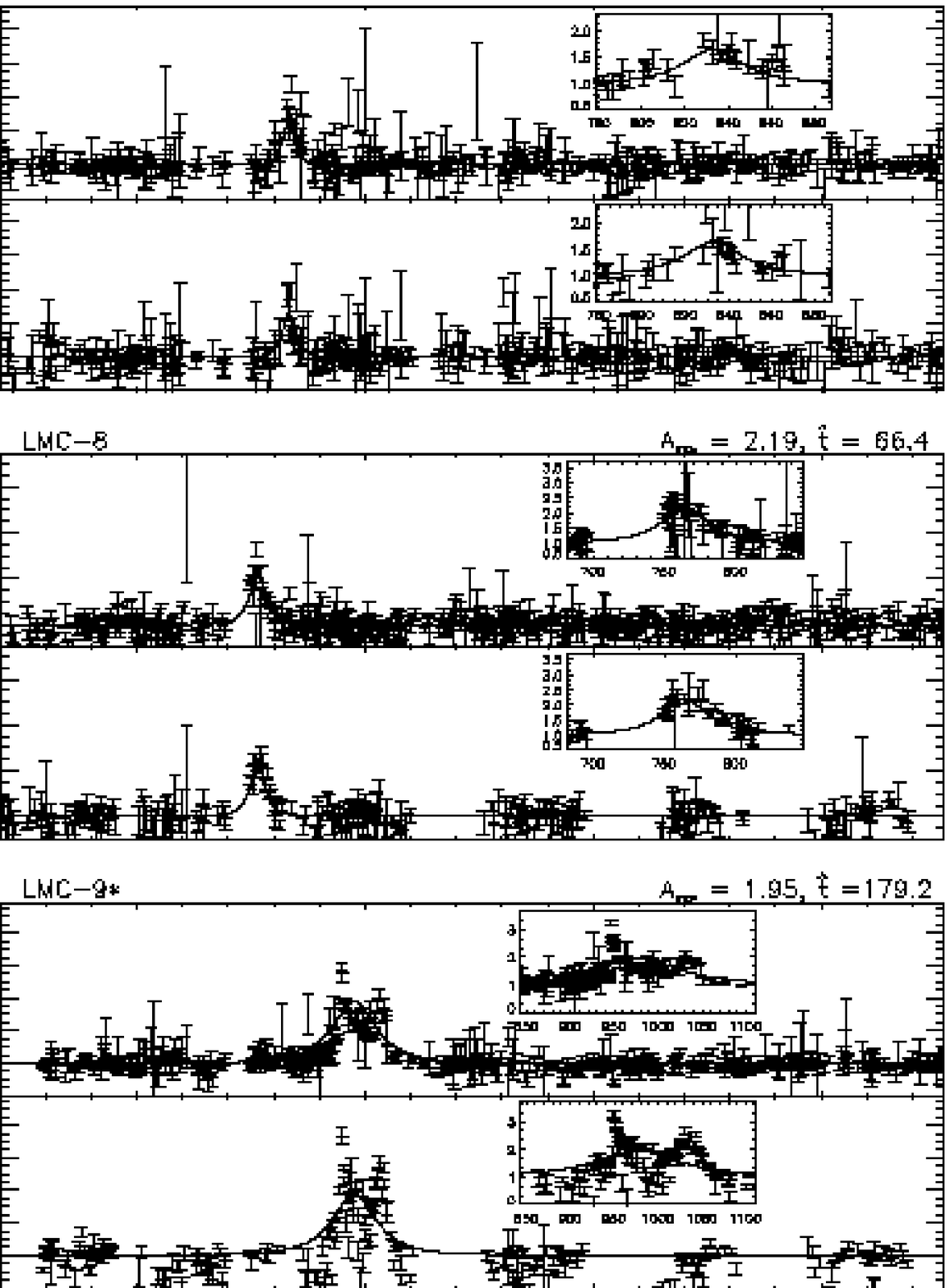}{8.0truein}{0}{100}{100}{-150}{50} 
%\plotone{fig_event_c.eps}
Figure~\protect\ref{fig-events} (continued)
\end{figure}

\begin{figure}
\plotfiddle{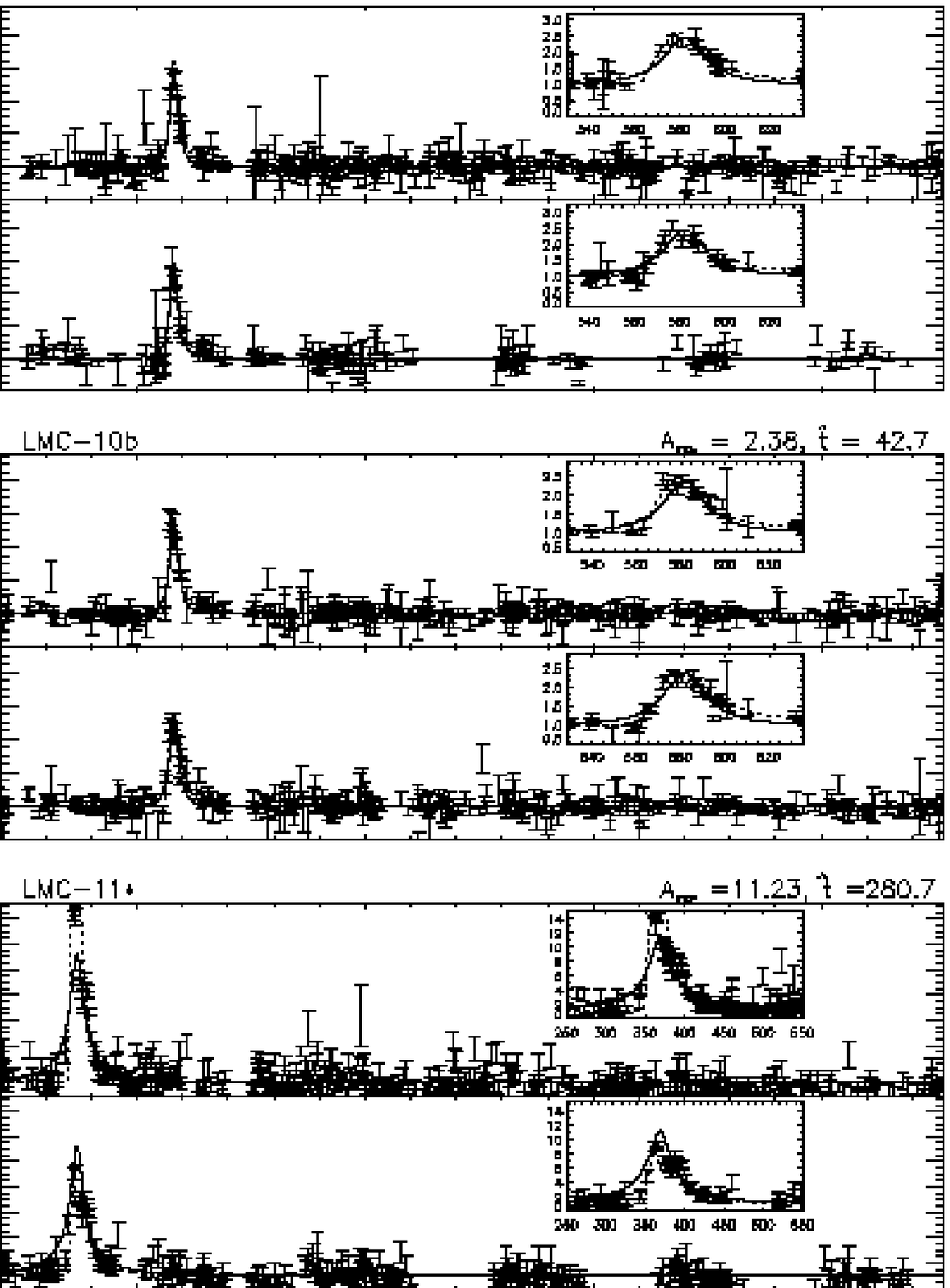}{8.0truein}{0}{100}{100}{-150}{50} 
%\plotone{fig_event_d.eps}
Figure~\protect\ref{fig-events} (continued)
\end{figure}

\begin{figure}
\plotfiddle{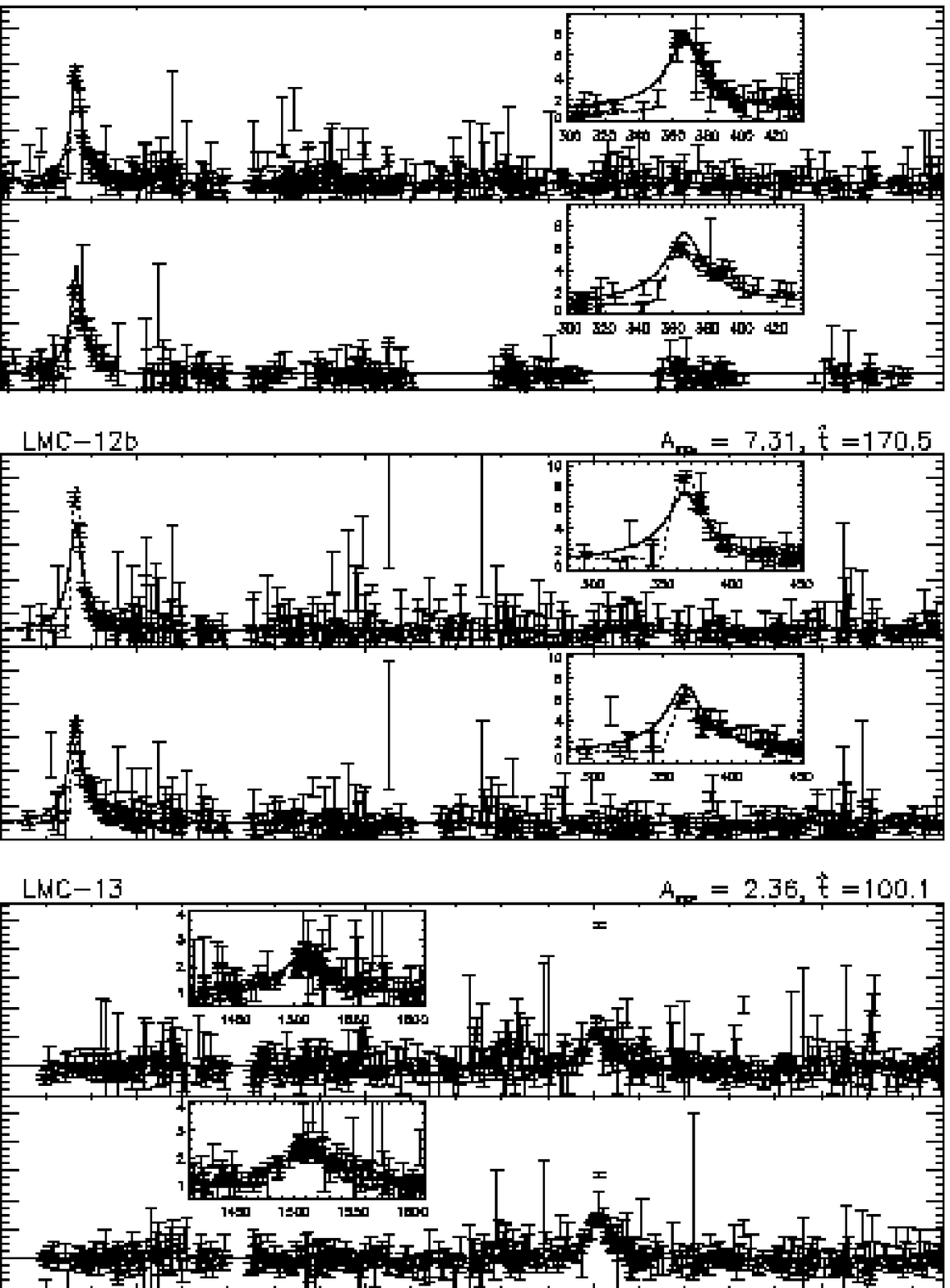}{8.0truein}{0}{100}{100}{-150}{50} 
%\plotone{fig_event_e.eps}
Figure~\protect\ref{fig-events} (continued)
\end{figure}

\begin{figure}
\plotfiddle{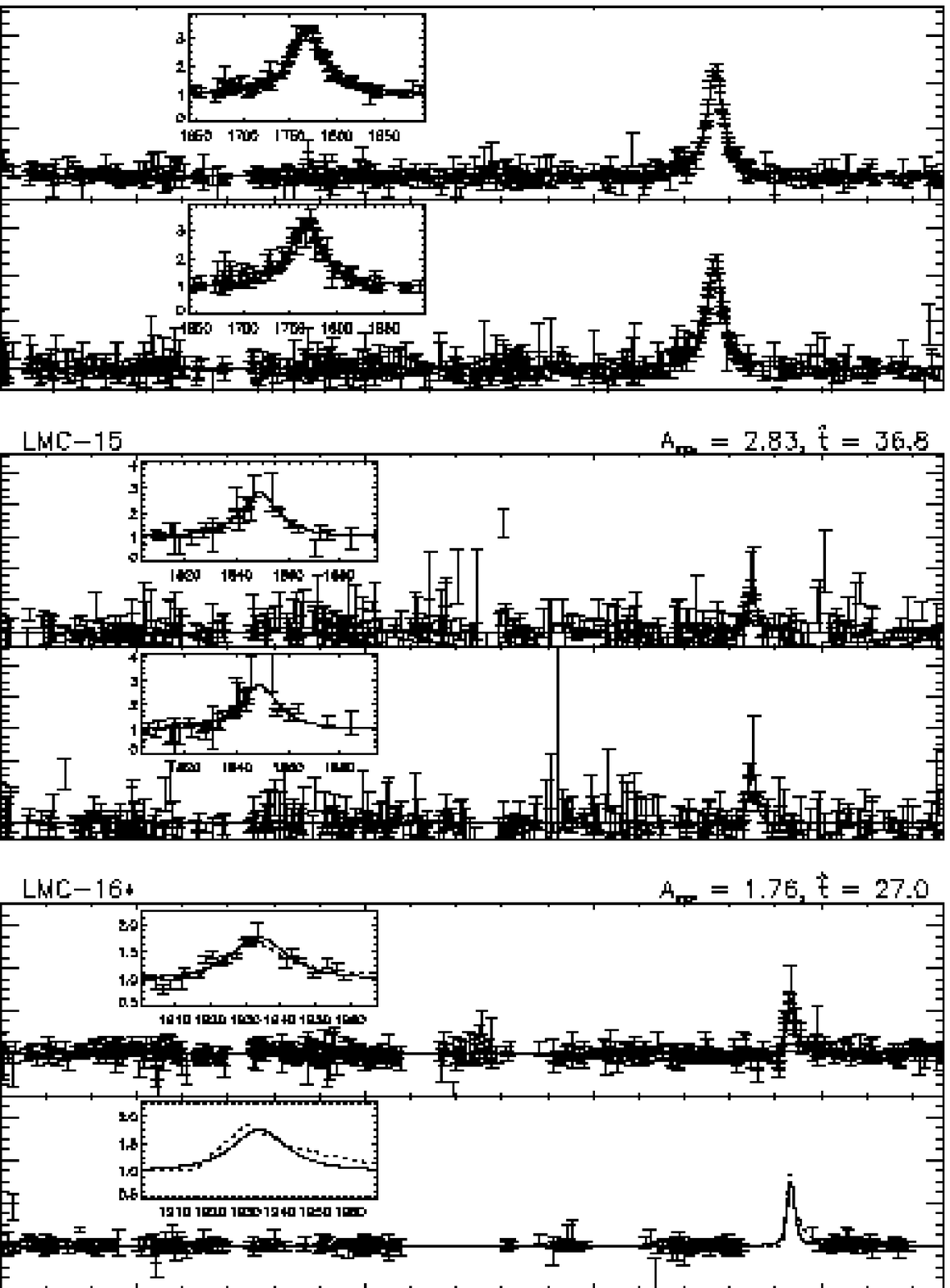}{8.0truein}{0}{100}{100}{-150}{50} 
%\plotone{fig_event_f.eps}
Figure~\protect\ref{fig-events} (continued)
\end{figure}

\begin{figure}
\plotfiddle{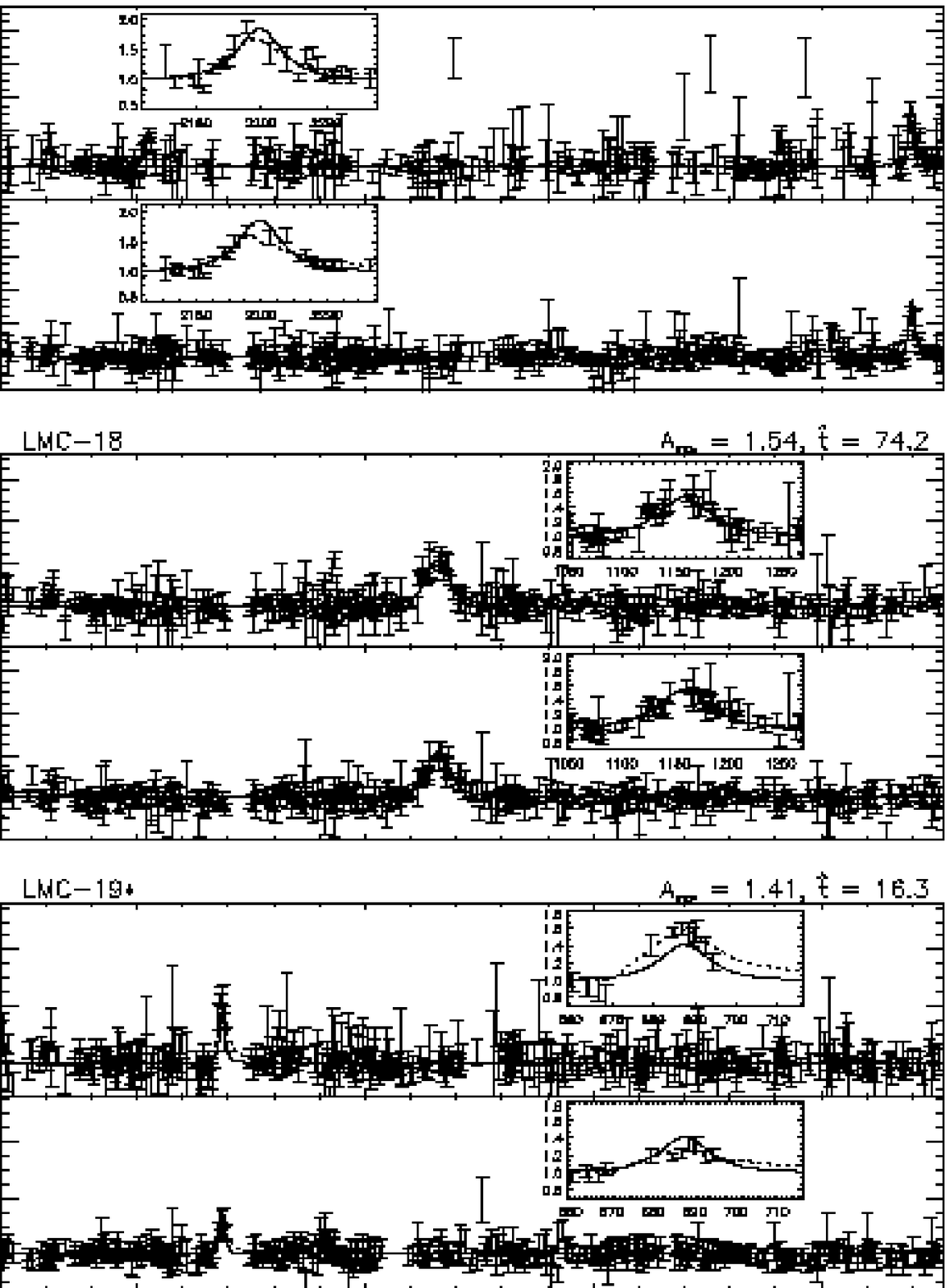}{8.0truein}{0}{100}{100}{-150}{50} 
%\plotone{fig_event_g.eps}
Figure~\protect\ref{fig-events} (continued)
\end{figure}

\begin{figure}
\plotfiddle{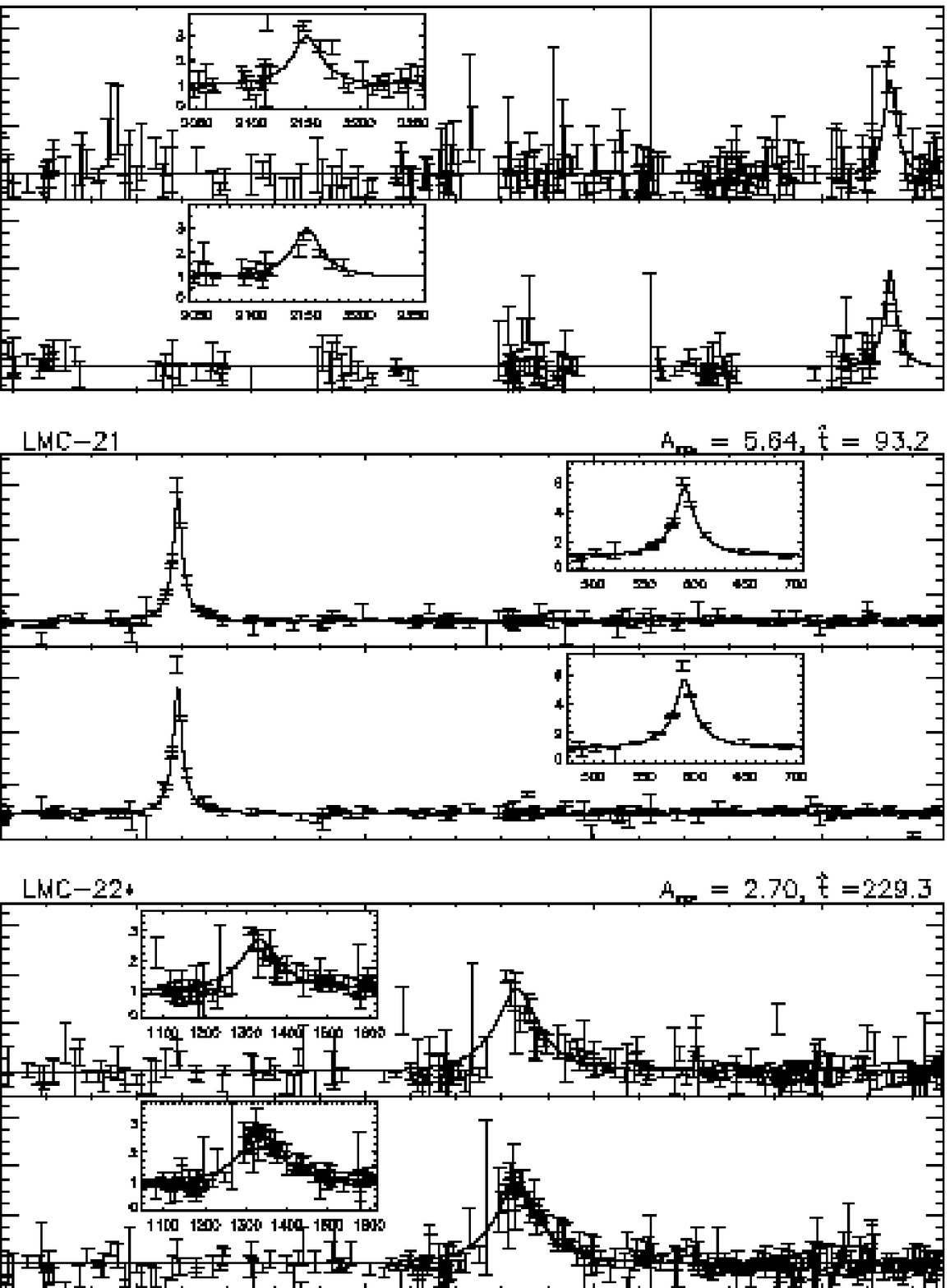}{8.0truein}{0}{100}{100}{-150}{50} 
%\plotone{fig_event_h.eps}
Figure~\protect\ref{fig-events} (continued)
\end{figure}

\begin{figure}
\plotfiddle{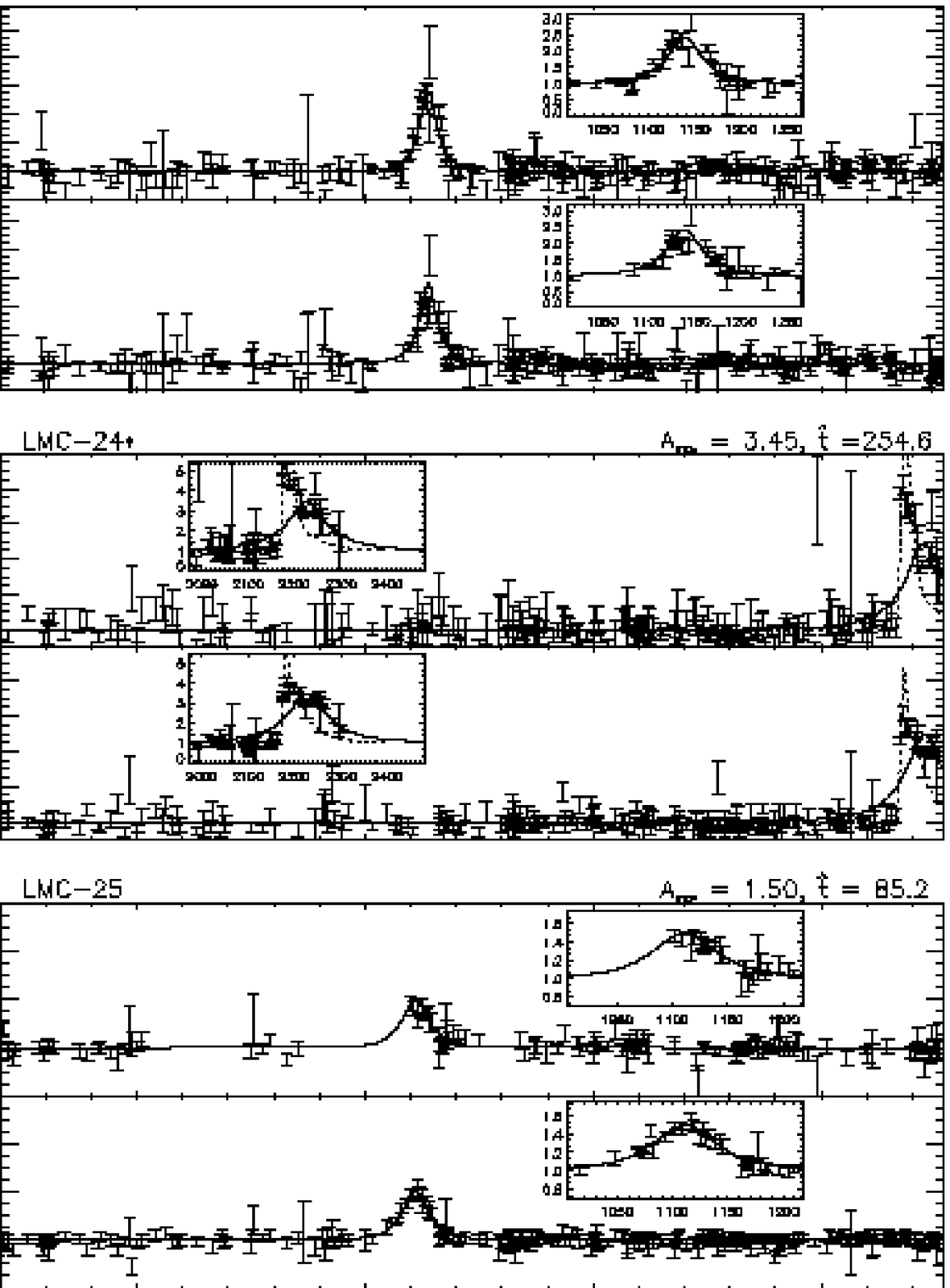}{8.0truein}{0}{100}{100}{-150}{50} 
%\plotone{fig_event_i.eps}
Figure~\protect\ref{fig-events} (continued)
\end{figure}

\begin{figure}
\plotfiddle{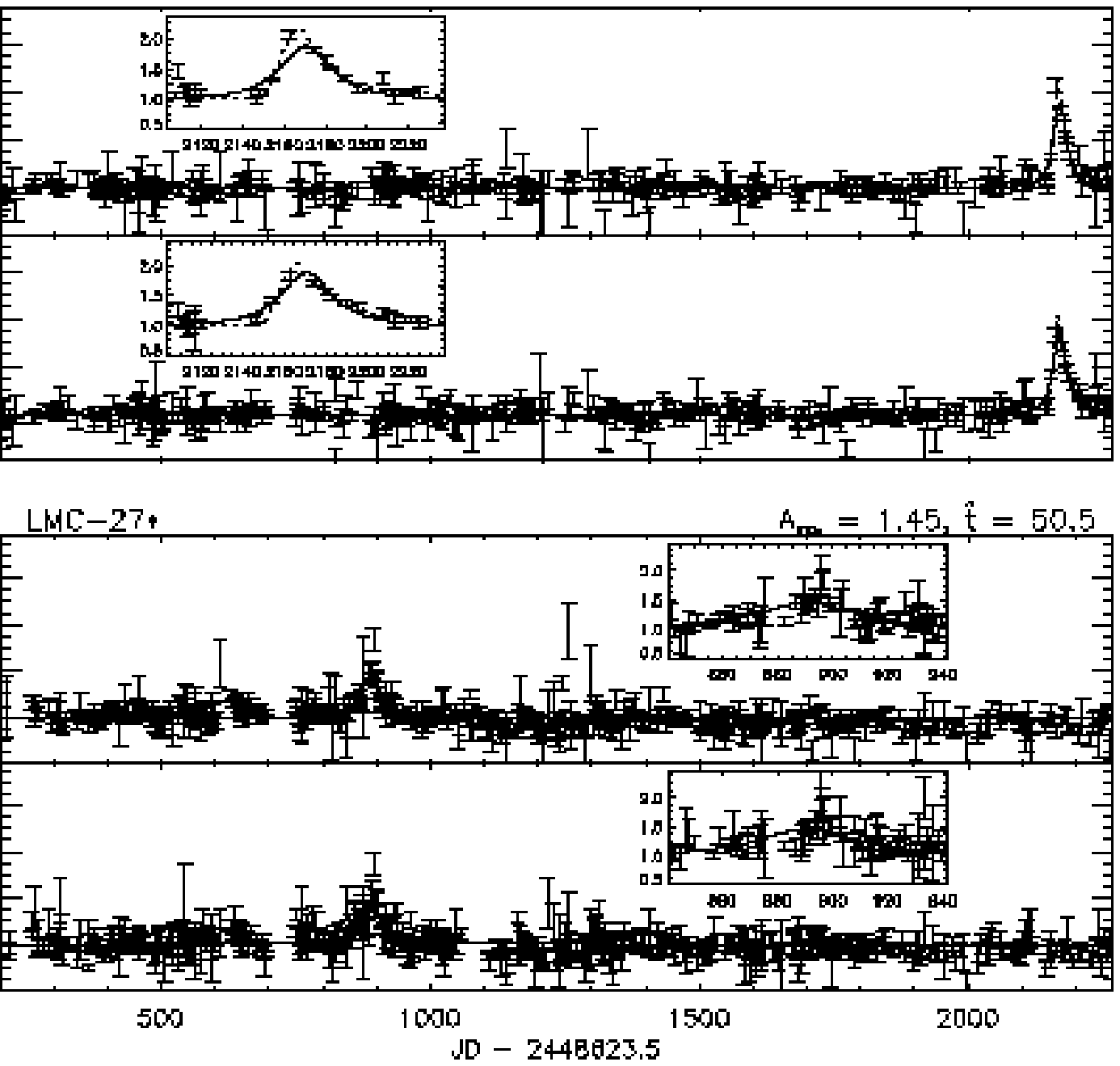}{8.0truein}{0}{100}{100}{-150}{50} 
%\plotone{fig_event_j.eps}
Figure~\protect\ref{fig-events} (continued)
\end{figure}

%% Back to side-by-side for portrait figs. 
\clearpage 
\twocolumn

%\begin{figure}
%\plotone{fig_lumin_dist.ps}
%\caption{The upper (lower) panel shows the cumulative distribution 
%in object magnitude $V_{\rm obj}$ for set A (set B). 
%The bin size is 0.5 magnitudes.
%The expected distribution is shown
%as a \textit{dashed line}: the LMC luminosity function modified by
%our efficiency.
%The results of K-S tests comparing the observed and expected distributions
%are also shown.
%\label{fig-lumin} } 
%\end{figure}

\clearpage 
\onecolumn

\clearpage 
\twocolumn

%\begin{figure}
%\plotone{fig_centroids_cumulative.ps} 
%\caption{
%The upper (lower) panel shows the cumulative distribution 
%in centroid offsets for set A (set B).  The expected distribution is shown
%as a \textit{dashed} line as determined using the Monte Carlo described in
%\S~\ref{sec-eff}.  The results of K-S tests comparing the observed 
%and expected distributions are also shown.
%\label{fig-centroids} }
%\end{figure}

\end{document}